\documentclass[12pt,a4paper]{article}
\usepackage{amsmath}
\usepackage{graphicx}
\usepackage{amsfonts}
\usepackage{amssymb}
\usepackage{epsfig}
\usepackage{psfrag}
\usepackage{wrapfig}
\usepackage{color}

\definecolor{yellow}{rgb}{1,0.9,0} % color values Red, Green, Blue
\definecolor{wine}{rgb}{0.5,0,0.4}
\definecolor{lightblue}{rgb}{0.3,0,0.7}
\definecolor{ancientrose}{rgb}{0.7,0,0.4}
\definecolor{cream}{rgb}{1.,1.,0.7}
\definecolor{violet}{rgb}{1.,0.9,0.95}
\definecolor{lightgreen}{rgb}{0.8,1,0.8}
\definecolor{darkgreen}{rgb}{0,0.6,0}
\newcommand{\be}{\begin{equation}}
\newcommand{\ee}{\end{equation}}
\newcommand{\beq}{\begin{eqnarray}}
\newcommand{\eeq}{\end{eqnarray}}

\def\nue{\mathrel{{\nu_e}}}
\def\numu{\mathrel{{\nu_\mu}}}
\def\nutau{\mathrel{{\nu_\tau}}}
\def\nux{\mathrel{{\nu_x}}}

\def\barnue{\mathrel{{\bar \nu}_e}}
\def\barnumu{\mathrel{{\bar \nu}_\mu}}
\def\barnutau{\mathrel{{\bar \nu}_\tau}}
\def\barnux{\mathrel{{\bar \nu}_x}}

\def \lta {\mathrel{\vcenter{\hbox{$<$}\nointerlineskip\hbox{$\sim$}}}}
\def \gta {\mathrel{\vcenter{\hbox{$>$}\nointerlineskip\hbox{$\sim$}}}}

\def\t13{\mathrel{{\sin^2 \theta_{13}}}}
\def\y12{\mathrel{{\tan^2 \theta_{12}}}}

\begin{document}

\begin{titlepage}

%\vspace*{0.2cm}
\hskip 11truecm 
{NSF-KITP-03-12}

\vspace*{1.2cm}
\begin{center}

{\Large \bf Probing the neutrino mass hierarchy and the 13-mixing with supernovae} \\

\vspace{0.8cm}
{\large
Cecilia Lunardini$^{a,b,1}$ and
Alexei Yu. Smirnov$^{c,d,2}$ } \\

\vspace{0.2cm}
{\em $^{a}$ Institute for Advanced Study, Einstein drive, 08540 Princeton,
New Jersey, USA}

\vspace{0.2cm}
{\em $^{b}$Kavli Institute for Theoretical Physics, University of California Santa Barbara, California, 93106-4030, USA}

\vspace{0.2cm}
{\em $^{c}$ The Abdus Salam ICTP, Strada Costiera 11, 34100 Trieste, Italy}

\vspace{0.2cm}
{\em $^d$ Institute for Nuclear Research, RAS, Moscow 123182, Russia.}

\end{center}
\begin{abstract}
\noindent We consider in details  the effects of the 13-mixing
($\sin^2\theta_{13}$) and of the type of  mass hierarchy/ordering
(sign[$\Delta m^2_{13}$]) on neutrino signals from the
gravitational collapses of stars. The observables 
(characteristics of the  energy spectra of $\nu_e$ and
$\bar{\nu}_e$ events) sensitive to $\sin^2\theta_{13}$ and
sign[$\Delta m^2_{13}$] have been calculated. They include  the
ratio of average energies of the spectra, $r_E \equiv \langle E
\rangle / \langle \bar{E}\rangle $, the ratio of widths of the
energy distributions, $r_{\Gamma} \equiv \Gamma/ \bar{\Gamma}$,
the ratios of total numbers of $\nu_e$ and $\bar{\nu}_e$ events at
low energies, $S$, and in the high energy tails, $R_{tail}$. We
construct and analyze scatter plots which show the  predictions
for the observables for different intervals of $\sin^2\theta_{13}$
and signs of $\Delta m^2_{13}$, taking into account  uncertainties
in the original neutrino spectra,  the star density profile, etc..
Regions in the space of observables $r_E$, $r_{\Gamma}$, $S$,
$R_{tail}$ exist in which certain mass hierarchy and intervals of  
$\sin^2\theta_{13}$ can be identified or discriminated. We elaborate
on  the method of the high energy tails in the spectra of events.
The conditions are formulated for which $\sin^2 \theta_{13}$ can be
(i) measured, (ii) restricted from below, (iii) restricted from
above.  We comment on the possibility to determine
$\sin^2\theta_{13}$ using  the time dependence of the signals due
to the propagation of the shock wave  through the resonance
layers of the star.   We show that the appearance of the {\it delayed} Earth 
matter effect in one of the channels
($\nu_e$ or $\bar\nu_e$) in combination with the undelayed effect in the other channel 
will allow to identify the shock wave appeareance and determine the mass hierarchy.

\end{abstract}



\vfil
\noindent
\footnoterule
{\small
%\noindent
$^{1}$ E-mail: lunardi@ias.edu\vskip-1pt\noindent}
{\small $^2$ E-mail: smirnov@ictp.trieste.it\vskip-1pt\noindent}

\thispagestyle{empty}
\end{titlepage}

\setcounter{page}{1}

%%%%%%%%%%%%% TEXT  %%%%%%%%%%%%%%%%%%%%%%%%%%%%%%%%%%%%%%%%%%%%%%%

%%%%%%%%%%%%%%%%%%%%%%%%%%%%%%%%%%%%%%%%%%%%%%%%%%%%%%%%%%%%%%%%%%%%%%%%%%
%%%%%%%%%%%%%%%%%%%%%%%%%%%%%%%%%%%%%%%%%%%%%%%%%%%%%%%%%%%%%%%%%%%%%%%%%%
\section{Introduction}
\label{sec:1}
%%%%%%%%%%%%%%%%%%%%%%%%%%%%%%%%%%%%%%%%%%%%%%%%%%%%%%%%%%%%%%%%%%%%%%%%%%

Collapsing stars (some of them appearing as supernova explosions) are the 
sources of neutrinos of different flavors  which can be used
for oscillation/conversion  experiments~\cite{Raffelt:2002tu}.
The structure of the neutrino mass spectrum and lepton mixing is imprinted
into the detected  signal   (see \cite{Mikheev:1986if}--\cite{Takahashi:2002cm} as an incomplete list of relevant works).  Therefore,  in principle, studying the
properties of a supernova neutrino burst one can get information
about~

\begin{itemize}

\item
the values of parameters relevant for the  solution of the solar neutrino problem,

\item
the type of the mass hierarchy/ordering,

\item
the 13-mixing parameter $\sin^2\theta_{13}$,

\item
the presence of sterile neutrinos,

\item
new neutrino interactions.

\end{itemize}

The first KamLAND results~\cite{Eguchi:2002dm} confirmed the LMA MSW as the dominant mechanism
of the solar neutrino conversion.
Further KamLAND measurements and solar neutrino studies
 will determine the corresponding oscillation parameters
with rather good accuracy ~\cite{Barger:2000hy,Barbieri:2000sv,Murayama:2000iq}. This confirmation implies, in particular, that already in
1987, with the detection of neutrinos from the supernova SN1987A,
significant conversion effects were observed on supernova neutrinos \cite{Arafune:1987kc}--\cite{Kachelriess:2001sg}.
The identification of the neutrino mass hierarchy and the determination of 13-mixing ($\sin^2 \theta_{13}$) have
become   the main issues of further studies.
Searches for sterile neutrinos and new neutrino interactions
are  also on the  agenda.

In this paper we will concentrate on the related subjects of the mass
hierarchy and $\sin^2 \theta_{13}$. We will consider a three neutrinos system,
assuming that sterile neutrinos, if they exist, produce negligible effect.\\

  The possibilities to study neutrino conversion effects using
supernovae are wide, but not exempt of problems. The main difficulty
originates from the fact that oscillation effects are proportional to
the difference of the electron and non-electron neutrino fluxes
originally produced inside the star, which are poorly known at the
moment.  The features of these fluxes depend on many details of the
neutrino transport inside the star and, in general, on the type of
progenitor star.  \\

\noindent
There are two approaches to resolve the problem:
\\

\noindent
1). perform a global fit of the data, determining both oscillation
parameters and the parameters of the original fluxes simultaneously.
However, the number of unknown parameters which describe the energy spectra of the
emitted neutrinos (temperatures, luminosities, pinching parameters) is
rather large, and moreover, these quantities change with time during the
burst.

Also degeneracies of parameters exist, so that
variations of the oscillation  parameters and of the parameters of the fluxes
can produce the same observable effect.
\\

\noindent
2). perform a (supernova) model independent analysis relying on some generic
(and model independent) qualitative features of the fluxes. Namely:

\noindent
- the inequality of average energies (temperatures) of the original fluxes of neutrinos of different flavors;

\noindent
- the dominance of the electron neutrino flux in the initial phase of the burst (neutronization peak)

\noindent
- the pinching of the energy spectra

\noindent
- the approximate equality of the original $\nu_{\mu}$,
$\bar{\nu}_{\mu}$, $\nu_{\tau}$, $\bar{\nu}_{\tau}$ fluxes.
\\

  The first approach has been used recently in~\cite{Barger:2001yx,Minakata:2001cd}, while the study   of ratios of numbers of events in specific energy intervals has been suggested in \cite{Takahashi:2001ep,Takahashi:2002cm}.

 In this paper we elaborate  the second type of approach, suggesting specific methods and taking into account all the possible uncertainties. Both analytical and numerical analyses are performed.  The paper is
organized as follows. In sect. 2 we  summarize our
knowledge of  the original neutrino spectra and of the  matter distribution along the
trajectory of the neutrinos in the star. We also  introduce the scheme of
neutrino masses and mixings.  In sect. 3. the dependence
of the conversion probabilities of supernova neutrinos on the mass hierarchy and the
13-mixing is considered. In sect. 4 we  introduce  observables
which are sensitive to $\sin^2\theta_{13}$ and sign[$\Delta m^2_{13}$].
In sect. 5 we  calculate these observables for normal and inverted hierarchy and
different intervals of $\sin^2\theta_{13}$,  identifying the  regions in the space of these observables
 in which the mass hierarchy and $\sin^2\theta_{13}$  can be determined.
In sect. 6 we elaborate on  the  method of  the high energy tails of the
spectra produced in the detectors by neutrinos and antineutrinos.
Sec. 7  is devoted to the study of the  possibility to restrict $\sin^2\theta_{13}$ using the time
variations  of signals induced by the shock wave propagation. Conclusions are given in sect. 8.

%%%%%%%%%%%%%%%%%%%%%%%%%%%%%%%%%%%%%%%%%%%%%%%%%%%%%%%%%%%%%%%%%%%%%%%%%%
%%%%%%%%%%%%%%%%%%%%%%%%%%%%%%%%%%%%%%%%%%%%%%%%%%%%%%%%%%%%%%%%%%%%%%%%%%
\section{Fluxes, density profile, neutrino mass spectrum}
\label{sec:2}
%%%%%%%%%%%%%%%%%%%%%%%%%%%%%%%%%%%%%%%%%%%%%%%%%%%%%%%%%%%%%%%%%%%%%%%%%%

%%%%%%%%%%%%%%%%%%%%%%%%%%%%%%%%%%%%%%%%%%%%%%%%%%%%%%%%%%%%%%%%%%%%%%%%%%
\subsection{Properties of supernova neutrino fluxes}
\label{sec:2.1}

In this section we  summarize our present knowledge of the
neutrino fluxes from the collapsing stars.

At a given time $t$ from the core collapse the original flux  of the neutrinos
of a given flavor, $\nu_\alpha$, can be described by a ``pinched''
Fermi-Dirac (F-D) spectrum,
\begin{eqnarray}
F^0_\alpha(E,T_\alpha,\eta_\alpha,L_\alpha,D) =
\frac{L_\alpha}{4\pi D^2 T^4_\alpha F_3(\eta_\alpha)} \frac{E^2}{e^{E/
T_{\alpha}-\eta_\alpha}+1}~,
\label{eq3}
\end{eqnarray}
where $D$ is the distance to the supernova (typically $D\sim 10$ kpc for a
galactic supernova), $E$ is the energy of the neutrinos, $L_\alpha$ is the
luminosity of the flavor $\nu_\alpha$, and $T_\alpha$ represents the
effective temperature of the $\nu_\alpha$ gas inside the neutrinosphere.
%%%%
Supernova simulations provide the indicative values of the average energies \cite{Keil:2002in}: 
\be
\langle E_{\bar e} \rangle = (14 - 22)~  {\rm MeV}, ~~~ \langle E_{x} \rangle/\langle E_{\bar e}\rangle = (1.1 - 1.6), ~~~
 \langle E_{e} \rangle/\langle E_{\bar e}\rangle = (0.5 - 0.8),~~~
\label{temp}
\ee
and the typical value  of the (time-integrated) luminosity in each flavor:
$L_\alpha \sim (1 - 5) \cdot 10^{52}~{\rm  ergs } $.  
The intervals in eq. (\ref{temp}) include the results of recent Monte-Carlo simulations \cite{Keil:2002in}, in which a difference between $\langle E_{x} \rangle$ and $\langle E_{\bar e}\rangle$  of about $10\%$ is given as typical, though differences as small as few per cent are not excluded.  We will briefly comment on this latter case in the discussion of our results. 
The $\numu$ and $\nutau$  ($\barnumu$ and $\barnutau$) spectra are equal with good approximation,
and therefore the two species  can be treated as a single one, $\nux$ ($\barnux$).  Conversely,
small differences exist between  the energy spectra of the non-electron neutrinos and
antineutrinos, $\nux$  and $\barnux$,  due to effects of weak magnetism, \cite{Horowitz:2001xf}.
In particular, the difference  
\be
\delta_x \equiv T_{\bar x} - T_{x} =  0.4-0.5~~{\rm MeV}
\label{deltax}
\ee
is found to be valid with $\sim 20\%$ accuracy  over a large variety of physical situations
\cite{Horowitz:2001xf}.   The luminosities of all neutrino species are expected to
be approximately equal, within a factor of two or so
\cite{Raffelt:2002tu,Raffelt:1996wa}:
\be
{L_e}/{L_x} = (0.5 - 2), ~~~L_{\bar e}/L_{\bar x} =  (0.5 - 2), ~~~L_{\bar x} \simeq L_x.
\label{fluxes}
\ee
%%%%
%
The pinching parameter $\eta_\alpha$ takes the values
\be
\eta_e \sim 0 - 3,~~~~\eta_{\bar e} \sim 0 - 3~~~~\eta_\mu = \eta_\tau \sim 0 - 2.
\label{pinch}
\ee 
  Notice that the $\nu_e$ and $\barnue$ spectra may have stronger pinching than the other flavors
\cite{Janka,Raffelt:1996wa}.

In eq. (\ref{eq3}) the normalization factor  $F_3(\eta_\alpha)$ equals:
\begin{equation}
F_3(\eta_\alpha)\equiv \int_0^{\infty} \frac{x^3}{e^{x-\eta_\alpha}+1} dx~.
\label{f3}
\end{equation}
In the absence of pinching, $\eta_\alpha=0$, one gets $F_3(0) =
7\pi^4/120\simeq 5.68$.
The average energy $\langle E_\alpha \rangle$ of the spectrum depends on both
$T_\alpha$ and $\eta_\alpha$; for $\eta_\alpha=0$
we have  $\langle E_\alpha \rangle \simeq 3.15 ~T_\alpha$.

The luminosity, temperature and pinching of the neutrino flux vary
with the time $t$; Time dependence can be different for different
stars~\cite{Keil:2002in}. If these variations occur over time scales larger
than the duration of the burst, $\tau \sim 10$ s, the integrated
$\nu_\alpha$ flux is can be  described  by the expression
(\ref{eq3}) but with smaller $\eta_\alpha$ (integration leads to
widening of spectra).

%%%%%%%%%%%%%%%%%%%%%%%%%%%%%%%%%%%%%%%%%%%%%%%%%%%%%%%%%%%%%%%%%%%%%%%%%%%%%%%%%%
\subsection{Matter profile}
%%%%%%%%%%%%%%%%%%%%%%%%%%%%%%%%%%%%%%%%%%%%%%%%%%%%%%%%%%%%%%%%%%%%%%%%%%%%%%%%%

In our study of neutrino conversions inside the star we use the following
matter density profile of the star:
\begin{eqnarray}
\rho(r)=10^{13}~ C  \left(\frac{10 ~{\rm km}}{r} \right)^3~{\rm g\cdot
cm^{-3}} ~,
\label{eq4}
\end{eqnarray}
with $C\simeq 1 - 15$ \cite{BBB,Notzold:1987vc,Kuo:1988qu,Janka}.  For
$\rho \sim (1 - 10^{5}) ~{\rm g\cdot cm^{-3}}$ expression (\ref{eq4})
provides a good approximation to  the calculated matter distribution during at least
the first few seconds of the neutrino burst emission.
For $\rho \lta 1~{\rm g\cdot cm^{-3}}$ the exact
shape of the profile depends on the details of evolution of the star,
its chemical composition, rotation, etc.. As it was recently pointed
out \cite{Schirato:2002tg}, after $\sim (2 -10)$ seconds from the core
collapse and bounce, the shock-wave propagating inside the star could
reach the regions relevant to neutrino resonant conversion
and modify the observed neutrino signal.

For three active neutrinos the difference of the $\nue$ and $\nux$
potentials in matter depends on the number density of electrons:
$n_{e} = Y_{e} \rho/m_N$, where $m_N$ is the nucleon mass and $Y_{e}$
is the number of electrons per nucleon. The transitions occur mainly
in the isotopically neutral region where $Y_{e} = 1/2$ with rather
good precision.

%%%%%%%%%%%%%%%%%%%%%%%%%%%%%%%%%%%%%%%%%%%%%%%%%%%%%%%%%%%%%%%%%%%%%%%%%%
\subsection{Neutrino mixing and mass spectra}
\label{sec:2.2}
%%%%%%%%%%%%%%%%%%%%%%%%%%%%%%%%%%%%%%%%%%%%%%%%%%%%%%%%%%%%%%%%%%%%%%%%%%%%%%%%%%%%%%%%%

We study the effects of neutrino flavor conversion in the star and in the
Earth in a three neutrino framework with mixing and mass splittings
allowed by the atmospheric and solar neutrino data.

The neutrino mass eigenstates, $\nu_i$ ($i=1,2,3$), with masses $m_i$
are related to the flavor eigenstates, $\nu_\alpha$
($\alpha=e,\mu,\tau$), by the mixing matrix $U$: $\nu_\alpha=\sum_{i}
U_{\alpha i} \nu_i$.  The standard parameterization of this matrix is
used here (see e.g. \cite{Krastev:1988yu}).
The atmospheric neutrino data determine \cite{Fukuda:2000np}:
\begin{equation}
m_3^2 - m_2^2 \equiv    \Delta m^2_{32} = \pm (1.5 - 4) \cdot 10^{-3}
{\rm eV}^2, ~~~~~~~ \tan^2 \theta_{23} = 0.48 - 2.1 ~.
\label{atmpar}
\end{equation}
The two possibilities, $\Delta m^2_{32} \approx \Delta m^2_{31} > 0$ and
$\Delta m^2_{32} \approx \Delta m^2_{31} < 0$, are
referred to as {\it normal}  and {\it inverted} mass hierarchies/ordering
respectively and will be denoted as n.h.  and i.h. in the text.

The solar neutrino data  and the KamLAND  results  select the large
mixing angle (LMA) MSW solution with parameters~(see e.g. \cite{Eguchi:2002dm}, \cite{Fogli:2002au}--\cite{deHolanda:2002iv}): 
\begin{equation}
m_2^2 - m_1^2 \equiv   \Delta m^2_{21} = (4  - 30) \cdot 10^{-5}
{\rm eV}^2, ~~~~~~\tan^2 \theta_{12}=0.25 - 0.85 ~.
\label{sunpar}
\end{equation}

The reactor experiments CHOOZ and Palo Verde give the upper bound
on 13-mixing~\cite{Apollonio:1999ae,Boehm:2000vp}:
\begin{equation}
\sin^2 \theta_{13} \lta 0.02~.
\label{ue3}
\end{equation}

%%%%%%%%%%%%%%%%%%%%%%%%%%%%%%%%%%%%%%%%%%%%%%%%%%%%%%%%%%%%%%%%%%%%%%%%%%
\section{Mass hierarchy, $U_{e3}$ and conversion effects}
\label{sec:3}
%%%%%%%%%%%%%%%%%%%%%%%%%%%%%%%%%%%%%%%%%%%%%%%%%%%%%%%%%%%%%%%%%%%%%%%%%%%%%%%%%%%

%%%%%%%%%%%%%%%%%%%%%%%%%%%%%%%%%%%%%%%%%%%%%%%%%%%%%%%%%%%%%%%%%%%%%%%%%%%%%%%%%%%%
\subsection{Permutation factors}
%%%%%%%%%%%%%%%%%%%%%%%%%%%%%%%%%%%%%%%%%%%%%%%%%%%%%%%%%%%%%%%%%%%%%%%%%%%%%%%%%%

The fluxes $F_{ e}$ and $F_{\bar e}$ of the electron  neutrinos and antineutrinos in the detector
can be expressed in terms of the permutation parameters $(1 - p)$ and $(1 - \bar p)$
and the original fluxes as follows \cite{Dighe:1999bi}:
\beq
&&F_{ e} = { p} F_{ e}^0 + (1-{ p}) F_{ x}^0 = F_{ e}^0 + (1-p) (F_{ x}^0 - F_{ e}^0) ~~,
\label{fluxes1}\\
&&F_{\bar e} = {\bar p} F_{\bar e}^0 + (1-{\bar p}) F_{\bar x}^0  =
F_{\bar e}^0 + (1-{\bar p}) (F_{\bar x}^0 - F_{\bar e}^0)~.
\label{fluxes2}
\eeq
The  factors  $p$ and ${\bar p}$  are the total $\nu_e$ and $\bar{\nu}_e$ survival
probabilities which describe the
conversion effects inside the
star, in the intergalactic medium, and, if the Earth is crossed, in the matter of the Earth.
As can be seen in eqs. (\ref{fluxes1}, \ref{fluxes2}), the conversion
effects are proportional to the difference of the original $\nue$ ($\barnue$) and $\nux$ ($\barnux$) fluxes.

As they propagate in the star, the neutrinos undergo two MSW
resonances (level crossings):\\

\noindent
(i) the first resonance (H) occurs at higher density, $\rho \sim
10^{3} ~{\rm g\cdot cm^{-3}}(10 {\rm MeV} / E)$, it is governed by the atmospheric mass
squared splitting, $\Delta m^2_{32}$, and by the $\theta_{13}$
angle. The H-resonance  is in the neutrino (antineutrino) channel if the mass
hierarchy is normal (inverted).
The probability of
transition between the eigenstates of the Hamiltonian (jumping
probability) in this resonance, $P_H$,
can be written for the density profile (\ref{eq4}) as \cite{Dighe:1999bi}:
\beq
&&P_H=\exp  \left[-\left( \frac{E_{na} }{ E }\right)^{2/3}\right] ~,
\label{phform} \\
&&E_{na}\simeq 1.08 \cdot 10^{7} ~{\rm MeV}  \left( \frac{\vert \Delta m^2_{32} \vert }{ 10^{-3} ~{\rm
eV}^2}\right)C^{1/2} \sin^3 \theta_{1 3}~.
\label{ena}
\eeq

The probability $P_H$ has been obtained on the basis of the Landau-Zener (LZ) formula
which holds for the linear density distribution.
It can be checked that for small mixing (as it is the case here) the formula
works well for the density  profile (\ref{eq4}) \cite{Friedland:2000rn,Kachelriess:2001bs,Fogli:2001pm}.    \\

\noindent
(ii) A second level crossing (L) determined by the ``solar'' parameters (\ref{sunpar}),
happens at lower density, $\rho \sim (30 - 140) (10 {\rm MeV} / E) ~{\rm
g\cdot cm^{-3}}$.  For $\Delta m^2_{21}$
and $\theta_{12}$ in the LMA
region the propagation through this resonance is adiabatic for all possible values of
$C$ \cite{Dighe:1999bi}. \\

Let us first consider  the effects of conversion in the star with no Earth crossing.
 We will use here the approximation of ``factorized  dynamics" in the  resonances,
which means that the dynamics of level crossing in the  two
resonances  are independent and the total survival probability is the product of the
survival probabilities in each resonance (see \cite{Dighe:1999bi} for details).
In this approximation the survival probabilities
$p$ and ${\bar p}$ equal \cite{Dighe:1999bi}:
\beq
&&p\simeq P_H |U_{e2}|^2 + (1 - P_H) |U_{e3}|^2,
\label{p3nh} \\
&&{\bar p}\simeq~ |U_{e1}|^2,
\label{p3nh2}
\eeq
for n.h. and
\beq
&&p\simeq |U_{e2}|^2
\label{p3ih2} \\
&&{\bar p} \simeq  P_H |U_{e1}|^2 + (1 - P_H) |U_{e3}|^2
\label{p3ih}
\eeq
for i.h.. The transition from normal to inverted hierarchy cases
corresponds to the interchange  of the permutation  factors, $p \leftrightarrow \bar p$, and of the
indices $1  \leftrightarrow 2$ in the mixing parameters.

Using the standard parameterization of the mixing matrix, according to which
\be
|U_{e1}|^2 = \cos^2 \theta_{12} \cos^2 \theta_{13}, ~~~
|U_{e2}|^2 = \sin^2 \theta_{12} \cos^2  \theta_{13} ,~~~
|U_{e3}|^2 = \sin^2 \theta_{13},
\ee
we can rewrite the survival probabilities in the following form
\beq
&&p\simeq P_H \sin^2 \theta_{12}~  + \left[ 1-P_H (1+\sin^2 \theta_{12}) \right]
\sin^2  \theta_{13}~,
\label{pue3nh} \\
&&{\bar p}\simeq \cos^2 \theta_{12} (1-\sin^2 \theta_{13})~,
\label{pue3nh2}
\eeq
for n.h. and
\beq
&&p\simeq \sin^2 \theta_{12} (1-\sin^2 \theta_{13})~,
\label{pue3ih2} \\
&&{\bar p} \simeq  P_H \cos^2 \theta_{12} + \left[ 1-P_H (1+\cos^2 \theta_{12})\right] \sin^2 \theta_{13}~,
\label{pue3ih}
\eeq
for i.h..

The permutation factors depend on $\sin^2 \theta_{13}$ via the jumping  probability
$P_H \equiv P_H (\sin^2 \theta_{13})$ and explicitly via the mixing parameters.
A further dependence of the permutation factors on $\theta_{13}$ is given by
small terms which are neglected in our approximation. They arise from:

1) generalizing the LZ expression for $P_H$:
a more precise  double exponential form of the jumping probability
$P_H$ (see e.g.  \cite{Kuo:1989qe}) gives  $\sin^2 \theta_{13}$ corrections to the  Landau-Zener
form. These corrections appear however with very small
coefficients and therefore can be neglected.

2)
relaxing the approximation of factorized dynamics: this
could  give linear (in $\sin \theta_{13}$) corrections of the form
$\sin \theta_{13} \Delta m^2_{21}/\Delta m^2_{31}$.  
For the large $\Delta m^2_{21}$ part of the  LMA
region we have $|\Delta m^2_{21}/\Delta m^2_{31}| \sim 0.1$, which is
large enough so that $\sin \theta_{13} |\Delta m^2_{21}/\Delta
m^2_{31}| \sim \sin^2 \theta_{13} $ and the linear terms can not be
neglected with respect to the $\t13$ ones.  In contrast, taking the
best fit values of the mass squared splittings one gets $|\Delta
m^2_{21}/\Delta m^2_{31}| \sim 10^{-2}$, therefore the $\t13$ terms
dominate for $\t13 \gta 10^{-3}$.
A detailed study of the deviation from the factorization will be given elsewhere~\cite{usprep}.\\

Two limits are important:

1). If $\theta_{13}$ is very small so that $P_H = 1$,  both mass hierarchies lead to the
same result:
\beq
p \simeq \sin^2 \theta_{12}, ~~~ {\bar p} \simeq  \cos^2 \theta_{12}~.
\label{small13}
\eeq

2).  For large enough  $\theta_{13}$ (in practice   for $\t13 \gta 10^{-3}$ see
the next subsection) when   $P_H = 0$ we have
\beq
p \simeq \sin^2  \theta_{13}~,~~~~  {\bar p}\simeq \cos^2 \theta_{12} (1-\sin^2 \theta_{13})
\approx \cos^2 \theta_{12}~~~~~~~{\rm ~~ (n.h.)}~,
\label{pue3nh2a}
\eeq
 and
\be
p \simeq \sin^2 \theta_{12} (1-\sin^2 \theta_{13}) \approx \sin^2 \theta_{12}~, ~~~
{\bar p} \simeq \sin^2 \theta_{13}~~~~~~~{\rm ~~ (i.h.)}~.
\label{pue3iha}
\ee
Let us underline that in this limit uncertainties related
to the density profile or with factorized dynamics disappear.
The conversion is strongly adiabatic and the results
(\ref{pue3nh2a}, \ref{pue3iha}) are exact.
Notice also that from the
point of view of present oscillation results (large or maximal 12 and 23 mixings)
such a possibility looks rather plausible.

%%%%%%%%%%%%%%%%%%%%%%%%%%%%%%%%%%%%%%%%%%%%%%%%%%%%%%%%%%%%%%%%%%%%%%%%%%%%%%%%%%%%%%%%%
\subsection{$P_H$ and $\sin^2 \theta_{13}$}
%%%%%%%%%%%%%%%%%%%%%%%%%%%%%%%%%%%%%%%%%%%%%%%%%%%%%%%%%%%%%%%%%%%%%%%%%%%%%%%%%%%%%%%%
 
\begin{figure}[hbt]
\begin{center}
\epsfig{file=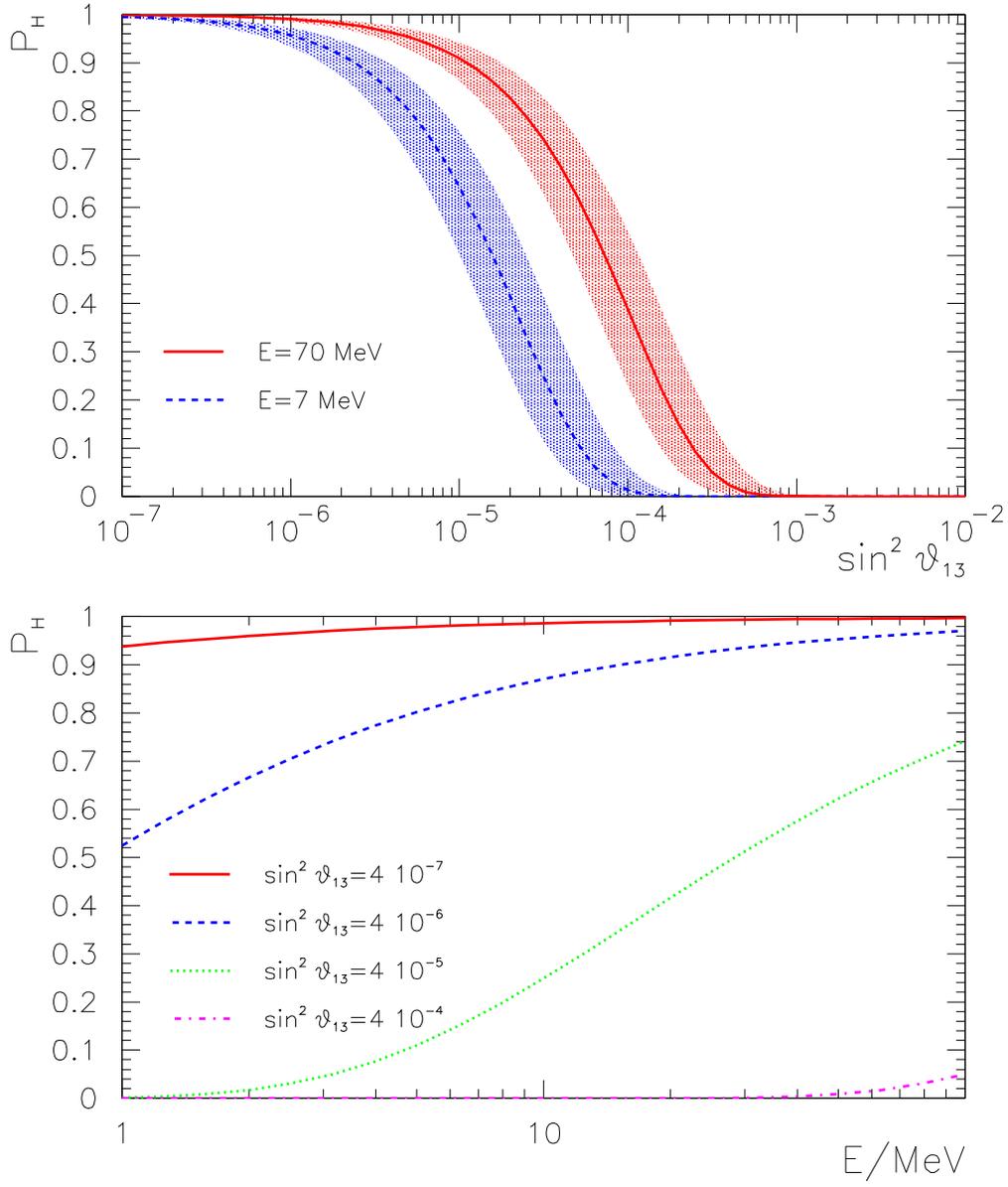, width=14.5truecm}
\end{center}
\caption{The transition probability in the H resonance, $P_H$, as a function of $\t13$ for two values of the neutrino energy (upper panel) and of the neutrino energy for different values of $\t13$ (lower panel). We took $C=4$ and $\vert \Delta m^2_{32} \vert =3\cdot 10^{-3}~{\rm eV^2}$. The shaded regions in the upper panel represent  the uncertainty associated to $C=1-15$.}
\label{fig:phfig}
\end{figure}

The jumping probability $P_H$ is shown in fig. \ref{fig:phfig} (upper panel, central lines) as a
function of $\t13$ for two extreme neutrino energies and
the parameters $\vert \Delta m^2_{32}\vert =3\cdot 10^{-3} {\rm eV^2}$ and $C=4$.
As follows from eq. (\ref{phform}), a constant value of $P_H$ corresponds to $\sin^2\theta_{13} \propto E^{2/3}$.
This determines the shift of
the curve $P_H$ in Fig. \ref{fig:phfig} along the axis $\sin^2 \theta_{13}$
with change of energy.

From the figure it follows  that, for all neutrino energies relevant for observations,
the conversion in the H  resonance is
adiabatic ($P_H=0$) for $\t13 \gta 4 \cdot 10^{-4}$. The adiabaticity is maximally
violated  ($P_H=1$) for $\t13 \lta 4 \cdot 10^{-6}$.  In the intermediate region $P_H$ depends
significantly on $\t13$, decreasing from 1 to 0 as $\t13$ increases.

Let us consider the uncertainties  in the relation between  $P_{H}$ and $\sin^2 \theta_{13}$.
The uncertainty in $P_{H}$ due to a  variation of $C$ by factor (1/4 - 4)  is shown in
fig.~\ref{fig:phfig} as a shadowed band.
 As given in eq.  (\ref{phform}), a constant value of $P_{H}$ corresponds to $\sin^2\theta_{13} \propto C^{-1/3}$.
Thus, for a given $P_{H}$, the  variation of $C$  corresponds  to a variation of $\sin^2\theta_{13}$
by factor  2.6.

Furthermore,  the profile (\ref{eq4})  is a
simplification and deviations from cubic dependence  can appear if \cite{clayton}:
\\

\noindent
1. the energy transfer in the star  is not purely radiative but also convective.
\\

\noindent
2. the luminosity $L$ is not constant with the radius $r$ \cite{clayton}.
\\

\noindent
3. the opacity $k$ has a spatial dependence: $k=k(r)$.
\\

In general, one can fit a realistic profile
which appears from numerical models of supernova progenitors
by $\rho \propto r^{-n(r)}$.  It can be seen
that  $n=3$ provides  the best power law fit of the realistic profile, however local
deviations can be significant
with variations of index in the interval  $n=1 \div 5$ \cite{Fogli:2001pm,shig}.

As discussed in ref. \cite{Dighe:1999bi},   if the position of the resonance, $r_{res}$,  is kept fixed,
the adiabaticity parameter $\gamma$ scales as $n^{-1}$: 
\beq
\gamma &\equiv& \frac{\vert \Delta m^2_{13}\vert }{2 E} \frac{\sin^2 2\theta_{13}}{\cos 2\theta_{13}}
\frac{1}{|(dn/dr)/n|_{res}}
%~\nonumber \\
\approx \frac{2 \vert \Delta m^2_{13}\vert  r_{res}}{ E} \frac{\sin^2 \theta_{13}}{n}~.
\label{gamscale}
\eeq 
So, the uncertainty $n=1 \div 5$  corresponds to a factor $\sim 2 - 3$
uncertainty in $\sin^2 \theta_{13}$
with respect to the $n=3$ case.

Thus,   the absolute limitation in determination of
$\sin^2\theta_{13}$ from measurements of $P_{H}$,
can be  characterized by the factor 1/3 - 3,
unless the knowledge of the star profile will be
substantially improved.
In principle, the uncertainty on the density profile  can be reduced if 
the progenitor of the star is identified and detailed observations
of the supernova optical signal (light curve, etc.) is done.

 The uncertainty  due to the error in  $\Delta m^2_{31}$ is numerically less important. We take it to be of $\sim 20\%$, in agreement with the expected precision of near future measurements \cite{Barger:2001yk}.\\

$P_H$ as a function of  energy for different values of $\t13$ is given in the bottom panel of fig. \ref{fig:phfig}.
In the observable part of the spectrum,  $E = (5 - 70)$ MeV,
a strong dependence of $P_H$ on $E$, and consequently
the strongest distortion  of the energy spectrum, is  expected if
$\sin^2\theta_{13} = (1 - 5)\cdot 10^{-5}$. For $\sin^2\theta_{13} = 4\cdot 10^{-5}$
the probability $P_H$ increases with $E$ by factor of 5 in the observable part of spectrum.
  Effects due to this strong change could give the possibility to probe $\theta_{13}$ is this region.  
The energy dependence is very weak for $\sin^2\theta_{13} > 4 \cdot 10^{-4}$ and
$\sin^2\theta_{13} < 4 \cdot 10^{-6}$.

%%%%%%%%%%%%%%%%%%%%%%%%%%%%%%%%%%%%%%%%%%%%%%%%%%%%%%%%%%%%%%%%%%%%%%%%%%%%%%%%%%%%%%%%
\subsection{Three regions}
%%%%%%%%%%%%%%%%%%%%%%%%%%%%%%%%%%%%%%%%%%%%%%%%%%%%%%%%%%%%%%%%%%%%%%%%%%%%%%%%%%%%%%%%%

\begin{figure}[hbt]
\begin{center}
\epsfig{file=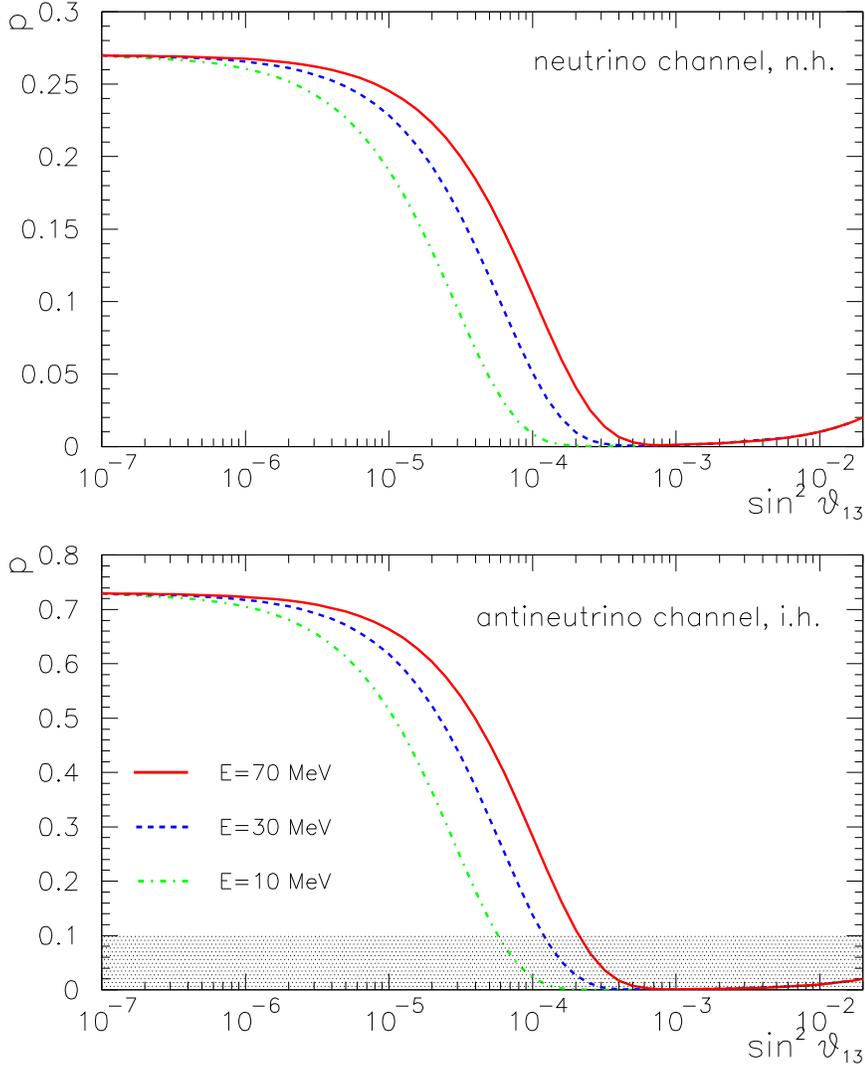, width=13truecm}
\end{center}
\caption{The survival probabilities $p$ for n.h.  (upper panel) and $\bar p$ for i.h. (lower panel) as functions of  $\t13$ for  $\tan^2 \theta_{12}=0.38$ and different values of the neutrino energy. Other parameters as in fig. \ref{fig:phfig}. The shaded region in the lower panel represents the error bar associated to a measurement  $\bar p=0$. The corresponding lower bound on $\t13$, $\t13 \gta 2 \cdot 10^{-4}$, is small enough to be calculated in the zeroth order approximation (i.e. neglecting the explicit $\t13$ dependence in eq. (\ref{pue3ih})). The error bar is much larger for the neutrino channel and normal hierarchy and was not drawn. }
\label{fig:regf}
\end{figure}

According to  eqs. (\ref{pue3nh})-(\ref{pue3nh2}) for  the normal mass hierarchy, the
probability in the antineutrino  channel, $\bar p$, does not depend on the neutrino energy
and only very weakly depends on $\theta_{13}$  via the  $\t13$ term.
In contrast, for the neutrino channel the permutation  factor, $1 - p$, depends on the energy via the
$P_H$  parameter and on the $\theta_{13}$ angle both  explicitly (the $\t13$ term in eq.
(\ref{pue3nh})) and implicitly through $P_H$ (eq. (\ref{phform})).   This dependence is
illustrated in the upper panel of fig. \ref{fig:regf} for different  values of the neutrino
energy, $\y12=0.38$ and the other
parameters  as in fig. \ref{fig:phfig}.  According to
the expression (\ref{pue3nh}) and to the figure we can distinguish three regions:
\\

\noindent
(i) {\it Adiabaticity breaking region}:
\beq
\t13 \lta 10^{-6} \left(\frac{E}{10 {\rm MeV}}\right)^{2/3}.
\label{range1}
\eeq
For these values of $\t13$ one gets $P_H \simeq 1$ and the first  term in eq. (\ref{pue3nh})
dominates. For $\t13 \rightarrow 0 $:  $p \approx \sin^2 \theta_{12}$ with  very good
approximation, independently of the values of  $\t13$ and of the neutrino energy.
In this region the conversion in the H resonance has little effect and
the permutation is  due to adiabatic conversion in the L-resonance.\\

\noindent
(ii) {\it  Transition region}:
\beq
\t13 \sim  (10^{-6} - 10^{-4})\cdot \left(\frac{E}{10 {\rm MeV}}\right)^{2/3}.
\label{range2}
\eeq
In this region  $P_H$
takes intermediate values between 1 and 0
(see fig. \ref{fig:phfig}).  Still $\t13$ corrections  to the permutation factor are negligible,
so that  $p \simeq P_H \sin^2 \theta_{12} $;  the
permutation $(1-p)$ increases   with $\t13$, following the jumping probability $P_H$.
\\

\noindent
(iii) {\it  Adiabatic region}:
\beq
\t13 \gta 10^{-4} \left(\frac{E}{10 {\rm MeV}}\right)^{2/3}.
\label{range3}
\eeq
Here $P_H\simeq 0$ and  therefore the two terms  in eq. (\ref{pue3nh})
are comparable near the border of the region and with increase of
$\t13$  the second term  dominates.
The survival probability  $p$ has a minimum at $ \t13 \sim 10^{-3}$, corresponding to  nearly maximal
permutation.
For $\t13 \gta 10^{-3}$ the term of order $\t13$ in the permutation  factor starts to dominate and
$p \simeq \t13$.

For inverted hierarchy similar results are found with the substitution
$\nu \leftrightarrow \bar \nu$:   $p$ has only a (small)
explicit dependence on
$\t13$, while $\bar p$ depends on $\t13$ both  explicitly and implicitly (see eq.
(\ref{pue3ih})). The same three regions discussed  here can be identified according to the
adiabaticity character of the H resonance.   The explicit dependence  of $\bar p$ on $\t13$
dominates for $\t13 \gta 10^{-3}$, where  $\bar p \simeq \t13$. These features are shown in the
lower panel of fig. \ref{fig:regf}.\\

 The effects of the explicit dependence of $p$ (or $\bar p$) on $\t13$ in
observable signals are
smaller than $\sim 2\%$   and  it will be very  difficult to study them due
to  larger experimental and theoretical uncertainties. However, their identification  may be  within the reach of the next generation large
volume detectors like HyperKamiokande \cite{HyperK}, UNO \cite{Jung:1999jq,UNO} and  TITAND \cite{Suzuki:2001rb}.

If the  accuracy of the experiments is not better than a few \% one can neglect the explicit dependence
of the probabilities on  $\theta_{13}$.
In this case the permutation  factor and therefore the observable
effects depend on $\theta_{13}$ only via the jumping  probability.
So one  can solve the problem in two steps:

- measure  $P_H$ immediately from experiments

- use the relation between $P_H$ and $\theta_{13}$ to determine
$\sin^2 \theta_{13}$. \\

Summarizing, according to (\ref{fluxes1}), (\ref{fluxes2}) and (\ref{pue3nh}) -
(\ref{pue3ih}) the effects of $\theta_{13}$ consist of

\begin{itemize}

\item
change of  the degree of permutation;

\item
distortion of the energy spectrum due to  dependence of the
jumping factor on energy.

\end{itemize}

For normal hierarchy a
change of the neutrino permutation factor, $(1 - p)$,  is smaller than  $30\%$:
from $\cos^2 \theta_{12} \approx 0.73$
at very small  $\theta_{13}$ to  $\approx 1$ at  $\theta_{13} > 10^{-3}$.
The change of antineutrino factor is negligible.
  Notice that, for any value of $\theta_{13}$,  the conversion inside the star leads
to the composite spectrum and $\theta_{13}$ changes
the level of this compositeness.

Thus, to determine $\theta_{13}$ in the case of normal hierarchy one needs to distinguish
between {\it completely} permuted  $\nue$ spectrum,
with  $F_{e} \approx F_{x}^0$,   and {\it strongly}  (3/4 or more) permuted spectrum.
In the first case one expects  a signal which corresponds to
the Fermi-Dirac spectrum, whereas in the second case, a composite spectrum
appears with the dominant hard component.

Clearly, to get an information about $\theta_{13}$ one needs, in general,
to know the original neutrino fluxes with better than $30\%$ accuracy
(including also substantial uncertainties of the detection procedure).

For inverted mass hierarchy the change of the antineutrino permutation
due to $\theta_{13}$
is more  significant: $(1 - \bar{p})$ increases  from  $\sin^2 \theta_{12} \sim 0.27$
for very small $\theta_{13}$ to  about 1 for $\theta_{13} > 10^{-3}$.
However, in this case the effect of permutation is suppressed due to smaller
difference of the original $\barnue$ and $\barnux$ fluxes.
The change of neutrino permutation is  negligible.\\

The permutation of $\nue$ and $\nux$ (or
$\barnue$ and $\barnux$)  spectra,  and therefore the observed neutrino signal, are mostly
sensitive to $\t13$ in the transition region; any dependence on $\t13$ outside this
interval is negligible.
In other words,
\\

\noindent
\begin{itemize}

\item
for $\t13 > 10^{-3}$ the effect of 1-3 mixing is strong in neutrino (antineutrino) channel if the mass hierarchy is normal (inverted).  This makes it possible to determine the hierarchy of the neutrino mass spectrum,  while  it is difficult
to measure  $\t13$. Observations of corresponding conversion effects will allow to
put a {\it lower} bound on $\t13$.

\item
in the region $\t13 \sim  (3 \cdot 10^{-4} - 3 \cdot 10^{-6})$ measurements of $\t13$ are possible,
or at least both upper and lower bounds on $\t13$ can be obtained;
values of $\t13$  in this region are at least one order
of magnitude below  the sensitivity of planned  terrestrial experiments;

\item
if  $\t13 < 10^{-6}$ no effects of $\t13$ should be seen and the observations are insensitive to the mass hierarchy. It follows that in this
case  the hierarchy can not be probed, while an upper bound on $\t13$ can be obtained.

\end{itemize}

%%%%%%%%%%%%%%%%%%%%%%%%%%%sect444%%%%%%%%%%%%%%%%%%%%%%%%%%%%%%%%%%%%%%%%%%%
%%%%%%%%%%%%%%%%%%%%%%%%%%%%%%%%%%%%%%%%%%%%%%%%%%%%%%%%%%%%%%%%%%%%%%%%%%
\section{Energy spectra. Observables}
\label{sec:3-obs}
%%%%%%%%%%%%%%%%%%%%%%%%%%%%%%%%%%%%%%%%%%%%%%%%%%%%%%%%%%%%%%%%%%%%%%%%%%

\subsection{Detected signals}
%%%%%%%%%%%%%%%%%%%%%%%%%%%%%%%%%%%%%%%%%%%%%%%%%%%%%%%%%%%%%%%%%%%%%%%%%%%%

Let us consider the effects of 13-mixing
on the energy spectra of the events induced by $\nu_e$  and $\bar{\nu}_e$
in the terrestrial detectors.

The number of  charged
current (CC) events produced by the $\nu_e$-flux  with electrons having the observed kinetic energy $E_e$
equals
\begin{eqnarray}
\frac{dN_{e} }{ dE_e} =
N_T\int_{-\infty}^{+\infty} dE_e' {\cal R}(E_e,E_e')
{\cal E}(E_e') \int dE F_e (E)
\frac{d\sigma (E_e', E) }{ dE_e'}~,
\label{eq:dnem}
\end{eqnarray}
where  $E_e'$ is the true energy of  the electron, $N_T$ is the number of target particles in the
fiducial volume and  ${\cal E}$ represents   the detection efficiency.
Here ${d\sigma (E_e', E)/dE_e'}$ is the differential cross section
of the detection reaction  and
${\cal R}(E_e, E_e')$ is the energy resolution  function.
The  $\nue$ flux in the detector,  $F_e$, is given in Eqs. (\ref{fluxes1},\ref{fluxes2}).

An expression analogous to (\ref{eq:dnem}) holds for the
$\barnue$ flux, $F_{\bar e}$.
\\

 The spectrum of the $\barnue$ induced events can be measured in  water cherenkov detectors, like  SuperKamiokande  (SK).  The relevant CC processes are:
\beq
\barnue + p \rightarrow n + e^+ ~,
\label{wat1}\\
\nue +O \rightarrow F + e^-~,
\label{wat2}\\
\barnue +O \rightarrow N + e^+~.
\label{wat3}
\eeq
They are isotropically distributed  and essentially indistinguishable from each other. The process
(\ref{wat1})
largely dominates the event rate because  of the much larger cross section: it produces $\sim 10^4$ events
for a supernova at $D=10$ kpc \cite{Beacom:1998ya,Takahashi:2001ep}.
Events from $\nue$ scattering on electrons  can be distinguished, and therefore subtracted, because of
their good directionality; other $\nu_\alpha + e^-$  scattering processes are neglected due to their
small cross section.\\

The spectrum of the $\nue$ induced events can be measured
in the heavy water cherenkov detector SNO.
The relevant CC reactions are:
\beq
\nue+ d \rightarrow p+p+e^- ~,
\label{hw1}\\
\barnue+ d \rightarrow n+n+e^+ ~.
\label{hw2}
\eeq
For a typical  galactic supernova at $D=10$ kpc (and LMA solar neutrino parameters) the processes
(\ref{hw1}) and (\ref{hw2})  give $\sim 300$ and $\sim 150$ events respectively \cite{Beacom:1998yb,Takahashi:2001ep}.  In the near
future, after the instrumentation  of SNO with the ${\rm ^3He}$ neutral current (NC) detectors, the events from (\ref{hw2})
will be  distinguished with $\sim 75-80\%$  efficiency due to the capture of at least one neutron
on ${\rm  ^3He}$ or  on deuterium in  coincidence
with the detection  of the charged lepton \cite{waltham}.

The light water volume (1.4 kt)   of SNO will give signals analogous to those discussed for SK (eqs.
(\ref{wat1})-(\ref{wat3})).\\

In what follows we consider $\nu_e$ and $\bar{\nu}_e$- events
at SNO and SK, assuming that they can be well distinguished and their
energy spectra can be separately measured.  Our methods of analysis
can in principle be extended and applied to other types of detectors,
e.g. scintillator and liquid argon experiments. For a discussion of
those in the context of supernova neutrinos, we refer to the papers by
the LVD \cite{Aglietta:2001jf}, ICARUS
\cite{Arneodo:2001tx,Strumia:bx} and LANNDD \cite{Cline:2001pt}
collaborations.

%%%%%%%%%%%%%%%%%%%%%%%%%%%%%%%%%%%%%%%%%%%%%%%%%%%%%%%%%%%%%%
\subsection{Observables}
\label{sec:obs}
%%%%%%%%%%%%%%%%%%%%%%%%%%%%%%%%%%%%%%%%%%%%%%%%%%%%%%%%%%%

Using the  differential spectra defined in (\ref{eq:dnem}) we can
introduce a set of observables
which are sensitive to the effects induced by the 13 mixing and depend on the mass hierarchy.\\

{\it 1. The average energy.} We define the average energy of events induced
by $\nu_e$ in a detector as
\be
\langle E \rangle =
\frac{1}{N_{tot}} \int_{E_{th}}^{\infty} dE_e  E_e  {dN_{e} \over dE_e}
\label{ave}
\ee
where
\be
N_{tot} =  \int_{E_{th}}^{\infty} dE_e  {dN_{e} \over dE_e}
\label{tot}
\ee
is the total number of events above the threshold energy $E_{th}$. We take
$E_{th} = 5$ MeV for neutrinos and $E_{th} = 7$ MeV for antineutrinos.

Notice  the different energy  dependences of the detection cross sections:
$\sigma \propto E^{2.25}$
for $\nue$ in heavy water \cite{Nakamura:2000vp}, while $\sigma \propto E^2$, with negative corrections  at high energy, for
$\barnue + p$ reaction in water \cite{Vogel:1999zy,Beacom:2001hr} (see also the recent calculation and discussion in \cite{Strumia:2003zx}).  This influences the observables.

The observed energy spectrum of events includes -- in addition to
the energy dependence of the neutrino  flux -- the energy
dependence of the detection  cross section and of the efficiency
and energy resolution function of the detector.  As a first
approximation, the effect of the efficiency and energy resolution
can be neglected and the cross section can be taken  as $\sigma(E)
\propto E^2$. For simplicity we can also put the threshold of
integration to be zero. Under these conditions  we find  the
average energy  of the observed spectrum of events  induced by
the  flux  (\ref{eq3}): 
 \begin{eqnarray}
&&\langle E_\alpha \rangle=\sigma (\eta_\alpha) T_{\alpha}~,
%\label{rel} ~, 
\hskip 1.6truecm \sigma (\eta_\alpha) \equiv 5 \frac{{\rm Li}_4(-e^{\eta_\alpha})}
{{\rm Li}_3(-e^{\eta_\alpha})}~,
\label{sigma}
\end{eqnarray}
where $\sigma (\eta_\alpha)$ is a function of the pinching
parameter expressed in terms of the polylogarithmic functions:
\begin{equation}
{\rm Li}_n(z)\equiv \sum_{k=1}^{\infty} \frac{z^k}{k^n}~.
\label{Li}
\end{equation}
For unpinched spectrum we find  $\sigma (0) = 31 \pi^6/(5670 \zeta(5)) \simeq 5.07$.  The quantity $\sigma (\eta_\alpha)$ increases with $\eta_\alpha$: e.g.
 $\sigma (2)  \simeq  5.33$
(this can be compared with characteristics of the Fermi-Dirac spectrum:
$\sigma(0)  = 3.15 $ and  $\sigma(2) \simeq 3.6 $).

 If the neutrino flux in the detector is an admixture of the colder $\nue$ (or $\barnue$) and the hotter $\nux$ (or $\barnux$) original fluxes, due to conversion effects, its average energy takes an intermediate value with respect to the average energies of the two component spectra.  It follows that conversion effects can be probed by studying average energies.

The analytical calculation with non-zero
threshold gives a more complicated result, which is less transparent and
is not very different numerically:  with $E_{th} = (5-7)$ MeV  the difference in the
average energies is typically  1 - 2 \%.

 An important test parameter is the ratio of  the average energies of the {\it observed} spectra of $\nue$ and $\barnue$ events:
\be
r_E = \frac{\langle E\rangle}{\langle \bar E\rangle} ~.
\label{r-e}
\ee \\

%%%%%%%%%%%%%%%%%%%%%%%%%%%%%%%%%%%%%%%%%%%%%%%%%%%%%%%%%%%%%%%%%%%%%%%%%%
{\it 2. The width of the spectrum.} The relative width of the spectrum can be characterized by the
dimensionless parameter
\be
\Gamma \equiv  \frac{\Delta E}{\langle E \rangle}~,
\label{wid}
\ee
where $\Delta E$ is defined as:
\be
\Delta E \equiv \sqrt{ \langle (E - \langle E\rangle)^2\rangle}= \sqrt{\langle E^2\rangle - \langle
E\rangle^2}~.
\label{2mom}
\ee
Here $\langle E \rangle$ is the average energy and $\langle E^2 \rangle$ is the average of the
energy squared. The latter is defined by the integral (\ref{ave}) with the substitution
$E_e  \rightarrow E^2_e $.

Measurements of the width  will allow to test the   compositness of the observed spectrum.
Performing the integration from $E_{th} = 0$, we find that  the relative width of the distribution induced in the detector by a Fermi-Dirac neutrino spectrum (\ref{eq3}) does not depend on  the
temperature  and is determined by the pinching parameter only: 
\be
\Gamma_{FD} (\eta_\alpha) = 
\left[
\frac{6}{5}
\frac{{\rm Li}_7(- e^{\eta_\alpha}){\rm Li}_5(-e^{\eta_\alpha})}
{({\rm Li}_6(-e^{\eta_\alpha}))^2} - 1
\right]^{1/2}~.
\label{gamp}
\ee 
This gives  $\Gamma_{FD} (0) \simeq 0.44$, while   $\Gamma_{FD}  (3) \simeq 0.38$,  corresponding to a $\sim 15\%$
narrowing of the spectrum with respect to the no-pinching case.
%The composite spectrum which appears as a result of
%partial permutation  has in general larger width. 
To get an idea about effect of the compositeness on the
width, we give its expression 
for small energy difference between the two original neutrino fluxes: $\epsilon \equiv  \langle E_x \rangle/\langle E_e \rangle -1 \ll 1 $ and for  equal pinching, $\eta_e=\eta_x =\eta$, and equal luminosities in the two flavors:
\be 
\Gamma (\eta) \simeq \Gamma_{FD} (\eta) \left[ 1 + \frac{\Gamma^2_{FD} (\eta)+1}{2 \Gamma^2_{FD} (\eta)} p (1-p) \epsilon^2 \right]~.
\label{widcomp}
\ee 
Moreover, this result is valid if the $\nue$ survival probability,
$p$, is independent of the neutrino energy. Eq. (\ref{widcomp})
shows that $\Gamma (\eta) \geq \Gamma_{FD} (\eta)$, if the permutation is partial ($0<p<1$).  In presence of strong pinching the width of
the composite spectrum could be smaller than the width of non-permuted unpinched spectra, i.e.  $\Gamma (\eta) \lta \Gamma_{FD} (0)$. For instance, for $\eta=3$, $ \epsilon=0.4$ and $p\simeq \sin^2 \theta_{12}\simeq 1/4$, eq. (\ref{widcomp}) gives  $\Gamma (3) \simeq 0.4$, to be compared with $\Gamma_{FD} (0) \simeq 0.44$.
It follows that 
an observation of
$\Gamma >  0.44$ will testify for the composite spectrum and partial permutation, while $\Gamma <  0.44$ will testify for pinched spectra without conclusive information on the amount of their permutation. 
%Complete permutation does not change the width if pinching parameters
%of the original $\nue$  and $\nux$ spectra are equal.

We introduce also the ratio of the widths of the observed energy spectra of $\nue$ and $\barnue$ events:
\be
r_{\Gamma}  \equiv  \frac{\Gamma}{\bar{\Gamma}}~.
\label{gamma-r}
\ee\\

%%%%%%%%%%%%%%%%%%%%%%%%%%%%%%%%%%%%%%%%%%%%%%%%%%%%%%%%%%%%%%%%%%%%%%%%%%%%%%%%%%%%%%%%%%%%%%%%
{\it 3. Numbers of events in the high energy tails}.  The physics and analysis become
simpler for  energies substantially above the average energy of the spectrum:
$E > (2 - 3) \cdot \langle E \rangle$. We will  refer to these parts of the spectra as 
the tails.

Let us define the number of CC events induced by $\nu_e$ in the detector with  visible energy above $E_L$:
\be
N_{e}(E > E_L) = \int_{E_L}^{\infty} dE_e \frac{dN_e}{dE_e} ~.
\label{intsp}
\ee
Similarly, one can define  $N_{\bar e}(E > \bar{E}_L)$  as the number of events induced by
$\bar{\nu}_e$ above the energy $\bar{E}_L$. 
For definiteness in our numerical estimations we will use
$E_L = 45$ MeV and $\bar{E}_L = 55$ MeV.  
A detailed discussion of the prescription for the choice of these thresholds, as well as the dependence of the results on their values, 
is given in sect. 6.2.
%(for a  motivation of these specific numbers see sect.~\ref{sec:secq}).

The following analytical study is useful for the interpretation of results.
If  $E_L \gg  \langle E \rangle$, the original  flux of a given flavor,
$F^0_{\alpha}$, above the cut is well approximated by the Maxwell -- Boltzmann distribution:
\be
F^0_{\alpha} \approx const~ \frac{L_\alpha}{T^4_\alpha} E^2 e^{-E/T_\alpha}~,
\label{mb}
\ee
where the normalization constant depends on pinching parameter.

An estimate of the number of  events $N^0_\alpha$ is given by the convolution of this expression  with the
detection cross-section  and the  detection efficiency; the energy resolution function can be neglected
in  a first approximation.  At high  energies the detection efficiency
does not depend on  energy, and therefore factors  out of the integration.
Taking the  cross section as $\sigma(E) \propto E^2$, from (\ref{mb}) one
gets:
\beq
N^0_\alpha &\approx & const~ L_\alpha T_\alpha \int^{\infty}_{E_L/T_\alpha} dx x^4 e^{-x}~ \nonumber \\
& = & const~ L_\alpha T_\alpha e^{-E_L/T_\alpha} P(E_L/T_\alpha)~,
\label{nevsimpl}
\eeq
where
\be
P(x)\equiv x^4 + 4 x^3 +12 x^2 + 24 x + 24 ~.
\label{px}
\ee
For very high cut, $E_L/T_\alpha \gg 1$, eq. (\ref{nevsimpl}) has the asymptotic behavior:
\be
N^0_\alpha \approx  const~ \frac{L_\alpha}{T^3_\alpha} E^4_L e^{-E_L/T_\alpha}~.
\label{nevasy}
\ee
Notice that the result (\ref{nevsimpl}) has an indicative characted only, especially in view of the fact
that the assumption  $\sigma(E) \propto E^2$ is a rather crude approximation.
In spite of that, however,  the form (\ref{nevsimpl}) turns out to be in acceptable agreement with the
numerical results, as will be discussed  later.

Let us define  the ratio of the  neutrino and antineutrino events in the tails:
\be
R_{tail}(E_L,\bar{E}_L)\equiv {N_{e}(E>E_L) \over N_{\bar e}(E>\bar{E}_L)}~.
\label{eq:rat}
\ee
This turns out to be very a powerful test parameter of the conversion
induced by  the 13-mixing. \\

{\it 4.  Low energy $\nu_e$ and $\bar{\nu}_e$ events.} Similarly
we introduce the numbers of events induced by $\nu_e$ and
$\bar{\nu}_e$ with visible energies below $E'_L$ and $\bar E'_L$
respectively. In what follows the equal values  $E'_L=\bar
E'_L=25$ MeV are taken for illustrative purpose, with  the low
energy thresholds  $E_{th} = 5$ MeV for the neutrino events and
$E_{th} = 7$ MeV for the antineutrino events.

We introduce also the ratio of numbers of the low energy neutrino and antineutrino events:
\be
S\equiv  \frac{N_e(E<E'_L)}{N_{\bar e}(E<\bar E'_L)}~.
\label{s-param}
\ee
In what follows we will calculate predictions for these observables
depending on the type of mass hierarchy and interval of $\theta_{13}$.

%%%%%%%%%%%%%%%%%%%%%%%%%%%%%%%%%%%%%%%%%%%%%%%%%%%%%%%%%%%%%%%%%%%%%%%%%%%%
\section{Identifying extreme  possibilities. Scatter plots}
%%%%%%%%%%%%%%%%%%%%%%%%%%%%%%%%%%%%%%%%%%%%%%%%%%%%%%%%%%%%%%%%%%%%%%%%%%%%

One can perform the analysis of the supernova data in two steps:

\noindent
1). Resolve the ambiguities related to hierarchy
``normal - inverted"  and to value of $\theta_{13}$: ``large - small".
At this point the bounds on $\theta_{13}$ can be obtained only.

\noindent
2). Once the hierarchy is determined and bounds on $\theta_{13}$
are found, one can proceed with a detailed analysis of data to measure $\theta_{13}$.

%%%%%%%%%%%%%%%%%%%%%%%%%%%%%%%%%%%%%%%%%%%%%%%%%%%%%%%%%%%%%%%%%%%
\subsection{Extreme cases}
\label{sec:extr}
%%%%%%%%%%%%%%%%%%%%%%%%%%%%%%%%%%%%%%%%%%%%%%%%%%%%%%%%%%%%%%%%%%%%

In this section we  consider the first step. We will
update the analysis of ref. \cite{Dighe:1999bi} taking the LMA solution
with the most plausible values of the oscillation parameters.
As in \cite{Dighe:1999bi}, we will denote by large $\theta_{13}$
the values of $\theta_{13}$ which lead to $P_{H} \approx 0$. The
interval is determined by eq. (\ref{range3}).  We  refer to small
$\theta_{13}$ to indicate values which satisfy eq. (\ref{range1}) so
that $P_{H} \approx 1$.  Then there are three  extreme possibilities: \\

\begin{itemize}

\item
A. Normal hierarchy - large $\theta_{13}$:
In this case the permutation factors for neutrinos and
antineutrinos equal: $(1 - p) \approx 1$ and
$(1 - \bar p) \approx \sin^2 \theta_{12} \sim 1/4$  (see eq. (\ref{pue3nh2a})).
One should observe unmixed completely permuted (and therefore hard)  $\nu_e$-spectrum, and
composite weakly mixed ($\sim 1/4$) $\bar{\nu}_e$-spectrum.

 These features can be quantified in terms of the avarage
energies and widths  of the $\nue$ and $\barnue$ fluxes in the
detectors: $\langle E (\nu_e) \rangle$, $\langle E (\bar{\nu}_e)
\rangle$ and $\Gamma (\nu )$. These quantities are defined
analogously to the corresponding ones for the observed spectra of
events, eqs. (\ref{ave}) and (\ref{wid}) \footnote{We mark
that, in contrast with sec. \ref{sec:obs},  the present discussion
refers to the average energies and widths of the spectra of the
neutrinos arriving at Earth, and {\it not} to the spectra of the
events induced by the neutrinos in the detectors. }
where  $\Delta E$ is determined similarly to the width of the observable spectrum (\ref{2mom}).
In general, we get: 
\begin{equation}
\langle E(\nu_e)\rangle > \langle E(\bar{\nu}_e)\rangle,~~~\Gamma (\nu_e) \lta \Gamma (\bar{\nu}_e).
\label{signA}
\end{equation}
The inequality of the widths can be violated in the particular case in
which compensations occur between pinching and permutation
effects. Indeed, as discussed in sec. \ref{sec:obs}, equal or slightly larger $\nue$ width could be
realized if the original $\nux$ spectrum (and therefore the $\nue$
spectrum arriving at Earth) has no pinching, while the original
$\barnue$ spectrum is strongly pinched and the difference between the $\barnue$ and $\nux$  average energies is small (see eq. (\ref{widcomp})). We consider this arrangement rather exotic, since similar pinching is expected in the different flavors \cite{Keil:2002in}. 

The Earth matter effect is expected in the antineutrino channel
and not in the neutrino channel (unless significant difference of the
$\nu_{\mu}$ and $\nu_{\tau}$ fluxes exists \cite{Akhmedov:2002zj}).

If this case is identified we will be able to determine
the mass hierarchy and put lower bound on $\sin^2\theta_{13}$.

\item
B. Inverted hierarchy - large $\theta_{13}$:
The permutation factors equal $(1 - p) \approx \cos^2\theta_{12} \sim 3/4$
and $1 - \bar p \approx 1$. In this case  $\bar{\nu}_e$ will have
unmixed completely permuted  F-D spectrum, whereas the $\nu_e$-spectrum,
should be composite with rather strong permutation.
Consequently, 
\begin{equation}
\langle E (\nu_e)\rangle <  \langle E(\bar{\nu}_e)\rangle,~~~\Gamma (\nu_e) \gta
\Gamma (\bar{\nu}_e).
\label{signB}
\end{equation} 
The Earth matter effect is expected in the neutrino channel.

If this is realized, we will conclude on the mass hierarchy and
put a lower bound on $\sin^2\theta_{13}$.

\item
C. Normal hierarchy - small $\theta_{13}$.  Inverted hierarchy - small $\theta_{13}$.

These two cases lead to identical consequences:
composite $\nu_e$-spectrum  with $\cos^2\theta_{12} \sim 3/4$
mixing (permutation), and composite $\bar{\nu}_e$-spectrum
with small $\sin^2\theta_{12} \sim 1/4$ mixing (permutation).

Since the permutation is stronger in the neutrino channel one expects:
\begin{equation}
\langle E(\nu_e)\rangle \gta  \langle E(\bar{\nu}_e)\rangle,~~~
\Gamma (\nu_e) \gta  \Gamma (\bar{\nu}_e);
\label{signC}
\end{equation}
however these  inequalities are not strict 
since the permutation effects are partially compensated by the fact that
the original $\nu_e$ spectrum is softer than the $\bar{\nu}_e$ spectrum.

The Earth effect is expected both in the neutrino and antineutrino channels.

In these case one  can put an upper bound on $\sin^2\theta_{13}$ only
and the hierarchy will not be identified.

\end{itemize}

Studies of the $\nu_e$-signal only allow,  in principle,  to disentangle
the  case A, for which the spectrum is hard and of the Fermi-Dirac type, and the cases
B, C   which lead to the same composite spectrum with permutation
$\cos^2\theta_{12} \sim 3/4$.

If the $\bar{\nu}_e$-signal is studied only, one can disentangle
the case B,  characterized by a hard Fermi-Dirac spectrum, from the cases  A, C  which give
the same composite spectra with mixture $\sin^2\theta_{12} \sim 1/4$.

The comparison of the properties of the $\nu_e$- and  $\bar{\nu}_e$-spectra
will allow to distinguish three possibilities: A, B, and C.\\

As we have marked in sect.~\ref{sec:obs}, the composite spectra should be wider than the
Fermi-Dirac spectrum (unless the parameters of neutrino radiation -- luminosity, temperature,
pinching --   strongly change with time during the burst).

To disentangle the  possibilities A-C, one can:

- search for  deviations of the spectral shapes from the Fermi-Dirac form,

- compare the average energies and the widths of the spectra for neutrinos and antineutrinos,

- search for the Earth matter effects in neutrino and antineutrino channels. \\

%%%%%%%%%%%%%%%%%%%%%%%%%%%%%%%%%%%%%%%%%%%%%%%%%%%%%%%%%%%%%%%%%%%%%%%%%%%%%%%
\subsection{Scatter plots}
%%%%%%%%%%%%%%%%%%%%%%%%%%%%%%%%%%%%%%%%%%%%%%%%%%%%%%%%%%%%%%%%%%%%%%%%%%%%%%

The criteria described in the previous section for neutrino spectra become
less strict when

(i) the spectra of the observed  events (and not of the neutrinos)  are considered (here the difference  of the
energy dependences of the neutrino and antineutrino cross-sections plays an important role);

(ii) uncertainties in the original neutrino fluxes are taken into account.

(iii) uncertanties in the 12-mixing are included.

Let us recall  that in  the cases of very small ($P_H=1$) or large  ($P_H=0$) $\t13$
the uncertainties on the $C$ parameter and on  $\vert \Delta m^2_{31} \vert$ have no effect on the physics.

The number of unknown parameters which describe the original
spectra is very large. For this reason we have constructed scatter
plots of the observables using the following procedure:

1).   We define the  space of the parameters   over which we perform scanning in the following way:  the average energies, luminosities and pinching parameters of the
original neutrino fluxes are taken in the intervals  (\ref{temp}),
(\ref{deltax}), (\ref{fluxes}) and (\ref{pinch}).
 We assume that $\y12$ will be known with $\sim 10\%$ accuracy and, as an example, use the interval:
\be
\y12=0.342 - 0.418.
\ee

2) We take a  grid of points in this parameter space. Depending on the case under investigation, the spacing of this grid is chosen conveniently, with a corresponding number of points  between $\sim 570$ and $\sim 10^4$.    The  number  of points used was smaller for the cases A and B. Indeed, 
the scenario A predicts complete permutation of the fluxes in the neutrino  channel and, as a consequence, the   $\nue$ flux at Earth does not depend on the original $\nue$  flux $F^0_e$ (see eq. (\ref{fluxes1})). It follows that 
 a scanning over the parameters of this flux can be avoided, resulting in a smaller number of points.  An analogous argument applies for the case B and the antineutrino channel: no scanning of the parameters of the original $\barnue $ flux is needed, since this flux cancels from the calculations (eq. (\ref{fluxes2})). 
In contrast, for the scenario C, all the original neutrino and antineutrino fluxes are relevant and a full scanning of all the parameters (with a larger number of points) is necessary.

 3) For each point of the grid we
calculated the observables $r_E$, $r_{\Gamma}$, $R_{tail}$, $S$. The calculations are performed for
 $\nue$  events at SNO (from the process (\ref{hw1})) and  $\barnue$ events at SK (from the
 reaction (\ref{wat1})).  We chose the energy cuts $E'_L = \bar E'_L =25$ MeV for the calculation of $S$ and the thresholds $E_L=45$ MeV and $\bar E_L= 55 $ MeV for $R_{tail}$. The prescription for the choice  of these values is given in sec. \ref{sec:secq}. 

4) We do not take into account possible correlations of parameters,
scanning the points within the intervals independently. This leads to the most
conservative conclusions.

\begin{figure}[hbt]
\begin{center}
\epsfig{file=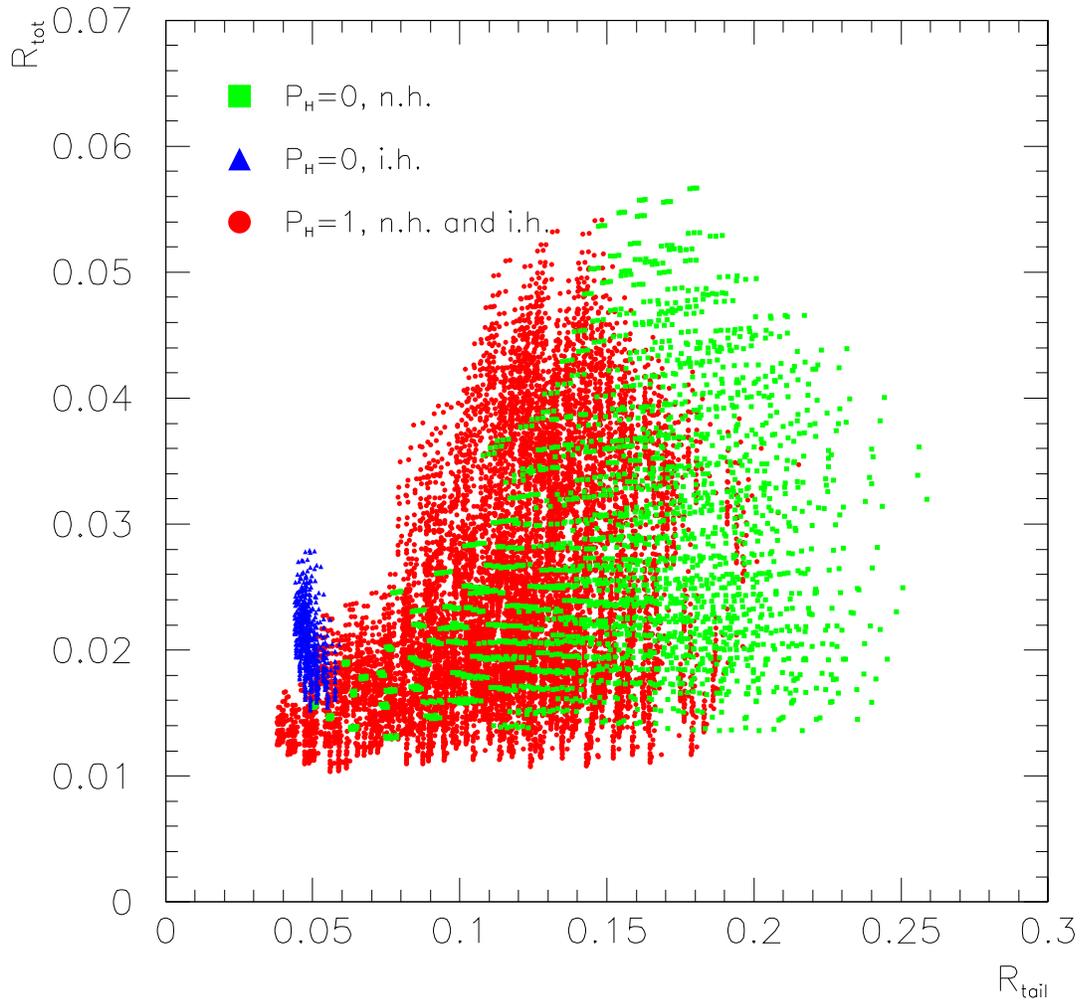, width=16truecm}
\end{center}
\caption{Scatter plot in the plane $R_{tail}-R_{tot}$ for the 
  cases A, B and C discussed in sec. \ref{sec:extr}.  The quantity $R_{tot}$
  is the ratio of the total rates of $\nue$ and $\barnue$ events, eq. (\ref{rtotdef}). The cuts
  $E_L=45$ MeV and $\bar E_L=55$ MeV were taken for $R_{tail}$. }
\label{fig:scatter1}
\end{figure}

\begin{figure}[hbt]
\begin{center}
\epsfig{file=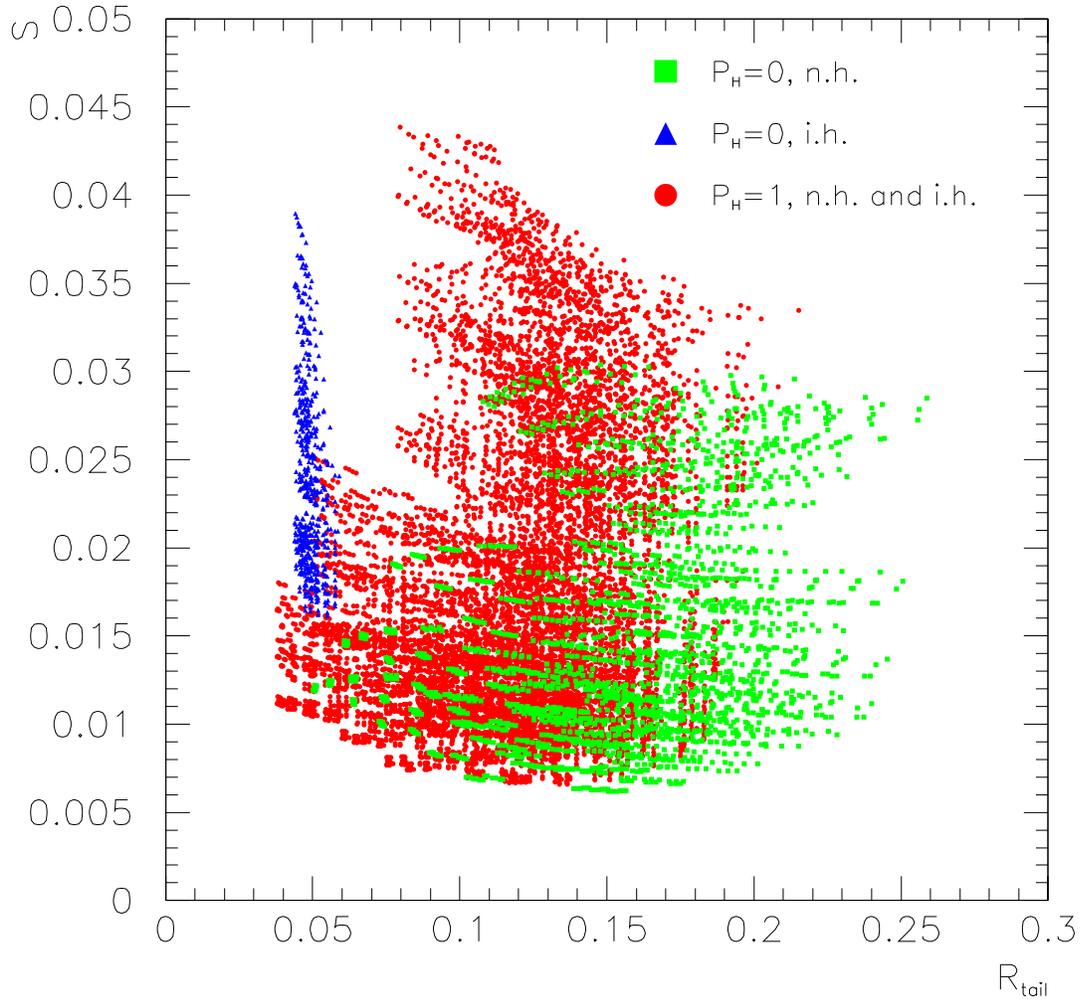, width=16truecm}
\end{center}
\caption{Same as fig. \ref{fig:scatter1} in the $R_{tail}-S$ plane,
  where $S$ is defined according to eq. (\ref{s-param}).     We  took
  the cuts $E'_L=\bar E'_L=25$ MeV for the calculation of $S$.}
\label{fig:scatter2}
\end{figure}

\begin{figure}[hbt]
\begin{center}
\epsfig{file=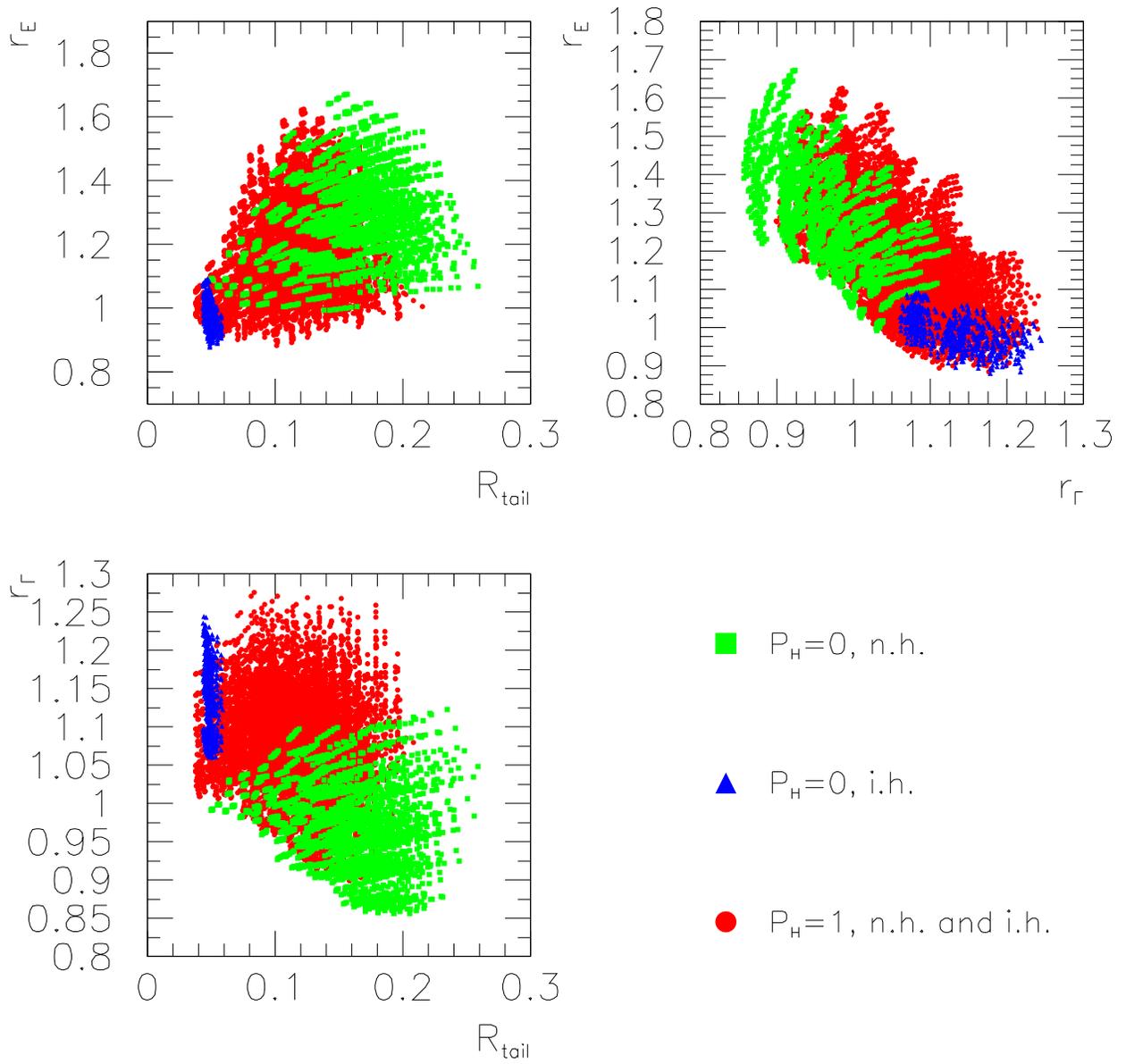, width=19truecm}
\end{center}
\caption{Scatter plots in the space of the variables: $R_{tail}$,  $r_E$, $r_\Gamma$, eqs. (\ref{eq:rat}), (\ref{r-e}) and (\ref{wid}).}
\label{fig:scatter4}
\end{figure}

The results  are shown in figs.~ \ref{fig:scatter1} -\ref{fig:scatter4}. \\

The fig. \ref{fig:scatter1} shows the ($R_{tot} - R_{tail}$)
scatter plot. The ratio of events in the tails, $R_{tail}$, is
defined in  (\ref{eq:rat}) and $R_{tot}$  is the ratio of the
total numbers of neutrino and antineutrino events:
\be R_{tot}
\equiv \frac{N_{tot}}{\bar{N}_{tot}}~,
\label{rtotdef}
\ee
where $N_{tot}$ for neutrinos is defined in (\ref{tot}).
From the  figure it follows that in the case B the ratio $R_{tot}$ can not  exceed $\sim 0.032$, while
for C $R_{tot}$ can  be nearly twice as large. 
The difference is explained by the different degree of
permutation in the various cases, according to what discussed in
sec. \ref{sec:extr}.  In case B the observed antineutrino spectrum is
completely permuted and therefore maximally hard.  This in turn
implies the largest rate of $\barnue$ events due to the larger
detection cross section at high energies.  In the case C the
permutation of $\barnue$ is partial, giving softer observed $\barnue$
spectrum and therefore smaller event rate and larger value of
$R_{tot}$. The $\nue$ spectrum is partially permuted in B and C, while
in case A it is totally permuted and therefore harder. This implies a
larger $R_{tot}$ with respect to C. Numerically the difference
between the two cases is small (see figure), because in C the amount of
permutation for $\nue$ is rather large: $1- p \simeq \cos^2
\theta_{12} \simeq 0.75$, making this case close to complete
permutation.

The parameter $R_{tail}$ has much higher discriminative power than $R_{tot}$.
$R_{tail} < 0.06$ for the case B and practically there is no overlap of the regions A and B.
The region C overlaps with both A and B, although for $R_{tail} > 0.22$ only A is allowed.

As follows from the figure, there are  certain regions in
the $R_{tail} - R_{tot}$ plane where only one possibility is realized.
For the case of inverted mass hierarchy and large $\theta_{13}$ (B), the region is defined as
$R_{tail} = (0.040 - 0.055)$, $R_{tot} = (0.02 - 0.03)$.
For the case of very small 13-mixing (C) there is a band around
$R_{tot} \simeq (0.3 \cdot R_{tail})$ with $R_{tail} < (0.05 - 0.14)$.
The normal mass hierarchy case A is the unique possibility for
$R_{tail} >  (0.14 - 0.22)$ where the border depends on value of $R_{tot}$.

Clearly it will be easy to identify or discriminate the case B.
It might be more difficult to disentangle A and C since
a significant overlap exists.
The  overlapping  areas
correspond to similar original fluxes  in the different neutrino flavors:
conversion effects are smaller
for smaller difference of the original fluxes,
making it difficult to disentangle different scenarios of
mass  hierarchy and 13-mixing.\\

The fig. \ref{fig:scatter2}  shows the  $S - R_{tail}$ scatter plot, where
the ratio of the numbers of low energy events, $S$,  is defined in (\ref{s-param}).
This figure is rather similar to  fig. \ref{fig:scatter1}, although the
$S$ parameter is more complementary to $R_{tail}$ than $R_{tot}$, and
experimentally $S$ and $R_{tail}$ are independent.
The case B has higher $S$ and smaller $R_{tail}$,
inversely, the scenario A predicts higher $R_{tail}$ and smaller $S$.
The case C is intermediate between the two.
As a consequence, the overlap of regions is smaller than in fig. \ref{fig:scatter1}.

We find the following  regions where only one possibility
is realized:
The case A is unique for $R_{tail} > 0.165$, the case B is unique
for $S > 0.017$ and $R_{tail} = (0.040 - 0.055)$, and C is unique
in the band $S \approx (0.2 - 0.3) \cdot R_{tail}$.

These
features are explained in terms of total  or partial  permutation, similarly to what we have dicussed for fig.  \ref{fig:scatter1}.
\\

The figure \ref{fig:scatter4}
shows  scatter plots  in the space of the variables
$R_{tail}$, $r_E$ (\ref{r-e}), and $r_{\Gamma}$ (\ref{wid}).  The
corresponding intervals of values of these observables for each of
the scenarios A, B, C are summarized in the Table  \ref{tab:obs}.
For comparison, the Table displays also the expected intervals of
the same variables for the neutrino spectra discussed in sec. \ref{sec:extr}. These parameters are not observable, however they allow to understand the features of the
observable spectra.

\begin{table}
\begin{center}
\begin{tabular}{lccc}
\hline
  parameter $\backslash$ scenario           &   A         &   B        &   C   \\
\hline
\hline
$r_E$(events)    &  1.0 - 1.7  & 0.87 - 1.1 &  0.9 - 1.6 \\
$r_E$(neutrinos) &    $>$ 1    &  $<$ 1     &  $\gta$  1 \\
\hline
$r_{\Gamma}$(events)     &  0.85 - 1.1 &  1.06 - 1.25 & 0.9 - 1.25 \\
$r_{\Gamma}$(neutrinos)  &  $\lta$ 1  & $\gta$ 1    &  $\gta$ 1 \\
\hline
$R_{tail}$        &  0.04 - 0.26  &   0.04 - 0.056 & 0.035 - 0.21  \\
\hline
\end{tabular}
%\caption{ The ranges of parameters of the  neutrino spectra 
%(``neutrinos")
%and spectra of observable events (``events") for three possible cases
%of mass hierarchy and 1-3 mixing. }
\caption{The ranges of parameters of the  spectra of observable events 
(``events") for three possible cases
of mass hierarchy and 1-3 mixing. For comparison we present also
the corresponding characteristics of the neutrino spectra (``neutrinos"), 
which are not immediate observables.}
\label{tab:obs}
\end{center}
\end{table}

 From the figure and the Table it follows  that  for the case B the allowed  region of parameters is relatively
small, while for A and C the allowed regions are larger and  expand over
$r_E \simeq 1 - 1.5 $. Furthermore,   $r_{\Gamma}$ is mostly larger than 1 in C and mostly smaller than 1 in A.   These results largely follow the expectations for the neutrino spectra, and can be interpreted  in terms of the amount of permutation according to the discussion in sec. \ref{sec:extr}.

One can see two slight deviations with respect to the predictions for neutrino spectra:
(1)
a significant number of points with $r_E>1$ in the case B and
(2)
an appreciable amount of points with $r_{\Gamma}>1$ for the case A.
The latter is in agreement with the possibility that the broadening of the spectrum due to compositeness is overcompensated by the effect of pinching, as commented in sec. \ref{sec:extr}. 
Moreover, the following reasons contribute to explain the deviations: 

%These deviations can be explained by the interplay of 

(i) the small contribution of the original $\nue$ flux  in the detected $\nue$ signal. This contribution
is -- depending on the value of $\theta_{13}$   -- not larger than
 $\sin^2 \theta_{12}\simeq 0.25$, therefore  the average energy of the observed $\nue$
spectrum is
dominated by the harder component due to the original $\nux$ flux.

(ii) the different energy dependence of the cross sections  of the $\barnue + p $ and $\nue+d$ detection
reactions:
the first grows like $E^2$ with suppressing corrections at  high energies \cite{Vogel:1999zy,Beacom:2001hr}, while the second has a
stronger
growth with energy, being proportional to $E^{2.25}$ \cite{Nakamura:2000vp}.   It follows that in presence of equal
$\nue$ and
$\barnue $ energy spectra, this difference leads to higher  average energy  and larger width of the observed $\nue$
spectrum.
\\

 Large  regions of the parameter space exist where only one among the scenarios A, B or C is
possible. Also  regions appear when two of these scenarios are realized.
If these regions are selected by the experiments, the third possibility
will be excluded. From the figs. \ref{fig:scatter1}-\ref{fig:scatter4} we conclude that:

\begin{itemize}

\item The case A  is excluded if  observations give   $R_{tail}
\lta 0.06$ and/or $r_\Gamma \gta 1.1$. If, in contrast,  the experiments give
$R_{tail} > 0.22$  the normal hierarchy would be identified. This
result would be further supported if $r_\Gamma \lta  0.95$ is also
found.

The possibility of getting information on $P_H$ and consequently  on $\theta_{13}$ depends on the
specific value of $R_{tail}$ (as illustrated in sec. \ref{sec:3.1}).

\item The case B is excluded by large  values of $R_{tail}$ and large values of  the ratio of average
energies: $R_{tail} \gta 0.06$ and   $r_E \gta 1.2$.    The identification of this scenario (and therefore of the inverted mass hierarchy)  appears difficult due to the almost complete overlap with the regions of the case C.
Another
indication of this scenario would be the result
 $r_E < 0.95$.
Again, conclusions on $\theta_{13}$ depend on the specific value of $R_{tail}$.

\item  The case C can not be easily identified. An indication of this possibility would be the
result  $r_E \gta 1.2$ and $r_\Gamma \gta 1.05$.
This combination would exclude  B and A, and therefore indicate $P_H > 0$, corresponding to
 small values of $\theta_{13}$: $\t13\lta few \cdot 10^{-4}$  (see fig. \ref{fig:phfig}).\\

\end{itemize}

The  scenarios in which $0< P_H<1 $ are not shown  in the figures. For n.h. and $0<P_H<1$
we expect the allowed region to be intermediate  between the  regions found for A and C. Similarly, for
i.h. and $0<P_H<1$ the region of possible values of parameters  is intermediate between the regions of
cases B  and C.  For this reason, the conclusions we derived from the figures \ref{fig:scatter1}-\ref{fig:scatter4}   have essentially an exclusion character and not the character of establishing one of the scenarios A, B, C.

It is clear that the potential of the method we have discussed 
depends on the statistics 
and therefore on the distance from the supernova. Some estimations are presented in sec. \ref{sec:res} and fig. \ref{fig:Rplot2}.
For a relatively close star  ($D\lta  4$ kpc) the error bars are substantially smaller than the field of points so that the 
discrimination of different possibilities becomes possible. 
\\

It is easy to understand the effect on the scatter plots of
choosing more conservative intervals of the parameters of the original
neutrino fluxes. In particular, for smaller differences of the average
energies in the different flavors the spectral distortions due to
conversion become smaller (see e.g. \cite{Dighe:1999bi}). The results
approach those expected in absence of conversion and therefore are the
same in the three secenarios, A, B, C. The corresponding points in the
scatter plots would be located where the regions for the three cases
are closer or overlap.  Clearly, any sensitivity to the neutrino
mixings and mass hierarchy is lost in this situation.

%%%%%%%%%%%%%%%%%%%%%%%%%%%%%%%%%%%%%%%%%%%%%%%%%%%%%%%%%%%%%%%%%%%%%%%%%%
\section{The method  of the high energy  tails}
\label{sec:3.1}
%%%%%%%%%%%%%%%%%%%%%%%%%%%%%%%%%%%%%%%%%%%%%%%%%%%%%%%%%%%%%%%%%%%%%%%%%%%%%

%%%%%%%%%%%%%%%%%%%%%%%%%%%%%%%%%%%%%%%%%%%%%%%%%%%%%%%%%%%%%%%
\subsection{ $R_{tail}$ and $r$.}
%%%%%%%%%%%%%%%%%%%%%%%%%%%%%%%%%%%%%%%%%%%%%%%%%%%%%%%%%%%%%%

The uncertainties related to the original neutrino fluxes can be substantially reduced
if

\begin{itemize}

\item
ratios of the electron neutrino and antineutrino signals are considered;

\item
the  high energy tails of spectra
are used.

\end{itemize}

The key point is that in the high energy tails the fluxes of
non-electron neutrinos dominate, and moreover, these fluxes are nearly equal.

In what follows we study the possibility to use
the ratio $R_{tail}(E_L, \bar{E}_L)$ introduced in (\ref{eq:rat}) to establish the mass
hierarchy and to probe $\theta_{13}$.
As we have seen in the previous sections
the dependences of the neutrino and antineutrino
signals on $\theta_{13}$ and on the sign of $\Delta m^2_{13}$ are different.

Let us consider the predictions for $R_{tail}$ in details.
This ratio can be written  in terms of the original neutrino fluxes
and the survival probabilities $p$ and $\bar p$ as:
\be
R_{tail}(E_L,\bar{E}_L) \simeq {1- \langle p \rangle  +  \alpha(E_L) \langle p \rangle
\over 1- \langle {\bar p}\rangle  +
\bar{\alpha}(\bar{E}_L) \langle \bar p \rangle }~ Q(E_L,\bar{E}_L)~.
\label{eq:simpl}
\ee
 Here the brackets  $\langle  \rangle$ denote the averaging over the corresponding energy intervals
(we  have taken into account the weak energy dependence of $p$ and $\bar{p}$).

The quantity $Q$ is defined as:
\be
Q (E_L,\bar{E}_L) \equiv \frac{N_{x}^0(E>E_L)}{N_{\bar x}^0(E>\bar{E}_L)}~,
\label{q}
\ee
where  $N_{x}^0$, $N_{\bar x}^0$ are the numbers of events calculated
according to eqs.  (\ref{intsp}) and (\ref{eq:dnem}) with the fluxes
$F_{ x}^0$ and $F_{\bar x}^0$ respectively.
Due to the near equality of the fluxes $F^0_x$ and
$F^0_{\bar x}$, the astrophysical uncertainties in $Q$ are almost cancelled for an optimized choice of the cuts $E_L$ and $\bar E_L$ (see sect. \ref{sec:secq}).

In (\ref{eq:simpl}) $\alpha$ and  $\bar \alpha$ are
the parameters which describe the relative contributions  of the original
$\nu_e$  and $\bar \nu_e$ fluxes to the numbers of events above the energy cuts:
\be
\alpha(E_L) \equiv \frac{N^0_e (E_L)}{N^0_x (E_L)} ~, \hskip 1truecm  \bar \alpha(\bar E_L) \equiv
\frac{N^0_{\bar e}(\bar{E}_L)}{N^0_{\bar x}(\bar{E}_L)}~.
\label{alphas}
\ee\\

Let us introduce the ratio:
\be
r(E_L,\bar{E}_L) \equiv \frac{R_{tail}(E_L,\bar{E}_L)}{Q(E_L,\bar{E}_L)}~,
\label{rsmall}
\ee
which should not depend substantially on the features of the detectors,
and is known once $R_{tail}$ is measured.
According to (\ref{eq:simpl}) we have:
\be
r (E_L,\bar{E}_L) \simeq {1- \langle p \rangle ( 1 -   \alpha(E_L)) \over
1 -  \langle {\bar p} \rangle (1 -  \bar{\alpha}(\bar{E}_L))}~.
\label{eq:rdef}
\ee
The ratio $R_{tail}$ is a measurable quantity, and, as we will argue in the next section, 
$Q(E_L, \bar{E}_L)$ can be reliably predicted. Therefore, the study of $R_{tail}$ can be reduced to that of the quantity $r(E_L,\bar{E}_L)$.

%In what follows we will consider
%$r(E_L,\bar{E}_L)$ as a measurable quantity.\\

%%%%%%%%%%%%%%%%%%%%%%%%%%%%%%%%%%%%%%%%%%%%%%%%%%%%%%
\subsection{The factor $Q$}
\label{sec:secq}
%%%%%%%%%%%%%%%%%%%%%%%%%%%%%%%%%%%%%%%%%%%%%%%%%%%%%%%

The factor $Q (E_L,\bar{E}_L)$ depends on (i) the  energy cuts $E_L$ and  ${\bar E_L}$,
(ii) the fluxes
$F^0_x$ and $F^0_{\bar x}$  and (iii) on the features of the detection method:  cross sections, efficiencies,
energy  resolutions, etc. (for the latter quantities we follow ref. \cite{Lunardini:2001pb}).
Since $F^0_x$ and $F^0_{\bar x}$ have  almost equal luminosities,
temperatures  and pinching parameters, the astrophysical uncertainties affect $Q$ only weakly.
Moreover, the effect of uncertainties can be further suppressed  by choosing different thresholds for
neutrino and antineutrino events, such to ``compensate" the  difference in  the temperatures.

This can be seen from
the  approximate form of the ratio  $Q$ (eq. (\ref{q})):
\beq
Q & \simeq & const~ \frac{T_x }{T_{\bar x}}  \frac{P(E_L/T_x)}{P(\bar E_L/T_{\bar x})} e^{(-E_L/T_x +
\bar E_L/T_{\bar x})}~,
\label{qappr} \\
& \sim & const~  \left( \frac{T_{\bar x}}{T_x} \right)^3  \left( \frac{E_L}{\bar E_L} \right)^4
e^{(-E_L/T_x +  \bar E_L/T_{\bar x})}~,
\label{qasy}
\eeq
which can be derived from the analytical formulas (\ref{nevsimpl}, \ref{nevasy}). In eq. (\ref{qasy})
we took $L_{\bar x} = L_{x}$ and $\eta_x =\eta_{\bar x}$. For simplicity we considered  the same energy dependence of the $\nue$ and $\barnue$ cross-sections,
$\sigma (\nu) \propto \sigma (\bar {\nu}) \propto E^2$, and neglected detector parameters like energy resolution, efficiency, etc..
 From eq. (\ref{qasy}) we see  that  if $T_x = T_{\bar x}$ the ratio  $Q$ reduces to  a constant   provided that  $E_L=\bar E_L$  is taken.
%as it is expected.
In the presence of  a difference $\delta_x$ between the $\nux$ and $\barnux$ temperatures, eq. (\ref{deltax}), the  dependence of $Q$ on $T_x$ does not
cancel exactly. However,  one can find  values of the energy cuts,
$E_L \neq \bar{E}_L$,
for
which this dependence becomes very  weak in the relevant range of $T_x$:
\be
\frac{\partial Q}{\partial T_x} \approx 0 \hskip 2truecm {\rm for} ~~~T_x = (5 \div  8)
~{\rm  MeV}~.
\label{mincond}
\ee
In the approximation $\delta_x/T_x \ll 1$ and $\bar E_L/E_L -1 \ll 1$, Eqs. (\ref{qasy})
and (\ref{mincond}) lead to 
\be
\frac{\bar E_L}{E_L} \simeq  1+ \frac{\delta_x}{T_x} \left( 2 - 3 \frac{T_x}{E_L}\right) ~.
\label{cutest}
\ee
Taking $E_L =45 $ MeV,  $T_x=7$ MeV and $\delta_x = 0.5$ MeV,  from   (\ref{cutest})   we get
$\bar E_L  \simeq 50 $ MeV.
A more detailed numerical study,  which takes into account also the different energy dependences  of the cross-sections, give $\bar E_L \simeq 55 $ MeV.

\begin{figure}[hbt]
\begin{center}
\epsfig{file=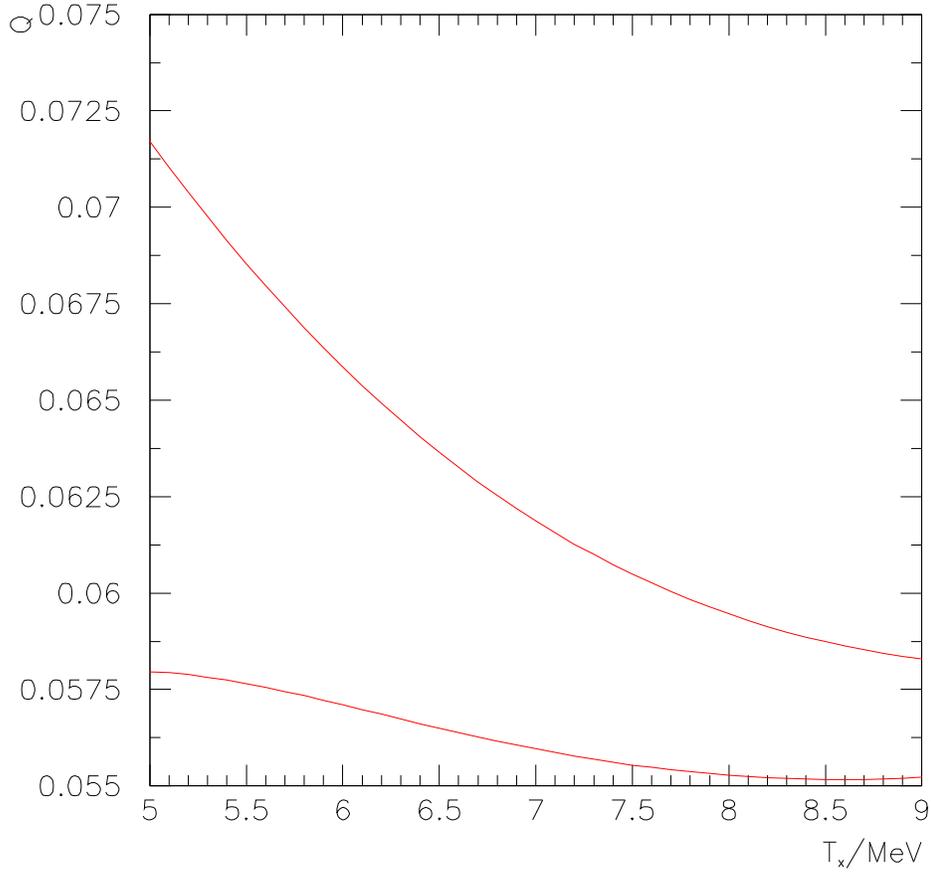, width=14truecm}
\end{center}
\caption{The factor $Q$ (region between  the lines), eq. (\ref{q}), as a function of the temperature $T_x$. The
  width of the region corresponds to the interval $T_{\bar x} -
  T_x=0.35 \div 0.5$ MeV.  The cuts $E_L=45$ MeV and $\bar E_L=55$ MeV
  have been taken.}  
\label{fig:q} 
\end{figure}

The dependence of the factor $Q$ on the temperature  $T_x$ is shown in fig. \ref{fig:q} (region between the lines), 
for $E_L=45$ MeV, $\bar E_L=55$ MeV and   $\delta_x = 0.35 \div 0.5 $ MeV.
 As it appears from the figure, over the interval  $T_x=5 \div 9 $ MeV, $Q$ varies by about
$\sim 10\%$  of its value,
therefore it can be taken as a constant with $\sim 10\%$ associated error:
\be
 Q=0.0635 (1 \pm 0.13)~ .
\label{qval}
\ee

 As it can be understood from eqs. (\ref{qasy}) and (\ref{cutest}), the error on $Q$ is dominated by  the uncertainty in the value of $\delta_x$: fixing $\delta_x=0.5$ MeV we get  $Q=0.0566 (1 \pm 0.05)$ in the relevant interval of values of $T_x$.

%%%%%%%%%%%%%%%%%%%%%%%%%%%%%%%%%%%%%%%%%%%%%%%%%%%%%%%%%%%%%%%%%%%%%%%%%%%%%%%%%%%%%%%%%%
\subsection{ $\alpha$ and $\bar \alpha$. Threshold Energies}
%%%%%%%%%%%%%%%%%%%%%%%%%%%%%%%%%%%%%%%%%%%%%%%%%%%%%%%%%%%%%%%%%%%%%%%%%%%%%%%%%%%%%%%%%

The ratios $\alpha$ and $\bar \alpha$, eq. (\ref{alphas}), depend
(i) on the features of the original electron and non electron
neutrino fluxes, (ii) on the energy cuts $E_L$, $\bar E_L$ and
(iii) on the characteristics of the detectors consider. Using the
Maxwell-Boltzmann  approximation, eq. (\ref{mb}), and the results
(\ref{nevsimpl},\ref{px}), one gets the following approximate
expression for $\alpha$: \be \alpha \simeq \frac{L_e T_e }{L_x
T_x} \frac{P(E_L/T_e)}{P(E_L/T_x)} e^{-E_L (1/T_e - 1/T_x)}~,
\label{alphappr} \ee where the dependence $\sigma (\nu)  \propto
E^2$ has been considered and other detection parameters
(efficiency, energy resolution, etc.) neglected.  The polynome
$P(x)$ is given in eq. (\ref{px}). The dependence of ratio
$P(E_L/T_e)/P(E_L/T_x)$  on $E_L$ is weak  and cancels in the
asymptotics $E_L\gg T_e,T_x $.  In this limit we have:
\be
\alpha \approx \frac{L_e}{L_x} \left( \frac{T_x}{T_e} \right)^3
e^{-E_L (1/T_e - 1/T_x)}~.
\label{alphasy}
\ee
From eqs.
(\ref{alphappr}, \ref{alphasy}) it follows that, for a given $T_x$, the ratio
$\alpha$ decreases with the decrease of $T_e$ and with the
increase of the cut $E_L$.  We have $\alpha \rightarrow 0$ in the
limit $E_L/T_e \rightarrow \infty$.  In particular, this implies
that the contribution of  $\alpha$ to the ratio $r$ (eq.
(\ref{eq:rdef})) can be reduced to a small correction provided
that a high enough cut is chosen.

Results analogous to eqs. (\ref{alphappr}, \ref{alphasy}) and similar considerations hold for $\bar{\alpha}$.
As a consequence of inequalities of the average energies, eq. (\ref{temp}), and of the nearly equality of luminosities
we  have
\be
\bar \alpha (\bar E_L)  > \alpha ( E_L)~,
\label{ineq}
\ee
%independently of $E_L$ and $\bar E_L$.
provided that   $\bar E_L $ and $E_L$ do not differ strongly.
%; this condition is satisfied by  the optimized choice of the cuts discussed in sec. \ref{sec:secq}.
\\

Though approximate, the expression (\ref{alphappr}) is in acceptable agreement with more accurate numerical calculations.
Scanning the intervals of parameters of the original neutrino
spectra (\ref{temp} - \ref{fluxes}) we find the
following ranges of $\alpha$ and $\bar \alpha$:
\be
\alpha (45 {\rm MeV}) \simeq 0 - 0.42~, ~~~~~ \bar \alpha (55 {\rm MeV}) \simeq 0 - 0.95~.
\label{avalues}
\ee
As an example, taking   the ``traditional'' scenario
with equal luminosities in the different flavors and hierarchical (unpinched) spectra -- with $T_x=7$
MeV, $T_{\bar e}=5$ MeV and $T_e = 3.5$ MeV --  we get
\be
\alpha (45 {\rm MeV})  \simeq  0.05, ~~~~~ \bar \alpha (55 {\rm MeV}) \simeq 0.16 .
\label{avalues1}
\ee

\begin{figure}[hbt] \begin{center}
\epsfig{file=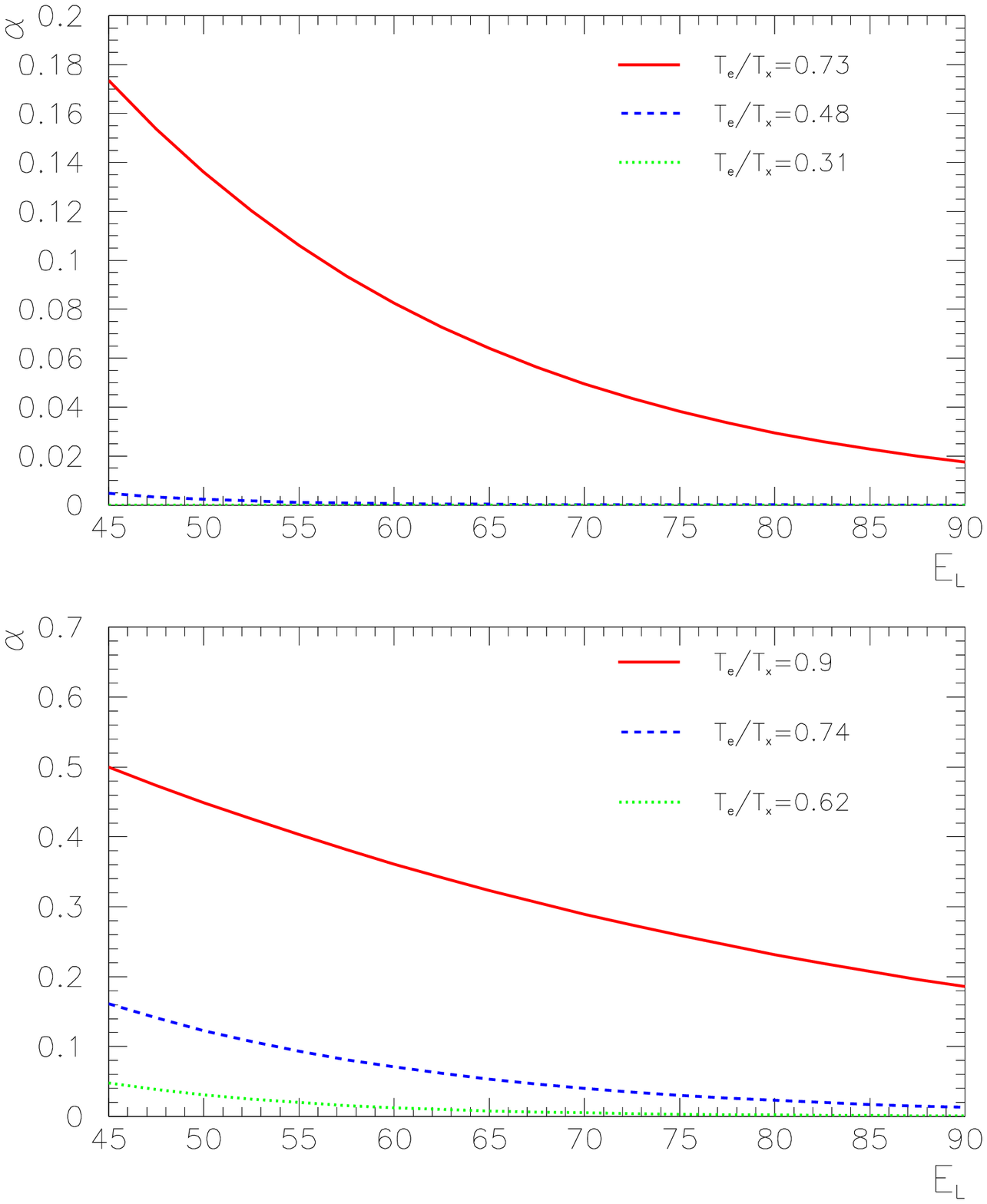,width=16truecm} 
\end{center} 
\caption{The dependence of the ratios
$\alpha$ (upper panel) and $\bar \alpha$ (lower panel), eq. (\ref{alphas}), on the energy cuts
$E_L$ and $\bar E_L$ respectively, for $T_x=7$ MeV, $L_{e}=L_{\bar e}=L_{x}$,
$\eta_e=\eta_{\bar e}=\eta_x = 0$ and different values of the $\nue $
and $\barnue$ temperature (expressed in term of ratios in the
legend).}  
\label{fig:alphex} 
\end{figure}

In fig. \ref{fig:alphex} we show the dependence of $\alpha$  and $\bar \alpha$  on the cut
energies, $E_L$ and $\bar E_L$, for $T_x=7$ MeV, $L_{e}=L_{\bar e}=L_{x}$, $\eta_e=\eta_{\bar e}=\eta_x = 0$ and different values of the $\nue $ and $\barnue$ temperature.
It can be seen that the decrease of $\alpha$ and $\bar \alpha$
 with the cuts is indeed exponential, in agreement with eq. (\ref{alphappr}). 
However, if  the
 hierarchy of the spectra is not strong (solid lines in the figure), for our choice of the cuts
 $\alpha$, and especially $\bar \alpha$, are not negligible with respect to the survival
 probabilities: $\alpha \lta \bar \alpha \sim p,\bar p\lta 1$.
In this case, the effects of the $\alpha$ and $\bar  \alpha$
 terms in $r$ (and therefore in $R_{tail}$) can be reduced by adopting higher energy cuts.
%Clearly, a smaller hierarchy
%of spectra requires higher cuts.
According to fig. \ref{fig:alphex}, requiring $\alpha \lta \bar \alpha \lta 0.1$ implies $E_L\lta 70$ MeV and $\bar E_L\lta 100-120$ MeV (we have considered the possibility, not shown in the figure, to have $L_e/L_x \simeq 2$ and/or $L_{\bar e}/L_x\simeq 2$, corresponding to twice as large values of $\alpha$ and $\bar \alpha$ with respect to what shown in the figure).
It is clear, however, that  the numbers of events in the tails
above these energies is strongly suppressed,  so that  no precise
measurements of $r$  are possible.

%%%%%%%%%%%%%%%%%%%%%%%%%%%%%%%%%%%%%%%%%%%%%%%%%%%%%%%%%%%%%%%%%%%%%%%%%%%%
\subsection{Large threshold energy limit}
%%%%%%%%%%%%%%%%%%%%%%%%%%%%%%%%%%%%%%%%%%%%%%%%%%%%%%%%%%%%%%%%%%%%%%

For $E_L \gg \langle E_e\rangle,  \langle E_{\bar e}\rangle $  parameters  $\alpha,~ \bar \alpha$  are  negligible and
expression (\ref{eq:rdef}) becomes
\be
r \approx {1- \langle p \rangle \over 1- \langle {\bar p} \rangle}~,
\label{r}
\ee
i.e. the astrophysical and oscillation parts of  $R_{tail}$ are factorized.

\begin{figure}[hbt]
\begin{center} 
\epsfig{file=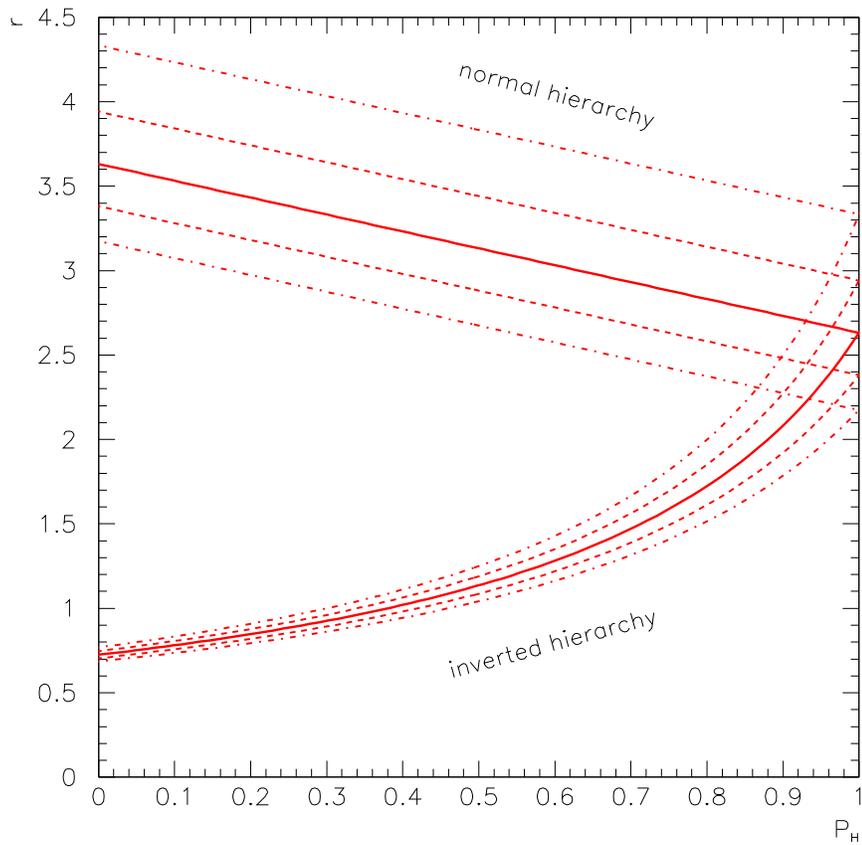, width=13truecm} 
\end{center}
\caption{The factor $r$, eq. (\ref{r}), as a function of $P_H$ for
    $\y12=0.38$ and both normal and inverted hierarchy (solid line).
    The bands within the dashed and dashed-dotted lines correspond to
    $10\%$ and $20\%$ uncertainty on $\y12$ respectively.}
\label{fig:rfact} 
\end{figure}

Inserting the explicit expressions for the permutation factors,
eqs. (\ref{pue3nh} - \ref{pue3ih}),
and neglecting  $ \sim \t13$ terms, we find
\be
r= \frac{1}{\sin^2 \theta_{12}} - \langle P_H~ \rangle  ~~~~{\rm (n.h.)},
\label{r2a}
\ee
\be
r = \frac{1}{\cos^{-2} \theta_{12} - \langle P_H~ \rangle}~ ~~~~{\rm (i.h.)},
\label{r2}
\ee
where $\langle P_H~ \rangle$ is the value of the jumping probability averaged
over the energy interval under consideration.
The ratio $r$ is shown in fig. \ref{fig:rfact} as  a function of $\langle P_H~ \rangle$ for
different values of $\y12$. In the limit of $\langle P_H \rangle \rightarrow 1$,  i.e.
very  small $\theta_{13}$, we get
\be
r_1 = \frac{1}{\tan^2\theta_{12}}~
\label{limit1}
\ee
for both hierarchies.
In general one finds
\be
r ~ \begin{cases}
\geq 1/\y12 & ~~~\text{(n.h.)} \\
\leq 1/\y12 & ~~~\text{(i.h.)}
\end{cases}~,
\label{hiercheck}
\ee
and in the limit $\langle P_H  \rangle \rightarrow 0$ (large $\theta_{13}$)
\be
r \rightarrow r_0 = \begin{cases}
1/\sin^2\theta_{12}    &~~    \text{(n.h.)} \\
\cos^2\theta_{12}      &~~    \text{(i.h.)}
\end{cases}~.
\label{hiercheck0}
\ee
For both the hierarchies $r$  decreases  with the increase of the mixing $\theta_{12}$.

Once (i) $R_{tail}$ is measured in  experiments,  (ii) $Q$ is calculated and (iii) $\y12$ is known
from the solar neutrino and the KamLAND experiments, 
%\cite{},
the inequalities (\ref{hiercheck}) can be used to establish the mass hierarchy.
Assuming that $\y12$ will be measured
with $20\%$ accuracy,  from the figure \ref{fig:rfact} (see dash-dotted lines;
 the central value $\y12=0.38$ has been taken as an example)
we find that:

\noindent
1). If $r < 2$ the inverted hierarchy will be selected. In particular,
if $r \sim 0.7$ even a poor accuracy in measurements of $r$, say $30 - 50\%$,  will be enough.
Moreover, the  (energy-averaged) jump probability  will be restricted:   $\langle P_H \rangle < 0.3$,  leading to a lower bound on $\t13$.

\noindent
2). If $r = (2.0 - 3.5)$ both types of hierarchy are possible.
For inverted hierarchy one can put a rather strong
bound on the jumping  probability: $\langle P_H \rangle > 0.8$, which would imply
an upper bound on $\tan^2 \theta_{13}$. In the case of normal  hierarchy
no bound on $P_H$ appears.

\noindent
3). For $r > 3.5$ the normal hierarchy with $\langle P_H \rangle  < 1$ is selected.\\

According to eqs. (\ref{r2a})-(\ref{r2}) we have: 
\be
\langle P_H \rangle = \frac{1}{\sin^2 \theta_{12}} - r ~~~~~{\rm (n.h.)},
\label{ph-norm}
\ee
\be
\langle P_H \rangle= \frac{1}{\cos^2 \theta_{12}} - \frac{1}{r}  ~~~~~{\rm (i.h.)}.
\label{ph-inv}
\ee
The result for $\langle P_H \rangle$ can be then
transferred into a result for $\tan^2 \theta_{13}$ using
fig. \ref{fig:phfig} and the expressions
(\ref{phform}, \ref{ena}).
As follows from
the fig. \ref{fig:phfig}, even for very precise measurements of $\langle P_H \rangle $ the
uncertainty  on the density profile will lead to a factor of 3
uncertainty on $\t13$.  To have a sensitivity to
$\sin^2 \theta_{13}$ in the adiabatic region one needs to measure  the permutation effects
at the level  2\% or smaller.
This looks practically impossible already in view of uncertainties on the
determination of $\tan^2 \theta_{12}$.\\

%%%%%%%%%%%%%%%%%%%%%%%%%%%%%%%%%%%%%%%%%%%%%%%%%%%%%%%%%%%%%
\subsection{The  general case}
\label{gencase}
%%%%%%%%%%%%%%%%%%%%%%%%%%%%%%%%%%%%%%%%%%%%%%%%%%%%%%%%%%%%%%

To have significant statistics in the tails the energy cuts
should not be too large. In this case $\alpha$, $\bar \alpha$
can not be neglected and for the ratio $r$ we should use the complete expression
(\ref{eq:rdef}). It can be rewritten
in terms of the average
 jumping probability
$\langle P_H \rangle$ and $\theta_{12}$ as
\be
r = \frac{1 + \tan^2 \theta_{12}  - \langle P_H \rangle \tan^2 \theta_{12}(1 - \alpha)}
{\tan^2\theta_{12}  + \bar \alpha } ~~~~~~ {\rm (n.h.)},
\label{rexpl-n}
\ee
\be
r = \frac{1 +    \alpha \tan^2 \theta_{12}}
{1 +  \tan^2 \theta_{12} -  \langle P_H \rangle (1 - \bar \alpha)} ~~~~~~{\rm (i.h.)} .
\label{rexpl-i}
\ee

For very small $\theta_{13}$  ($P_H = 1$)  both hierarchies  give the same result:
\be
r_1 =  \frac{1 +  \tan^2 \theta_{12} \alpha}
{\tan^2\theta_{12} + \bar \alpha} ~.
\label{rexpl-1}
\ee
Comparing $r_1$ with the general expressions (\ref{rexpl-n}, \ref{rexpl-i}) we get the
inequalities
\be
r > r_1~~ {\rm (n.h.)}, ~~~~~~~ r <  r_1~~{\rm  (i.h.)} ~,
\ee
which  provide a test of the type of mass hierarchy. \\

In the adiabatic case ($P_H \approx 0$):
\be
r_0^{(n)}  = \frac{1 +  \tan^2 \theta_{12}}
{\tan^2\theta_{12} +  \bar \alpha} ~~~~~~ {\rm (n.h.)},
\label{rexpl-na}
\ee
\be
r_0^{(i)} = \frac{1 +   \alpha \tan^2 \theta_{12}}
{1 + \tan^2\theta_{12}} ~~~~~~{\rm (i.h.)} .
\label{rexpl-ia}
\ee
These quantities turn out to be  the upper (for n.h.) and the lower (for i.h.) bounds on $r$:
\be
r_1 \leq  r  \leq  r_0^{(n)} ~~~~~ {\rm (n.h.)},
\ee
\be
r_0^{(i)} \leq   r \leq  r_1~~~~~ {\rm  (i.h.)} .
\ee\\

The averaged probability, $\langle P_H \rangle$, can be expressed in terms of measurable
quantities as:
\be
\langle P_H \rangle = \frac{1}{(1 - \alpha)}
[1 + \cot^2\theta_{12} - r(1 + \bar\alpha \cot^2\theta_{12} )]~~~~~
{\rm  (n.h.)}
\label{nh}
\ee
\be
\langle P_H \rangle = \frac{1}{(1 - \bar \alpha)}
[1  + \tan^2\theta_{12}   - \frac{1}{r}(1 +  \alpha \tan^2\theta_{12})] ~~~~~
{\rm  (i.h.)}.
\label{ih}
\ee\\

We can  define the  energy $E_{av}$ in such a way that
\be
\langle P_H \rangle \equiv   P_H (E_{av}).
\ee
Using the Landau-Zener formula for $P_H$ we find  expression for
$\sin^2 \theta_{13}$ in terms of observables:
\be
\t13 \simeq
- 2.05 \cdot 10^{-5}  \left({E_{av} \over {\rm MeV} } \right)^{2/3}
\left({10^{-3} ~{\rm eV}^2  \over |\Delta m^2_{32}|}\right)^{2/3} C^{-1/3}
\ln(\langle  P_H \rangle)  ~~~~(a = n, i),
\label{gfun}
\ee
where $\langle  P_H \rangle$ are given in Eqs. (\ref{nh}, \ref{ih}).

%%%%%%%%%%%%%%%%%%%%%%%%%%%%%%%%%%%%%%%%%%%%%%%%%%%%%%%%%%%%%%%%%%%%%%%%%%%%%%%%%%%%%
\subsection{Results}
\label{sec:res}
%%%%%%%%%%%%%%%%%%%%%%%%%%%%%%%%%%%%%%%%%%%%%%%%%%%%%%%%%%%%%%%%%%%%%%%%%%%%%%%%%%%%

Let us first estimate the influence of $\alpha$ and $\bar\alpha$ on
the ratio $r = r(\alpha, \bar\alpha)$. We denote here and later
by  $x^+$ and $x^-$ the upper and lower edges
of the uncertainty interval  of the variable $x$ ($x = \alpha, ~ \bar\alpha, ...$).
Using  the general expressions for
$r$, eq. (\ref{rexpl-n}), and taking into account the restriction (\ref{ineq})
we find the upper ($r_{max}$) and the lower ($r_{min}$)  bounds on $r$ for
given values of $\theta_{12}$ and $P_H$. In the case of the n.h.  we get
\be
r_{min} (P_H)  \approx  r(0,  \bar \alpha^+) =
\frac{1 + \tan^2 \theta_{12}  - \langle P_H \rangle \tan^2 \theta_{12}}
{\tan^2\theta_{12}  + \bar \alpha^+},
\label{rmin-nh}
\ee
\be
r_{max}(P_H)  \approx  r(0, 0) =
\frac{1}{\sin^2 \theta_{12}}  - \langle P_H \rangle
\label{rmax-nh}
\ee
which coincides with expression in the high energy limit.
 Here the minimum values $\alpha^-$ and $\bar \alpha^-$ have been set to zero for simplicity.

For inverted hierarchy we find the lower limit:
\be
r_{min} (P_H) = r(0,  \bar \alpha^+)  =
\frac{1}{1 +  \tan^2 \theta_{12} -  \langle P_H \rangle (1 - \bar \alpha^+)}~,
\label{rmin-ih}
\ee
while,
due to the restriction (\ref{ineq}), the upper limit has a more complicated
dependence  
\be
r_{max}(P_H) \approx
\begin{cases}
r(\alpha^+, \alpha^+) =
 \left[ 1 + \alpha^+ \tan^2 \theta_{12}  \right]
\left[ 1 +  \tan^2 \theta_{12} -  \langle P_H \rangle (1 - \alpha^+) \right]^{-1}
& \text{ $P_H\leq \y12 $} \\
r(0, 0)  =  \left[ 1 +  \tan^2 \theta_{12} -  \langle P_H \rangle \right] ^{-1}
& \text{ $P_H> \y12 $} 
\end{cases}~.
\label{rmax-ih}
\ee
These limits are shown in  fig. \ref{fig:rfactgen}   for $\tan^2 \theta_{12}=0.38$ and $\alpha^+=\bar \alpha^+=1$, together  with predictions for
 $\alpha=0.17$ and $\bar\alpha=0.34$ as an example.
As follows from the figure, with the present  knowledge of the original spectra
the mass hierarchy can be identified from the tail method only if
\be
r > 2.6
\ee
will be found. 
%hese values are on the upper side of predictions interval.
In this case also the upper bound
on $\langle P_H \rangle$, and consequently a lower bound on $\sin^2 \theta_{13}$
can be obtained:
\be
\langle P_H \rangle \leq \frac{1}{\sin^2 \theta_{12}} - r.
\label{upper}
\ee

For $r < 2.6$ neither the hierarchy nor $\sin^2 \theta_{13}$  can be found.
For the most plausible scenario (normal mass hierarchy,
large $\sin^2 \theta_{13}$ ($P_H = 0$) we get $r \sim 2$,  that is, below
the identification interval.

More can be said if the hierarchy is identified by
some other method ({\it e.g.}, from the Earth matter effect or the  shock wave effects, see secs. \ref{sec:9} and \ref{sec:10}).
Thus in the case of normal hierarchy, in addition to the upper bound
(\ref{upper}), one can put a lower bound on $\langle P_H \rangle$ if $r = 0.7 -1$:
\be
\langle P_H \rangle \geq \frac{1}{\sin^2 \theta_{12}} -
r\left(1 + \frac{\bar \alpha^+}{\tan^2\theta_{12}}\right),
\label{lower}
\ee
and, correspondingly, one gets an upper bound on $\sin^2 \theta_{13}$.

If  the inverted hierarchy is identified and  $r = 1 - 2.6$ is found, one can put
a lower bound on $P_H$
(upper bound on $\sin^2 \theta_{13}$):
\be
\langle P_H \rangle \geq 1 + \tan^2 \theta_{12} - 1/r.
\label{lower-ih}
\ee
The discrimination power increases with the increase of the low energy cut (i.e., decreases with $\alpha^+ $ and $\bar \alpha^+ $),
however the statistics decreases fast correspondingly, so that such a possibility can
be  realized only in the case of  supernova  at small distances.

\begin{figure}[hbt]
\begin{center} 
\epsfig{file=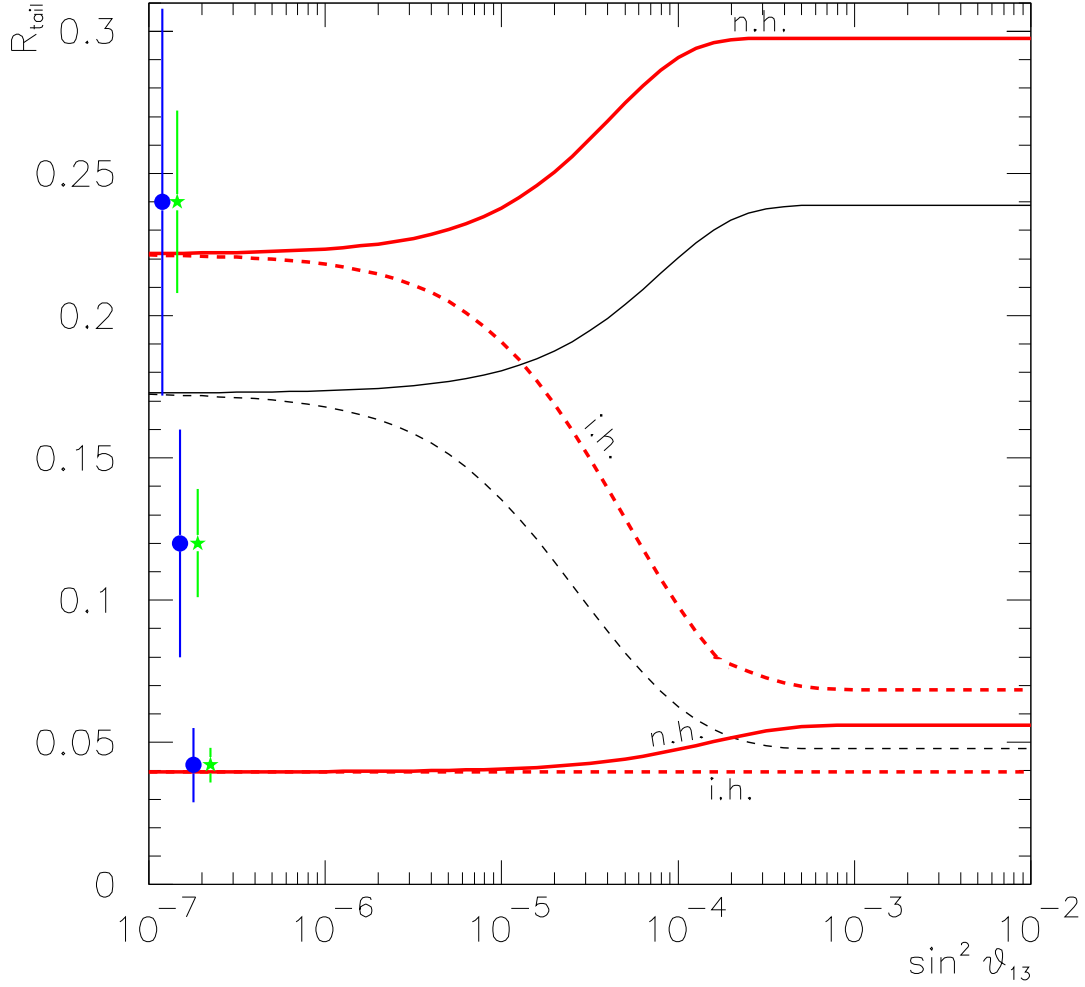, width=16truecm} 
\end{center}
  \caption{ The ratio $R_{tail}$ as a function of $\t13$ for normal mass
    hierarchy (solid lines) and inverted hierarchy (dashed lines),   in absence of Earth crossing. The
    bands within the thick lines represent the values that $R_{tail}$ can
    take once all the possible theoretical uncertainties are
    considered (see text). The thin lines refer to the ``ideal'' case
    $\alpha=\bar \alpha=0$.    We have taken $\y12=0.38$, $E_L=45$ MeV and  ${\bar E}_L=55$ MeV. 
Three examples of experimental results for
    $R_{tail}$ are shown with two different (98\% C.L.) error bars (marked by
    bullets and stars) corresponding to $T_x=7$ MeV,  $L_e=L_{\bar e}=L_x=L_{\bar x}=5\cdot 10^{52}$ ergs, and $D=8.5$ kpc
    and $D=4$ kpc respectively. }  
\label{fig:Rplot2} 
\end{figure}

The features of $r$ are reproduced also in  the
observable $R_{tail}$, eq. (\ref{eq:rat}). We find the minimal and maximal values of
$R_{tail} = R_{tail}(E_{av},(|\Delta m^2_{32}|),C,\alpha,\bar \alpha,\theta_{12},\theta_{13},Q)$
for a given  $\theta_{13}$
taking into account also uncertainties on $\Delta m_{23}$, $\theta_{12}$,
$C$, $E_{av}$.   For the normal mass hierarchy, eqs. (\ref{rmin-nh},\ref{rmax-nh}) and (\ref{phform},\ref{ena}) give:
\beq
&& R_{max}(\theta_{13})= R(E^{-}_{av},(|\Delta m^2_{32}|)^{+},C^{+},0,0,\theta^{-}_{12},\theta_{13},Q^{+})
\nonumber \\
&&R_{min}(\theta_{13})= R(E^{+}_{av}, (|\Delta m^2_{32}|)^{-},C^{-},0,\bar
\alpha^+,\theta^{+}_{12},\theta_{13},Q^{-})~.
\label{possRnh}
\eeq
Similarly, from eqs. (\ref{rmin-ih},\ref{rmax-ih}) we get for inverted mass hierarchy:
\beq
&& R_{max}(\theta_{13})=\begin{cases}
R(E^{+}_{av},(|\Delta m^2_{32}|)^{-}, C^{-},\alpha^+, \alpha^+,\theta^{-}_{12},\theta_{13},Q^{+})  &
\text{ $P_H\leq \y12 $}\\
R(E^{+}_{av},(|\Delta m^2_{32}|)^{-}, C^{-},0~~, 0~~,\theta^{-}_{12},\theta_{13},Q^{+})   & \text{ $P_H>
\y12 $}
\end{cases}
\nonumber \\
&&R_{min}(\theta_{13})=R(E^{-}_{av}, (|\Delta m^2_{32}|)^{+},C^{+},0,\bar
\alpha^+,\theta^{+}_{12},\theta_{13},Q^{-}).
\label{possRih}
\eeq
$R_{max}$ and $R_{min}$ as
functions of $\t13$ for n.h. and i.h. are shown in fig. \ref{fig:Rplot2}.
In our calculations  we have used the following uncertainties:

\begin{itemize}

\item The effective energy $E_{av}$  practically coincides with the value energy cut, $E_L$ for n.h. or
$\bar E_L$ for i.h.. A $10\%$ uncertainty on its value was adopted to be conservative.

\item A $20\%$ error was taken on $|\Delta m^2_{32}|$.

\item The parameter $C$ was taken to be in the interval $C= 1 -15$, so that $C^{-1/3}=0.4 -1$.

\item $\y12$ was assumed to be known with  $10\%$ accuracy.

\item the parameter $Q$ was taken according to eq. (\ref{qval}) .

\item  $\alpha$ and  $\bar \alpha$ were taken in the intervals (\ref{avalues}).

\end{itemize}

We mark that these assumptions on the uncertainties are consistent with the intervals (\ref{temp},\ref{fluxes}), which have been taken to generate the scatter plots \ref{fig:scatter1}-\ref{fig:scatter4}.

Notice that for i.h.  $R_{min}$ is independent of $\theta_{13}$. This is due to the fact
that $R_{min}$ is realized  in our analysis at  $\bar \alpha = \bar \alpha^+\simeq 1$ (equal permuted fluxes), for which
no conversion effect appears and the dependence of $r$ on $P_H$ cancels (see eq. (\ref{rexpl-i})). \\

To estimate the sensitivity of the method we have simulated three
possible experimental results for two different distances to the supernova, 
$D=4$ and $D=8.5$ kpc and the integrated luminosity $L_\beta = 5 \cdot 10^{52}~{\rm ergs}$ in each of the six neutrino (and antineutrino) species.  The error bars correspond to $99\%$ C.L..

%The study of $R_{tail}$ in the case of no Earth crossing is shown in  
%fig. \ref{fig:Rplot2}.  and use the notation $(1+\epsilon)$, where the
%number $\epsilon$ represents the relative shift due to theoretical
%uncertainties.  
In absence of Earth crossing, The following relevant intervals for $R_{tail}$ are found (see fig.
\ref{fig:Rplot2}): if 
\be 
R_{tail} \simeq Q^+/\tan^2 \theta^-_{12}~
\div~ Q^+/\sin^2 \theta^-_{12}
\ee 
the n.h. is established and a lower
bound on $\t13$ is put.  If this condition is not fulfilled it is not
possible to establish the mass hierarchy and conditional bounds on
$\t13$ can be obtained.  In particular, for

\begin{itemize}

\item  $R_{tail} \simeq  Q^+~ \div~ Q^+/\tan^2 \theta^-_{12}~ $:  an upper bound on $\t13$ is found in
the hypothesis of inverted mass hierarchy.

\item $R_{tail} \simeq  Q^-~ \div~Q^+$: no information is obtained on $\t13$.

\item $R_{tail} \simeq  Q^- \cos^2 \theta^+_{12}~  \div~ Q^- $:  an upper bound on $\t13$ is established
in the hypothesis of normal mass hierarchy.

\end{itemize}

No measurement of $\theta_{13}$ (that is, no lower and upper bound) is possible.  The sensitivity to  the n.h.
(i.h.) increases (decreases) with the increase of $\theta_{12}$.

Clearly, conclusions depends  on the statistical error on $R_{tail}$ as
given by the experiments. This in turn is determined by  the distance  to the supernova, by the neutrino
luminosities, the volumes of the detectors, etc. (see fig. \ref{fig:Rplot2}).\\

%%%%%%%%%%%%%%%%%%%%%%%%%%%%%%%%%%%%%%%%%%%%%%%%%%%%%%%%%%%%%%%%%%%%%%%%%%
\subsection{Including the  Earth matter effects}
\label{sec:9}
%%%%%%%%%%%%%%%%%%%%%%%%%%%%%%%%%%%%%%%%%%%%%%%%%%%%%%%%%%%%%%%%%%%%%

If the neutrino burst crosses the Earth before  detection the
regeneration effects in the matter of the Earth should be taken into account.
The effects increase with neutrino energy and can reach 30 - 50\% at
$E > 50$ MeV \cite{Dighe:1999bi,Takahashi:2000it,Lunardini:2001pb,Takahashi:2001dc}. With  these effects the survival probabilities  $p$ and $\bar p$ become:
\beq
p\simeq P_H P_{2e}~,
\hskip 2truecm {\bar p}\simeq {\bar P}_{1e}~, ~~~~({\rm n.h.})
\label{pearthnh}
\eeq
\beq
p\simeq P_{2e}~,
\hskip 2truecm
{\bar p} \simeq  P_H {\bar P}_{1e}~, ~~~~({\rm i.h.}),
\label{pearthih}
\eeq
where $P_{2e}$ and ${\bar P}_{1e}$ are the probabilities of $\nu_2 \rightarrow \nue$ and
${\bar \nu}_1 \rightarrow \barnue$  conversion in the Earth respectively.   They have an oscillatory
dependence on the neutrino energy and
can be written in terms of the {\it regeneration factors} \cite{Dighe:1999bi}:
\be
P_{2e} \equiv \sin^2 \theta_{12} + f_{reg},
\hskip 2truecm {\bar P}_{1e}\equiv \cos^2 \theta_{12} + {\bar f}_{reg}~.
\label{regf}
\ee
According to (\ref{pearthnh},   \ref{pearthih},  \ref{regf})
the generalization of results to  the case of the Earth matter effect is straightforward:
In the formulas of sect. \ref{gencase} one should substitute
$\sin^2 \theta_{12}  \rightarrow \sin^2 \theta_{12} + f_{reg}$,
$\cos^2 \theta_{12} \rightarrow \cos^2 \theta_{12} + {\bar f}_{reg}$
and use the averaged regeneration factor \footnote{ Rigorously, $f$ depends also on $\alpha$,
and not only on the $\nux$ original flux, as given eq.
(\ref{effregf}). However, it can be checked that the corrections due to $\alpha$ are always
negligible  for the energy cuts in consideration. An analogous conclusion holds for the
$\bar \alpha$ corrections to $\bar f$.}:
\be
f \equiv \frac{\int_{E_L}^{+\infty} \int  dE_e' {\cal R}(E_e,E_e')
{\cal E}(E_e') \int dE f_{reg}(E) F^0_x (E)
(d\sigma (E_e', E) / dE_e')}{\int_{E_L}^{+\infty} \int  dE_e' {\cal R}(E_e,E_e')
{\cal E}(E_e') \int dE (E) F^0_x (E)
(d\sigma (E_e', E)/ dE_e')}~,
\label{effregf}
\ee
for neutrino channel and  $\bar f$ (with an analogous definition)   for the antineutrino channel.

From eqs. (\ref{pearthnh})-(\ref{effregf}) one finds the following generalization of eqs. (\ref{rexpl-n},\ref{rexpl-i}):
\be
r =   \frac{1 - \langle P_H \rangle (\sin^2 \theta_{12} + f)(1-\alpha)}{
1-(\cos^2 \theta_{12}+{\bar  f})(1-\bar \alpha)} ~~~~~~ \text{(n.h.)}
\ee
\be
r = \frac{1- (\sin^2 \theta_{12} + f) (1-\alpha)}{1 -  \langle P_H \rangle (\cos^2 \theta_{12}+{\bar f}) (1-\bar
\alpha)} ~~~~~~ \text{(i.h.)}
\label{r2earth}
\ee
The quantities $f$ and  $\bar f$ depend on  $\Delta m^2_{21}$, $\theta_{12}$, on the direction to the
supernova,  on the Earth density profile, on  $T_x$ and $T_{\bar x}$ and on the energy cuts $E_L$ and $\bar E_L$.
We calculated the values of the averaged  regeneration factors using a realistic density profile of the
Earth \cite{PREM} and  taking the nadir  angles $\theta_n=84.3^\circ$ at SNO and $\theta_n=24.5^\circ$ at SK
\cite{Lunardini:2001pb}, with   $E_L=45$ MeV, ${\bar E}_L=55$ MeV,  $\Delta m^2_{21}=5 \cdot 10^{-5}~{\rm eV^2}$ and
$\y12=0.38$.  To estimate the uncertainties on  $f$  and  $\bar f$  a $10\%$ error was considered on
$\Delta m^2_{21}$ and $\y12$, while  $T_x$ ($T_{\bar x}$)  was allowed to vary in the -- rather
large  -- interval  $4.5 \div 8$ MeV ( $5 \div 8.5$ MeV).
We get
\be
f= 0.21 \div 0.24, ~~~~~{\bar f}= 0 \div 0.06.
\ee
 For different nadir angles, comparable uncertainties on $f$ and $\bar f$ are found.

\begin{figure}[hbt] 
\begin{center} 
\epsfig{file=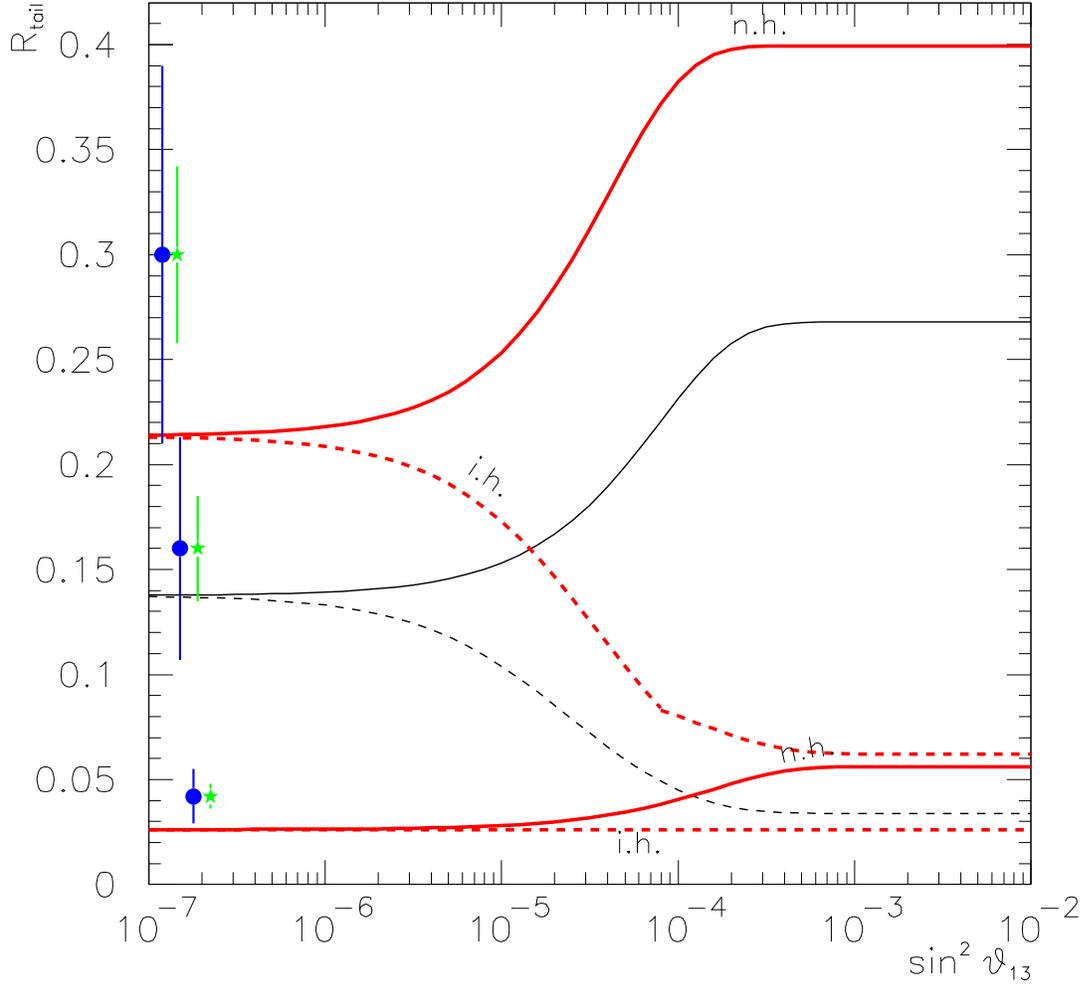, width=16truecm} 
\end{center} 
\caption{The same as fig.
  \ref{fig:Rplot2} for Earth crossing trajectories. We took  $\Delta m^2_{21} = 5\cdot 10^{-5} ~{\rm eV^2}$ 
and  the nadir angles $\theta_n=84.3^\circ$ at SNO and $\theta_n=24.5^\circ$ at  SK.}  
\label{fig:Rplotearth} 
\end{figure}

Similarly to what discussed in secs. \ref{gencase} and \ref{sec:res},  using (\ref{r2earth})
we calculate the interval of possible values of $R_{tail}$, $R_{min} \div R_{max}$,
taking into account  all the
uncertainties, including those on $f$ and  $\bar f$. The results are shown in
fig. \ref{fig:Rplotearth}. Notice that  with respect to  the no-crossing case, a stronger
dependence of $R_{tail}$ on $\theta_{13}$ is seen for n.h.,
implying a larger  sensitivity of the data to the normal mass hierarchy.
This is related to  the fact that the Earth regeneration effect is stronger in the $\nue$ than in the $\barnue$ channel.

Let us mark that the observation of oscillatory distortions due to Earth  matter effects in the energy
spectrum of the $\nue$ ($\barnue$) signal can establish the normal  (inverted) mass hierarchy \cite{Dighe:1999bi,Lunardini:2001pb} 
Once the information on the hierarchy is known, the method discussed  here will provide a
measurement or bound on $\theta_{13}$.

%%%%%%%%%%%%%%%%%%%%%%%%%%%%%%%%%%%%%%%%%%%%%%%%%%%%%%%%%%%%%%%%%%%%%%%%%%%%
\subsection{Refining  the method}
%Analytic approach for  $\alpha$'s}
%%%%%%%%%%%%%%%%%%%%%%%%%%%%%%%%%%%%%%%%%%%%%%%%%%%%%%%%%%%%%%%%%%%%%%%%%%%

\begin{figure}[hbt] \begin{center}
\epsfig{file=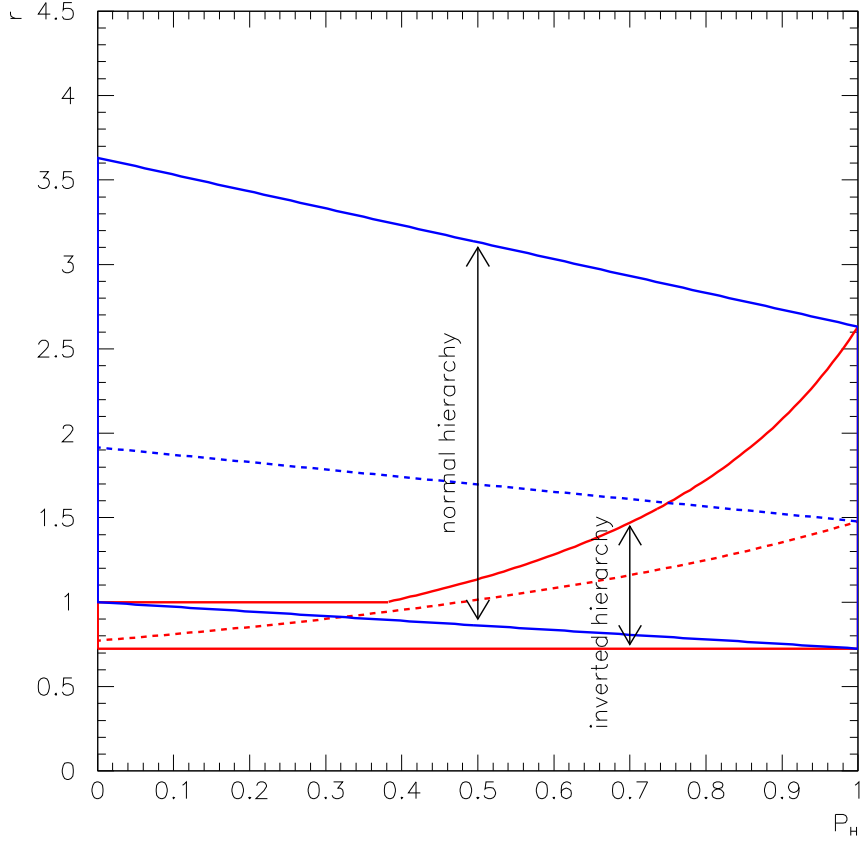, width=13truecm} 
\end{center} 
\caption{The same as fig. \ref{fig:rfact} where the assumption of
  hierarchy 
of the energy
  spectra has been relaxed to $\alpha \equiv N^0_e/N^0_x \leq 1$ and
  $\bar \alpha \equiv N^0_{\bar e}/N^0_{\bar x}\leq 1$ (eqs.
  (\ref{rexpl-n})-(\ref{rexpl-i})). The possible values of $r$ are
  those within the solid contours; the dashed lines correspond to
  $\alpha=0.17$ and $\bar \alpha=0.34$. We took $\y12=0.38$; for
  simplicity the effects of uncertainties on this parameter are not
  shown.}  
\label{fig:rfactgen} 
\end{figure}

As we saw in the previous sections appearance of the $\alpha$-terms
substantially reduces the identification power of the method
(cf. fig \ref{fig:rfactgen} and \ref{fig:Rplot2}). Let us consider the possibilities
to  refine it.

1. Further studies of the physics of supernovae can lead to more precise
predictions for the original neutrino spectra. Furthermore,
one can use results of studies of the  whole neutrino and
antineutrino spectra to reduce the uncertainty intervals of
the average energies  and fluxes, which then can be used in the tail method.

2. One can take larger  cut energies $E_L$ and  $\bar E_L$.
With the increase of $E_L$ and  $\bar E_L$ the
$\alpha$-terms decrease and the picture shown in fig. \ref{fig:rfactgen}
will approach that in fig. \ref{fig:rfact}.
For higher $E_L$ and  $\bar E_L$
the hierarchy can be identified in
 a larger interval of $r$ (and $R_{tail}$) and the 13-mixing can be measured
(especially in the case of inverted hierarchy).
This will be possible in the case the distance to the supernova is relatively small, so that the
statistics is  high enough.

3. One can determine $\alpha$, $\bar \alpha$ or other relevant parameters of the original neutrino spectra immediately from the
experimental data.
This can be done by studying the  dependences of the numbers of events $N_{e}(E_L)$, $N_{\bar e}(\bar{E}_L)$,
as well as their ratio $R_{tail}$, on the threshold energies $E_L$,  $\bar E_L$.   Indeed, the number of events $N_{e}(E_L)$ depends on $E_L$ via the quantities $N^0_x$ and $\alpha$:
\be N_{e}(E_L) =
(1 - p) N_{x}^0 (E_L) (1 - p + p \alpha (E_L)) ~.
\ee
 These two contributions   in principle can be disentangled, since they are different functions
of  $E_L$.   As follows from eq. (\ref{alphasy}), $\alpha$ has an exponential dependence on $E_L$ of
the form: $\alpha(E_L) = A \exp (-E_L/T_{ex}) $ (with $1/T_{ex}\equiv 1/T_e - 1/T_x$).
A fit of data could provide the unknown parameters $A$ and $T_{ex}$, allowing to fully reconstruct
the function $\alpha(E_L)$.

%%%%%%%%%%%%%%%%%%%%%%%%%%%%%%%%%%%%%%%%%%%%%%%%%%%%%%%%%%%%%%%%%%%%%%%%%%
\section{Time dependence of the signal: shock-wave effects}
\label{sec:10}
%%%%%%%%%%%%%%%%%%%%%%%%%%%%%%%%%%%%%%%%%%%%%%%%%%%%%%%%%%%%%%%%%%%%%%%%%%

%%%%%%%%%%%%%%%%%%%%%%%%%%%%%%%%%%%%%%%%%%%%%%%%%%%%%%%%%%%%%%%%%%%%%%%%%%%%
\subsection{Shock wave and H-resonance}
%%%%%%%%%%%%%%%%%%%%%%%%%%%%%%%%%%%%%%%%%%%%%%%%%%%%%%%%%%%%%%%%%%%%%%%%%%%%%

As it was pointed out  in ref. \cite{Schirato:2002tg}, the shock-wave propagating inside the star may
reach the region of  densities relevant for neutrino conversion during the neutrino emission period ($10 - 20$  seconds). It modifies  the density profile of the star, thus affecting the pattern of neutrino conversion \cite{Schirato:2002tg,Takahashi:2002yj}.  

\begin{figure}[hbt]
\begin{center} 
\epsfig{file=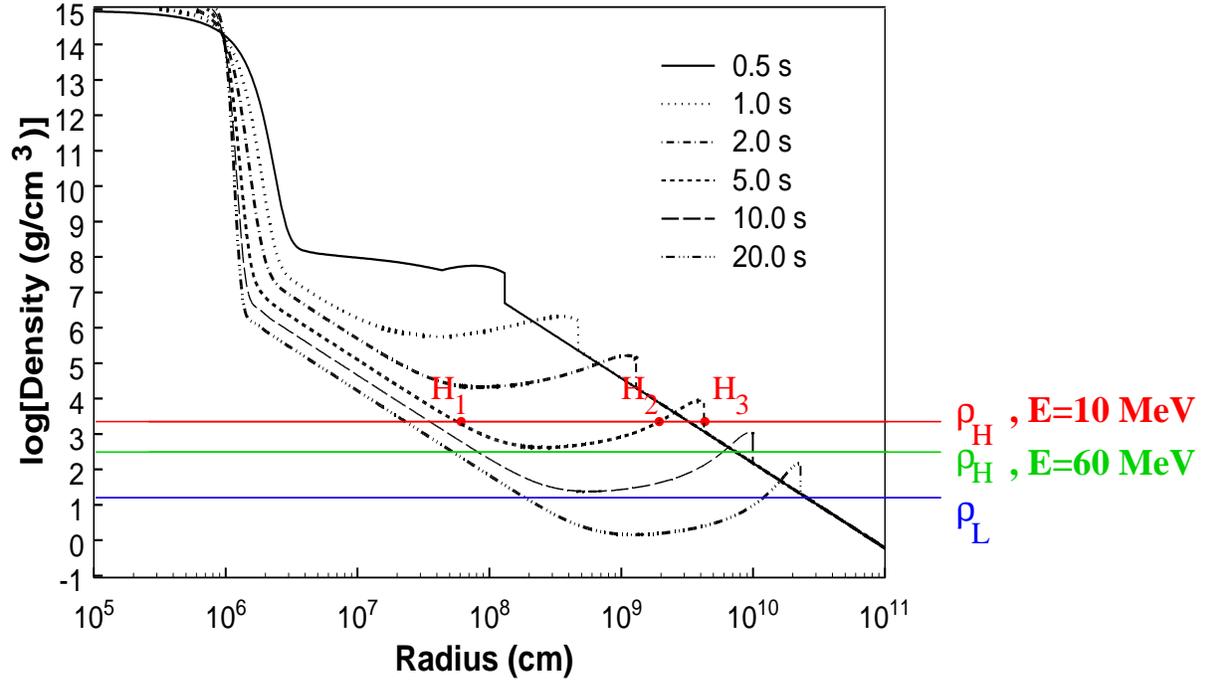, width=16truecm} 
\end{center}
  \caption{The density profile at various times post-bounce as adopted
    in ref. \cite{Schirato:2002tg}.  The horizontal lines represent
    the resonance densities $\rho_H$ (for two values of the neutrino
    energy) and $\rho_L$; the three high-density resonances
    $H_1,H_2,H_3$ are marked for $t=5$ s and $E=10$ MeV. 
 }
\label{fig:shock} 
\end{figure}

An indicative description  of the  time dependence of the matter density profile is given in fig.
\ref{fig:shock} (adapted  from \cite{Schirato:2002tg}).  As shown in the figure, at a given time $t$
post-bounce, the density  distribution presents an overdense region corresponding to the shock-front. The
outer border of this region (front)  consists in  a sharp ``step'' in which
the density increases from $\rho_{i}$  to $\rho_{f}$.
According to \cite{Schirato:2002tg}  the relative increase of the density in the front, $\xi$, can be of about
an order of magnitude:
\be
\xi \equiv \rho_{f}/\rho_i \sim 10.
\label{xi10}
\ee
Above the shock front  the profile coincides with that  of the progenitor star.  Below the front a rarefaction  region, a ``hot bubble'', is produced.

According to \cite{Schirato:2002tg}, the velocity of the shock front, $v_s$,  increases with the distance from the center of the star.
For distances  $d = (1 - 5) \cdot 10^{3}$ km it changes
in the interval
\be
v_{s} \sim (0.8 - 1.2) \cdot 10^4~{\rm Km/s},
\label{speed}
\ee
and for $d >  5 \cdot 10^{3}$ km, $v_s$ is nearly constant: $v_{s} \approx 1.2 \cdot 10^4~{\rm Km/s}$.

The shock wave influences the neutrino conversion when it reaches the resonance  layer. Given the mass squared splitting $\Delta m^2$,  the neutrino level crossing (resonance)
is realized at the  density
\be
\rho_{res}= \frac{m_N \Delta m^2}{\sqrt{2} G_F E}  \cos 2\theta_{13}
\simeq 1.4 \cdot 10^{3} {\rm g\cdot cm^{-3}}
\left(\frac{\Delta m^2}{10^{-3}~{\rm eV^2}} \right)  \left(\frac{10~{\rm MeV}}{E} \right),
\label{resdens}
\ee
where $m_N$ is the nucleon mass and $G_F$ the Fermi constant
and we have taken the electron fraction  $Y_e=1/2$.
  From  the fig. (\ref{fig:shock}) and eq. (\ref{eq4}) we find the resonance radius \footnote{  In ref. \cite{Schirato:2002tg} the density profile $\rho(d)\propto d^{-n}$ is used, with $n=2.4$. Here we use $n=3$ for simplicity. The difference between  these two values of $n$ is within the uncertainty quoted in sec. 3.2.}: 
\be
d_{res} = 5 \cdot 10^{4} {\rm km} \left(\frac{10^{-3}~{\rm eV^2}}{\Delta m^2}\right)^{1/3}
\left(\frac{E}{10~{\rm MeV}} \right)^{1/3}~,
\label{rres}
\ee
and -- taking a constant speed  $v_s$ with values (\ref{speed}) -- one gets  the time
$t_s \sim d_{res}/v_s$ after which the shockwave reaches  the resonance:
\be
t_s(E) =  \frac{d_{res}}{v_s} =  (4 -  5) ~{\rm s}  \left(\frac{E}{10~{\rm MeV}} \right)^{1 \over 3}~.
\label{numrel}
\ee
For the H resonance ($\Delta  m^2\simeq 3\cdot 10^{-3}~{\rm eV^2} $)
and $E=10$ MeV we have $\rho_{res}\simeq 4000  ~{\rm g\cdot cm^{-3}}$,
$d_{res} \simeq 5\cdot 10^{4}$ Km  (see fig. \ref{fig:shock}) and $t_s \simeq 5$ s.
The period
$\Delta t_s$ during which the H resonance is on the shock front (i.e. $\rho_i \leq \rho_{res} \leq \rho_f$)  can be estimated as  the time
which the shock wave takes to propagate from the position
of the resonance, $d_{res}$  ($\rho_{i} = \rho_{res}$)  to the position
$d_f$ at which $\rho_{f} =  \rho_{res}$. The latter can be found
using  the density profile (\ref{eq4}):
\be
d_f = d_{res} \xi^{1/3}~.
\ee
Then the interval $\Delta t_s$ equals:
\be
\Delta t_s = \frac{d_{res}}{v_s} (\xi^{1/3} -1).
\ee
Taking $\xi \approx 10$ (eq. (\ref{xi10})) and constant velocity  $v_s$, (\ref{speed}), we get
\be
\Delta t_s \approx (5 - 6) ~{\rm sec}  \left(\frac{E}{10 {\rm MeV}}\right)^{1/3}
\label{interv}
\ee
which is comparable   with  the arrival time $t_s$.

 According to eqs.  (\ref{numrel}) and (\ref{interv}),  $t_s$ and $\Delta t_s$  increase
with energy: for the lowest detectable energies in the spectrum one gets
$t_s(6 {\rm MeV}) = 3 - 5$ s,   while   for the highest energies $t_s(70 {\rm MeV}) = 7 - 10$ s
is obtained.

Let us  mark, however, that the details of the shock-wave propagation
depend on the specific characteristics
of the progenitor and therefore  even large variations of parameters are expected between
different SN models.\\

Let us consider the conversion in the H  resonance and its modifications due to the passage of the shock wave. For a given neutrino energy
 three time intervals can be defined: \\

\noindent
1. The pre-shock phase, $t(E) \lta t_{s}(E) \sim 3 - 5$ sec, in which the shock has
not yet reached the resonance region and therefore has no  effects
on the  neutrino conversion.
\\

\noindent
2. The shock phase, $t (E) =  t_{s}(E)  - (t_{s}(E) +  \Delta t_s (E))$,
when  the resonance condition $\rho = \rho_{res}$ is
fulfilled in the step of the shock front.

As it follows  from the fig.  \ref{fig:shock}, in this phase three resonances appear at three different
radii: two of them are on the walls of the hot bubble, while  the third is at the shock front. We denote these
resonances as $H_1,H_2,H_3$ from the inner to the outer.

For a given value of $\t13$,  the conversion  pattern due to the presence of these resonances exhibits  a
variety of possibile scenarios, depending on  the adiabaticity character of these resonances, and
therefore on the details of the density profile of the bubble and of the shock front.
The shock-wave  propagation has  no effect if  $\t13 \lta 10^{-6}$,
corresponding to non--adiabatic neutrino conversion  in all the relevant resonances during both the
pre-shock and the shock phases. Therefore in what follows
we discuss the case in which  $\t13$ is in the adiabatic region of the H resonance in the
pre-shock regime, that is, $\t13 \gta 10^{-4}$  (see fig. \ref{fig:phfig}), and  the $H_1$ and $H_2$
resonances are
adiabatic, while the adiabaticity is  broken in the $H_3$ resonance. In this case the permutation
parameters $p$ (for n.h.) and $\bar p$  (for i.h. \footnote{The conversion in the non-resonant channel
($\barnue$ fon n.h. and $\nue$ for i.h.)  can be modified by the shock-wave propagation if the density
profile at the shock front is very steep  so that the adiabaticity of the conversion is broken. However it
can be checked that this requires unphysically  large slopes of the shock front and therefore  this
possibility will not be discussed here. })  are  given by eqs. (\ref{pue3nh})-(\ref{pue3ih}) with
$P_H$ being the jumping probability in the $H_3$ resonance. For maximal adiabaticity breaking
\be
P_H \approx P_{H_3} \sim 1~.
\label{ph3}
\ee
The effect of the arrival of the shock-wave  consists in a sudden change of the character of the H level
crossing from adiabatic to maximally non--adiabatic,  with significant modifications of the observed
neutrino signal.
\\

\noindent
3. The post-shock phase,  $t (E) > t_{s}(E) +  \Delta t_s (E)$,
when the matter density  at the shock front is smaller than the resonance
density $\rho_{res} \geq \rho_f$,  so that only one level crossing, $H_1$,  survives. The conversion effects
are described by the transition probability in the $H_1$  resonance:  $P_H\simeq P_{H_1}\sim 0$.  In the
transition between the shock and the post-shock phases, the resonance  points $H_2$ and $H_3$ become closer
until they merge and disappear \footnote{Three low density (L)  resonances may appear at these later times, $t
\sim 20$ s, as shown in the figure \ref{fig:shock}.  The  one at the shock front could have a non-adiabatic
character, so that the condition of adiabatic conversion ($P_L=0$) adopted here is  not valid in this case.  However this happens in the
latest part of the signal, when the neutrino luminosity  is small. For this reason the effects of the
shockwave on the L resonance are not discussed in detail here.}.\\

For the scenario discussed here  we expect  the following time dependence of the effective
jumping probability $P_H$:
\be
P_H \approx \begin{cases}
0  & \text{for}~~  t\lta t_{s}(E) \\
1  & \text{for}~~ t =  t_{s}(E) - (t_s + \Delta t_s(E)) \\
0  &  \text{for}~~ t\gta t_{s}(E) + \Delta t_s (E)~.
\end{cases}
\label{phtstruc}
\ee

In general, the shock wave effect is proportional to  $P_{H_3}$,
which in turn depends on the neutrino energy.
Therefore not only the time but also the size of the effect depend on the energy.
 The ratio of the adiabaticity parameters of a resonance at and above the shock front (where the density profile equals that of the progenitor star) equals the  ratio of the corresponding density gradients:
\be
\frac{\gamma_f}{\gamma_i} \equiv k_{grad} = \frac{(d\rho/dr)_f}{(d\rho/dr)_0}~.
\ee
From this and eqs. (\ref{phform},\ref{ena}) it follows that for
$\sin^2 \theta_{13} = 0.01$ the adiabaticity  is violated at
all relevant energies if $k_{grad} = 10^{4}$. That is the gradient of density
in the shock wave is 4 orders of magnitude larger than the gradient of the progenitor profile
(at the same density). If $k_{grad} = 10^{3}$, we get $P_{H_3} \approx 1$  for high energies:
$E \sim 70$ MeV but $P_{H_3} \sim 0.5$ for $E \sim 6$ MeV. For smaller gradient:
$k_{grad} = 10^{2}$, the adiabaticity is not violated for low energies, and therefore there is
no shock wave effect, but it is still violated for high energies: $P_{H_3} \sim 0.5$ for $E \sim 70$ MeV.
In general one expects larger effects for high energies.
Results for other values of $\sin^2 \theta_{13}$ can be obtained immediately by
rescaling eqs.  (\ref{phform},\ref{ena}).

%%%%%%%%%%%%%%%%%%%%%%%%%%%%%%%%%%%%%%%%%%%%%%%%%%%%%%%%%%%%%%%%%%%%%%%%%%%%%%%%%%%%%%%%%%%
\subsection{Shock wave effects on the neutrino spectra}
%%%%%%%%%%%%%%%%%%%%%%%%%%%%%%%%%%%%%%%%%%%%%%%%%%%%%%%%%%%%%%%%%%%%%%%%%%%%%%%%%%%%%%%%%%%%%

Let us consider the influence of the shock wave on the neutrino spectra in more details.
For a given energy $E$ the effect of the shock wave consists
in a  change of the flux when the front of
shock wave reaches the corresponding resonance
layer with resonance density $\rho_{res}(E)$  and then restoration of the
original (undisturbed) flux after the front shock shifts to smaller densities.
In what follows for simplicity we will assume that the density gradient
in the shock wave is large enough so that the adiabaticity is
strongly broken for all observable energies.

In the case of normal mass hierarchy (H-resonance in the neutrino channel)
 during the interval $\Delta t_s$ the shock wave leads to a change of the $\nue$ flux exiting the star from completely to partially permuted:
\be
F_x^0(E) \rightarrow F_s (E) =
\cos^2 \theta_{12} F_x^0(E)  + \sin^2 \theta_{12}  F_e^0(E).
\label{soft}
\ee
Let us introduce  critical energy $E_c$ such that  $F_x^0 (E_c) =  F_e^0(E_c)$.
Then for $E < E_c$ we have  $F^0_x(E) < F_s (E)$,  and for
$E >  E_c$ the inequality  $F^0_x(E) > F_s (E)$ holds.
So,  the pattern of perturbation of the whole spectrum
can be described as follows: first the perturbation reaches
the lowest energies of the spectrum and then propagates to high energies.
We can say that the wave of perturbation of the spectrum propagates in the energy scale.
It leads to an increase
of the flux of the electron neutrinos
%(in the case of normal hierarchy)
according to eq. (\ref{soft}) when the wave propagates up to $E_c$.
Then for  $E > E_c$ the pertubation decreases the flux
according to (\ref{soft}).
 So, we can say that the shock-wave effects consist in  a wave of softening of the spectrum
which propagates from  low to high energies (softening wave).
When the wave reaches the highest energies
of spectrum, the perturbation start to disappear at low energies, restoring
the original spectrum:
\be
F_s (E) \rightarrow F^0_x(E)~.
\label{hard}
\ee
The spectrum becomes hard again  and the restoration moves again from
low to high energies.

The antineutrino conversion  will not be affected by the shock wave.\\

In the case of inverted mass hierarchy the H-resonance
is in the antineutrino channel and therefore the shock wave will modify the antineutrino
conversion. Similarly to (\ref{soft})  $\bar{\nu}_e$ spectrum  changes as
\be
F_{\bar x}^0(E) \leftrightarrow F_{\bar s} (E) =
\sin^2 \theta_{12} F_{\bar x}^0(E)  + \cos^2 \theta_{12}  F_{\bar e}^0(E).
\label{softan}
\ee
Although now the change of the permutation factor is stronger
(from complete permutation to weak permutation: $(1 - \bar p) = 1 \leftrightarrow \sin^2 \theta_{12}$),
the effect on the spectrum can be similar to the normal hierarchy case due to the
smaller difference of  $F_{\bar x}^0(E)$ and $F_{\bar e}^0(E)$ fluxes.
The neutrino flux will not be  affected by the shock wave.

%%%%%%%%%%%%%%%%%%%%%%%%%%%%%%%%%%%%%%%%%%%%%%%%%%%%%%%%%%%%%%%%%%%%%
\subsection{Shock wave and the Earth matter effect}
%%%%%%%%%%%%%%%%%%%%%%%%%%%%%%%%%%%%%%%%%%%%%%%%%%%%%%%%%%%%%%%%%%%%

The shock wave propagation can influence the Earth matter effect.
Indeed, for  the normal mass hierarchy  the Earth matter effect in $\nue$ channel vanishes for adiabatic H transition while it  is maximal for maximal violation of the adiabaticity in the H resonance, $P_H=1$ \cite{Dighe:1999bi,Lunardini:2001pb}.
The effect  increases fast with energy, reaching 30 - 50 \%
at $E \sim 50 - 70$ MeV.

Taking this into account, for the scenario (\ref{phtstruc}) we  get the following time dependence:

in the pre-shock phase, no regeneration effect is realized in the $\nue$ channel  due to the adiabaticity character of the H resonance.
The  effect is small in the first few seconds of the shock wave phase:
in this interval the adiabaticity is broken at  low energies, where the
 Earth matter effect is small.
After 5 - 7 sec the softening wave reaches the high energy region,
where the Earth matter effect is large.
So one expects the absence of the Earth matter effects in the
first 5 - 7 sec of the neutrino burst and then its fast increase
with time in the high energy part of the spectrum.   We will call this the delayed Earth matter effect.

Being unaffected by the H resonance, the $\barnue$ spectrum exhibits the Earth matter effects
for the whole duration of the neutrino signal.

In contrast, if the mass hierarchy is inverted, the Earth matter effect
is observed in the neutrino channel during the whole burst and it appears in the antineutrino
channel in the late stage (after 5 - 7 sec).

%%%%%%%%%%%%%%%%%%%%%%%%%%%%%%%%%%%%%%%%%%%%%%%%%%%%%%%%%%%%%%%%%%%%
\subsection{Shock wave effects, mass hierarchy and $\theta_{13}$}
%%%%%%%%%%%%%%%%%%%%%%%%%%%%%%%%%%%%%%%%%%%%%%%%%%%%%%%%%%%%%%%%%%%%

Precise measurements of the energy spectra in different
moments of time can, in principle, reveal  spectrum distortions
which depend on time and propagate from low
to high energies. For this however very high statistics is needed.

Another possibility is to use some global characteristics of spectra and their change with time
to look for  shock wave effects. In particular, one expects

\begin{itemize}

\item
decrease of the average energy of the spectrum $\langle E \rangle$
which reflects its  softening during the shock wave phase;

\item
increase of the relative width $\Gamma$, as a consequence of the appearance of a composite spectrum
during the time $\Delta t_s$;

\item
appearance of the Earth matter effect in the late stage of the burst;

\item 
  change in the total even rates \cite{Schirato:2002tg}.

\end{itemize}

A problem of dealing with global characteristics is the possible degeneracy between the changes of the neutrino energy spectra
due to shock wave effects and those due to astrophysical factors (cooling of the energy spectra, decay of the luminosity, etc.).

The shock-wave effects on the $\nue$ (or $\barnue$)  signal could be identified  by studying the time
dependence of the ratios of $\nue$ (for n.h.) or $\barnue$ (for i.h.)   CC to NC event rates at SNO
\cite{Schirato:2002tg}.  Since the NC rate is not affected by  shock-induced conversion effects, the CC to
NC ratio represents a particularly ``clean'' quantity, in which   various time-dependent features other
than the shock effect (e.g. cooling of energy spectra) are at  least partially subtracted.    Using this
method, in ref. \cite{Schirato:2002tg} it is found  that the  distortion due to the shock-wave amounts to
$\sim 30\%$, whose statistical significance has not been clarified, however.

 An alternative approach to probe effects of the passage of the shock wave could be to study the time dependence  of the ratio $R_{tot}$ of the total number of $\nue$ and $\barnue$ events (eq. (\ref{rtotdef})), where still an at least partial subtraction of  cooling effects is realized. In this quantity  the
uncertainties related to the poor knowledge of astrophysical parameters can by no means mimic a  specific time structure and therefore should be distinguishable from the shock effects.

The observation of the  delayed Earth matter effect is another signature of the shock wave effects. \\ 

Let us summarize the possibilities to identify the mass hierarchy and measure (or restrict) $\theta_{13}$.

The observation of shock-effects in $\nue$ ($\barnue$) channel would select the normal (inverted) mass hierarchy and  would tell  that $\t13$ is relatively large, $\t13 \gta 5 \cdot 10^{-4}$,
so that $P_H<1$ significantly  in the  pre-shock phase. Moreover, it would allow to study the physical
features of the shock-propagation (speed, etc.).
The exclusion of shock-effects would imply
that $\t13$ is in the adiabaticity breaking region in the pre-shock times or that the properties of the shock wave differ significantly from the predictions (e.g.   that the shock stalls, or travels with smaller  velocity, so that the shock does not reach the resonance region during the duration of the burst,  or  that the shock front is not steep enough to change the adiabaticity character of the conversion).

Quantitative estimates of the  bounds  on $\t13$ that can be obtained from the study of
shock-induced time dependences of the  signal
will be given elsewhere \cite{CAprep}.

%%%%%%%%%%%%%%%%%%%%%%%%%%%%%%%%%%%%%%%%%%%%%%%%%%%%%%%%%%%%%%%%%%%%%%%%%%
%%%%%%%%%%%%%%%%%%%%%%%%%%%%%%%%%%%%%%%%%%%%%%%%%%%%%%%%%%%%%%%%%%%%%%%%%%
\section{Discussion and conclusions}
\label{sec:11}
%%%%%%%%%%%%%%%%%%%%%%%%%%%%%%%%%%%%%%%%%%%%%%%%%%%%%%%%%%%%%%%%%%%%%%%%%%

I. We have studied the effects of the 13-mixing on the
conversion of the supernova neutrinos.\\

 If $\theta_{13}$ is negligible ($\t 13 \lta 10^{-5}$), at the cooling stage a
 partial permutation of the original $\nue$ and $\nux$, $\barnue$ and $\barnux$ energy spectra
 occurs in the star due to
the large 12-mixing.
For larger values of $\theta_{13}$,  $\sin^2 \theta_{13} \geq 10^{-5}$, these conversion effects are modified in the $\nue$ ($\barnue$) channel for normal (inverted ) mass hierarchy.

For normal mass hierarchy
the degree of permutation in $\nue$ channel increases
from $(1 - p) = \cos^2 \theta_{12} \sim 0.7 - 0.8$ to
$(1 - p) = \cos^2 \theta_{12} + (1 -  P_H) \sin^2 \theta_{12}$, resulting in an hardening of the observed $\nue$ spectrum.
If $\sin^2 \theta_{13} > 10^{-4}$  the transition driven by $\theta_{13}$ is adiabatic and  the
change is maximal: $(1 - p) \approx 1$. This  corresponds to a complete permutation and therefore
maximally hard $\nue$ spectrum.

Some additional distortion of the energy spectrum is expected
for $\sin^2 \theta_{13} <  10^{-4}$ due to the dependence of the jumping probability
$P_H$ on energy.

In the case of inverted mass hierarchy the 13 mixing leads to an increase of the
degree
of permutation in the antineutrino channel:
from $(1 - \bar p) = \sin^2 \theta_{12}$ to $(1 - \bar p) =
\sin^2 \theta_{12} + (1 -  P_H) \cos^2 \theta_{12}$.
Again for  $\sin^2 \theta_{13} > 10^{-4}$ the effect is maximal:
$(1 - \bar p) \approx 1$.
\\

 In this paper we have elaborated  methods to study the effects of the $\theta_{13}$ mixing in presence and absence of Earth matter effects.  Various possibilities to probe the mass hierarchy and measure (or restrict) $\theta_{13}$ have been discussed as well.\\

II. The permutation due to $\theta_{13}$ changes the average energy and width of the $\nue$
(for n.h.) or $\barnue$ (for i.h.) spectrum.
Signatures of these modifications can be obtained from the  comparison of the features of
neutrino and antineutrino spectra.

In particular, for large 13-mixing and the normal mass hierarchy one
expects
$\langle E(\nu) \rangle > \langle E(\bar \nu) \rangle$ and
$\langle \Gamma (\nu) \rangle < \langle \Gamma (\bar \nu) \rangle$.
Opposite inequalities are expected in the case of inverted hierarchy.

In this connection we have studied
the observed spectra of events from $\nue+d$ and $\barnue +p$ CC reactions at SNO and SK respectively; in particular, the following observables  have been considered:
the ratio of the average energies of the $\nue$ and $\barnue$ events,  $r_E = \langle E \rangle/\langle \bar E
\rangle$, the ratio of the widths of
the energy spectra,  $r_\Gamma = \Gamma/{\bar \Gamma} $, and the ratio $R_{tail}$ of the
numbers of events  in the  high energy tails (above certain energy cuts $E_L$ and
$\bar E_L$) at  the two
experiments.  These quantities
depend on  $\theta_{13}$ via the jumping probability in the H resonance,
$P_H(\theta_{13})$.

Although the 13-mixing produces a strong conversion, the corresponding observable effects
may not be so large due to presence of neutrino fluxes of all types
and also due to effects of the 12-mixing effect (see item I ).
Furthermore, substantial astrophysical uncertainties make the  identification of the 13-mixing effects difficult, due to possible
degeneracies between the  parameters of the original neutrino spectra and the
oscillation parameters.

We have studied the effects of uncertainties on:
(i) the energy spectra and luminosities of the
neutrino fluxes originally produced inside the star,
(ii) the density profile of the star, (iii) the mass square splitting
$\Delta
m^2_{31}$, (iii) the mixing  angle $\theta_{12}$.
\\
Methods to take these uncertainties into account in the data analysis have been elaborated.
\\

III. We have considered
the cases of adiabatic transition ($P_H=0$) with (A) of normal mass hierarchy  and
(B) inverted mass hierarchy  and we have
compared
them with  the case (C) of maximally non-adiabatic conversion
(i.e. negligibly small $\theta_{13}$:   $\sin^2 \theta_{13} < 10^{-5}$).

The scatter plots of the observables ($r_E$,
$r_\Gamma$, $R_{tail}$,...) for these three cases show
that there are regions -- in the space of these parameters -- where only one or two possibilities
exists,
implying the exclusion of  the remaining ones.

In particular,

\begin{itemize}
\item
The case A  is excluded if observations give $R_{tail} \lta
0.06$ and/or $r_\Gamma \gta 1.1$. If, in contrast, the
experiments give $R_{tail} > 0.22$ the normal
hierarchy would be identified. This result would be further supported if
$r_\Gamma \lta 0.9$ is also
found.

\item
The case B is excluded by (i)  large values of $R_{tail}$, $R_{tail }\gta 0.07$, (ii)
large values of the ratio of average
energies $r_E \gta 1.2$ and (iii)
$r_\Gamma \lta 1$.  In
contrast, the identification of this scenario appears difficult.

\item
The case C is difficult to single out.
%would be identified by
The result $r_\Gamma
\gta 1.1$ and $0.06 \lta R_{tail}
\lta 0.2$  would exclude B and A, and therefore
indicate that $P_H>0$, corresponding to small
values of $\theta_{13}$: $\t13\lta few \cdot 10^{-4}$.

\end{itemize}

IV. The high energy tail method allows, in principle, to establish the mass
hierarchy and to restrict
or to measure $\sin^2 \theta_{13}$.
Taking the thresholds $E_L=45$ MeV and $\bar E_L =55$ MeV and $\y12=0.38$ as an example, we find that
 if $R > 0.22$  the normal mass hierarchy can be identified and
a lower bound on $\sin^2 \theta_{13}$
can be obtained. For smaller values of the mixing the hierarchy can
not be established; however  strong bounds on $\sin^2 \theta_{13}$
are put if the mass
hierarchy is know from other results
(e.g. by measurements of other parameters of spectra, by studying the  Earth regeneration
effects etc.).   A measurement of $\theta_{13}$ will be possible only if  a reduction
of the uncertainties on the original neutrino fluxes is achieved.

 The Earth matter effect enhances the
identification power of the method. Since the regeneration is larger in
neutrino than in antineutrino channel, the
sensitivity of $R_{tail}$ to the n.h. (i.h.) is enhanced (reduced) with respect
to the no Earth-crossing case. In
particular, in presence of Earth effects the method of analysis discussed
in this paper would be complementary to
the information that can be extracted from the observation of oscillatory
distortions in the energy spectra of
$\nue$ and/or $\barnue$ events in the detectors. \\

The method of the high energy tails can be improved if (i)  additional
information about parameters of the
neutrino fluxes is obtained from the studies of the whole spectra (and not
only of the high energy tails) (ii) if the dependence of the ratio $R_{tail}$ on the energy cuts is studied, (iii)
higher energy cuts are used.  These improvements can
be implemented if the statistics is large enough.\\

V. The study of shock-wave effects  on the neutrino burst, which can be
realized at late times ($t\gta 5$
s) could provide additional information  on the $\theta_{13}$ angle and
the mass hierarchy (in support of
the methods discussed in sections 5. and 6., which  refers to early times, $t\lta 5$
s). In particular, the presence
of time dependent spectral distortions  produced by the shock-wave passage
through the resonance layers
would lead to the lower bound  $\t13\gta  few \cdot 10^{-4}$ and establish
the n.h. (i.h.) if observed in
the $\nue$ ($\barnue$) channel.

An important signature of the shock wave is the appearance of the delayed Earth
matter
effect in the neutrino or antineutrino channel at the late stages of the burst
($t > 5 - 7$ sec).
The channel of appearance will identify the mass hierarchy.

%%%%%%%%%%%%%%%%%%%%%%%%%%%%%%%%%%%%%%%%%%%%%%%%%%%%%%%%%%%%%%%%%%%%%%%%%%
\subsection*{Acknowledgements}
\label{sec:ackn}
C.L. acknowledges support from the Keck fellowship and the grants
PHY-0070928 and PHY99-07949.  She would like to thank J.~N.~Bahcall,
A.~Friedland, C.~Pe\~na--Garay and H.~Minakata for fruitful discussions,
T.~Totani for useful communications
and the Kavli Institute
of Theoretical Physics (KITP), Santa Barbara (CA), as hosting
institution where part of this work was prepared.  The authors are grateful to S.~Palomares-Ruiz, O.~L.~G.~Peres and F.~Vissani for useful comments on the preliminary version of this paper.

%%%%%%%%%%%%%%%%%%%%%%%%%%%%%%%%%%%%%%%%%%%%%%%%%%%%%%%%%%%%%%%%%%%%%%%%%%

%%%%%%%%%%%%%%%%%%%%%%%%%%%%%%%%%%%%%%%%%%%%%%%%%%%%%%%%%%%%%%%%%%%%%%%%%%
%%%%%%%%%%%%%  BIBLIOGRAPHY %%%%%%%%%%%%%%%%%%%%%%%%%%%%%%%%%%%%%%%%%%%%%%%

%\newpage

\bibliography{paper}

\providecommand{\href}[2]{#2}\begingroup\raggedright\begin{thebibliography}{10}

\bibitem{Raffelt:2002tu}
See e.g. the review: G.~G. Raffelt, {\it Physics with supernovae},  {\em Nucl. Phys. Proc. Suppl.}
  {\bf 110} (2002) 254--267,
  [\href{http://xxx.lanl.gov/abs/hep-ph/0201099}{{\tt hep-ph/0201099}}], and references therein.


\bibitem{Mikheev:1986if}
S.~P. Mikheev and A.~Y. Smirnov, {\it Neutrino oscillations in a
  variable-density medium and neutrino bursts due to the gravitational collapse of
  stars},  {\em Sov. Phys. JETP} {\bf 64} (1986) 4--7.

\bibitem{Wolfenstein:1987pj}
L.~Wolfenstein, {\it Effects of matter oscillations on supernova neutrino
  flux},  {\em Phys. Lett.} {\bf B194} (1987) 197.

\bibitem{Dutta:1999ir}
G.~Dutta, D.~Indumathi, M.~V.~N. Murthy, and G.~Rajasekaran, {\it Neutrinos
  from stellar collapse: effects of flavour mixing},  {\em Phys. Rev.} {\bf
  D61} (2000) 013009, [\href{http://xxx.lanl.gov/abs/hep-ph/9907372}{{\tt
  hep-ph/9907372}}].

\bibitem{Dighe:1999bi}
A.~S. Dighe and A.~Y. Smirnov, {\it Identifying the neutrino mass spectrum from
  the neutrino burst from a supernova},  {\em Phys. Rev.} {\bf D62} (2000)
  033007, [\href{http://xxx.lanl.gov/abs/hep-ph/9907423}{{\tt
  hep-ph/9907423}}].

\bibitem{Dutta:2000zq}
G.~Dutta, D.~Indumathi, M.~V.~N. Murthy, and G.~Rajasekaran, {\it Neutrinos
  from stellar collapse: comparison of the effects of three and four flavor
  mixings},  {\em Phys. Rev.} {\bf D62} (2000) 093014,
  [\href{http://xxx.lanl.gov/abs/hep-ph/0006171}{{\tt hep-ph/0006171}}].

\bibitem{Dutta:2001nf}
G.~Dutta, D.~Indumathi, M.~V.~N. Murthy, and G.~Rajasekaran, {\it Neutrinos
  from stellar collapse: comparison of signatures in water and heavy water
  detectors},  {\em Phys. Rev.} {\bf D64} (2001) 073011,
  [\href{http://xxx.lanl.gov/abs/hep-ph/0101093}{{\tt hep-ph/0101093}}].

\bibitem{Takahashi:2001ep}
K.~Takahashi, M.~Watanabe, K.~Sato, and T.~Totani, {\it Effects of neutrino
  oscillation on the supernova neutrino spectrum},  {\em Phys. Rev.} {\bf D64}
  (2001) 093004, [\href{http://xxx.lanl.gov/abs/hep-ph/0105204}{{\tt
  hep-ph/0105204}}].

\bibitem{Fogli:2001pm}
G.~L. Fogli, E.~Lisi, D.~Montanino, and A.~Palazzo, {\it Supernova neutrino
  oscillations: a simple analytical approach},  {\em Phys. Rev.} {\bf D65}
  (2002) 073008, [\href{http://xxx.lanl.gov/abs/hep-ph/0111199}{{\tt
  hep-ph/0111199}}].

\bibitem{Barger:2001yx}
V.~Barger, D.~Marfatia, and B.~P. Wood, {\it Inverting a supernova: neutrino
  mixing, temperatures and binding energy},  {\em Phys. Lett.} {\bf B547}
  (2002) 37--42, [\href{http://xxx.lanl.gov/abs/hep-ph/0112125}{{\tt
  hep-ph/0112125}}].

\bibitem{Minakata:2001cd}
H.~Minakata, H.~Nunokawa, R.~Tomas, and J.~W.~F. Valle, {\it Probing supernova
  physics with neutrino oscillations},  {\em Phys. Lett.} {\bf B542} (2002)
  239--244, [\href{http://xxx.lanl.gov/abs/hep-ph/0112160}{{\tt
  hep-ph/0112160}}].

\bibitem{Barger:2002px}
V.~Barger, D.~Marfatia, and B.~P. Wood, {\it Supernova 1987A did not test the
  neutrino mass hierarchy},  \href{http://xxx.lanl.gov/abs/hep-ph/0202158}{{\tt
  hep-ph/0202158}}.

\bibitem{Takahashi:2002cm}
K.~Takahashi and K.~Sato, {\it Effects of neutrino oscillation on supernova
  neutrino: inverted mass hierarchy},
  \href{http://xxx.lanl.gov/abs/hep-ph/0205070}{{\tt hep-ph/0205070}}.

\bibitem{Eguchi:2002dm}
{\bf KamLAND} Collaboration, K.~Eguchi {\em et.~al.}, {\it First results from
  KamLand: Evidence for reactor anti- neutrino disappearance},  {\em Phys. Rev.
  Lett.} {\bf 90} (2003) 021802,
  [\href{http://xxx.lanl.gov/abs/hep-ex/0212021}{{\tt hep-ex/0212021}}].

\bibitem{Barger:2000hy}
V.~D. Barger, D.~Marfatia, and B.~P. Wood, {\it Resolving the solar neutrino
  problem with KamLand},  {\em Phys. Lett.} {\bf B498} (2001) 53--61,
  [\href{http://xxx.lanl.gov/abs/hep-ph/0011251}{{\tt hep-ph/0011251}}].

\bibitem{Barbieri:2000sv}
R.~Barbieri and A.~Strumia, {\it Non standard analysis of the solar neutrino
  anomaly},  {\em JHEP} {\bf 12} (2000) 016,
  [\href{http://xxx.lanl.gov/abs/hep-ph/0011307}{{\tt hep-ph/0011307}}].

\bibitem{Murayama:2000iq}
H.~Murayama and A.~Pierce, {\it Energy spectra of reactor neutrinos at
  KamLand},  {\em Phys. Rev.} {\bf D65} (2002) 013012,
  [\href{http://xxx.lanl.gov/abs/hep-ph/0012075}{{\tt hep-ph/0012075}}].

\bibitem{Arafune:1987kc}
J.~Arafune, M.~Fukugita, T.~Yanagida, and M.~Yoshimura, {\it Neutrino mass and
  mixing constrained from the LMC supernova burst},  {\em Phys. Lett.} {\bf
  B194} (1987) 477.

\bibitem{Walker:1987xd}
T.~P. Walker and D.~N. Schramm, {\it Resonant neutrino oscillations and the
  neutrino signature of supernovae},  {\em Phys. Lett.} {\bf B195} (1987) 331.

\bibitem{Notzold:1987vc}
D.~Notzold, {\it MSW effect analysis for SN1987A sets severe restrictions on
  neutrino masses and mixing angles},  {\em Phys. Lett.} {\bf B196} (1987)
  315--320.

\bibitem{Minakata:1988cn}
H.~Minakata and H.~Nunokawa, {\it Neutrino flavor conversion in supernova
  SN1987A},  {\em Phys. Rev.} {\bf D38} (1988) 3605.

\bibitem{Jegerlehner:1996kx}
B.~Jegerlehner, F.~Neubig, and G.~Raffelt, {\it Neutrino oscillations and the
  supernova 1987A signal},  {\em Phys. Rev.} {\bf D54} (1996) 1194--1203,
  [\href{http://xxx.lanl.gov/abs/astro-ph/9601111}{{\tt astro-ph/9601111}}].

\bibitem{Lunardini:2000sw}
C.~Lunardini and A.~Y. Smirnov, {\it Neutrinos from SN1987A, earth matter
  effects and the LMA solution of the solar neutrino problem},  {\em Phys.
  Rev.} {\bf D63} (2001) 073009,
  [\href{http://xxx.lanl.gov/abs/hep-ph/0009356}{{\tt hep-ph/0009356}}].

\bibitem{Minakata:2000rx}
H.~Minakata and H.~Nunokawa, {\it Inverted hierarchy of neutrino masses
  disfavored by supernova 1987A},  {\em Phys. Lett.} {\bf B504} (2001)
  301--308, [\href{http://xxx.lanl.gov/abs/hep-ph/0010240}{{\tt
  hep-ph/0010240}}].

\bibitem{Kachelriess:2000fe}
M.~Kachelriess, R.~Tomas, and J.~W.~F. Valle, {\it Large lepton mixing and
  supernova 1987A},  {\em JHEP} {\bf 01} (2001) 030,
  [\href{http://xxx.lanl.gov/abs/hep-ph/0012134}{{\tt hep-ph/0012134}}].

\bibitem{Kachelriess:2001sg}
M.~Kachelriess, A.~Strumia, R.~Tomas, and J.~W.~F. Valle, {\it SN1987A and the
  status of oscillation solutions to the solar neutrino problem},
  \href{http://xxx.lanl.gov/abs/hep-ph/0108100}{{\tt hep-ph/0108100}}.

\bibitem{Keil:2002in}
See e.g. M.~T. Keil, G.~G. Raffelt, and H.-T. Janka, {\it Monte carlo study of supernova
  neutrino spectra formation},
  \href{http://xxx.lanl.gov/abs/astro-ph/0208035}{{\tt astro-ph/0208035}}, and references therein.

\bibitem{Horowitz:2001xf}
C.~J. Horowitz, {\it Weak magnetism for antineutrinos in supernovae},  {\em
  Phys. Rev.} {\bf D65} (2002) 043001,
  [\href{http://xxx.lanl.gov/abs/astro-ph/0109209}{{\tt astro-ph/0109209}}].

\bibitem{Janka}
H.~T.~Janka and W.~Hillebrandt, {\it Monte Carlo simulations of neutrino
  transport in type II supernovae}, Astron.\ Astroph.\ Suppl.\ {\bf 78} (1989)
  375; H.~T.~Janka and W.~Hillebrandt, {\it Neutrino emission from type II
  supernovae - an analysis of the spectra}, Astron.\ Astrophys.\ {\bf 224}
  (1989) 49.

\bibitem{Raffelt:1996wa}
See also G.~G.~Raffelt, ``Stars as laboratories for fundamental physics: The
 astrophysics of neutrinos, axions, and other weakly interacting particles,''
  {\it Chicago, USA: Univ. Pr. (1996) 664 p}.


\bibitem{BBB}
G.~E.~Brown, H.~A.~Bethe and G.~Baym, Nucl. \ Phys. \ A {\bf 375} (1982) 481.

\bibitem{Kuo:1988qu}
T.~K. Kuo and J.~Pantaleone, {\it Supernova neutrinos and their oscillations},
  {\em Phys. Rev.} {\bf D37} (1988) 298.

\bibitem{Schirato:2002tg}
R.~C. Schirato and G.~M. Fuller, {\it Connection between supernova shocks,
  flavor transformation, and the neutrino signal},
  \href{http://xxx.lanl.gov/abs/astro-ph/0205390}{{\tt astro-ph/0205390}}.

\bibitem{Krastev:1988yu}
P.~I. Krastev and S.~T. Petcov, {\it Resonance amplification and t violation
  effects in three neutrino oscillations in the earth},  {\em Phys. Lett.} {\bf
  B205} (1988) 84--92.

\bibitem{Fukuda:2000np}
{\bf Super-Kamiokande} Collaboration, S.~Fukuda {\em et.~al.}, {\it Tau
  neutrinos favored over sterile neutrinos in atmospheric muon neutrino
  oscillations},  {\em Phys. Rev. Lett.} {\bf 85} (2000) 3999--4003,
  [\href{http://xxx.lanl.gov/abs/hep-ex/0009001}{{\tt hep-ex/0009001}}].

\bibitem{Fogli:2002au}
G.~L. Fogli {\em et.~al.}, {\it Solar neutrino oscillation parameters after
  first KamLand results},  \href{http://xxx.lanl.gov/abs/hep-ph/0212127}{{\tt
  hep-ph/0212127}}.

\bibitem{Bahcall:2002ij}
J.~N. Bahcall, M.~C. Gonzalez-Garcia, and C.~Pena-Garay, {\it Solar neutrinos
  before and after KamLand},
  \href{http://xxx.lanl.gov/abs/hep-ph/0212147}{{\tt hep-ph/0212147}}.


\bibitem{Nunokawa:2002mq}
H.~Nunokawa, W.~J.~Teves and R.~Zukanovich Funchal,
{\it Determining the oscillation parameters by solar neutrinos and KamLand},
   \href{http://xxx.lanl.gov/abs/hep-ph/0212202}{{\tt hep-ph/0212202}}.



\bibitem{deHolanda:2002iv}
P.~C. de~Holanda and A.~Y. Smirnov, {\it LMA MSW solution of the solar neutrino
  problem and first KamLand results},
  \href{http://xxx.lanl.gov/abs/hep-ph/0212270}{{\tt hep-ph/0212270}}.

\bibitem{Apollonio:1999ae}
{\bf CHOOZ} Collaboration, M.~Apollonio {\em et.~al.}, {\it Limits on neutrino
  oscillations from the Chooz experiment},  {\em Phys. Lett.} {\bf B466} (1999)
  415--430, [\href{http://xxx.lanl.gov/abs/hep-ex/9907037}{{\tt
  hep-ex/9907037}}].

\bibitem{Boehm:2000vp}
F.~Boehm {\em et.~al.}, {\it Results from the Palo Verde neutrino oscillation
  experiment},  {\em Phys. Rev.} {\bf D62} (2000) 072002,
  [\href{http://xxx.lanl.gov/abs/hep-ex/0003022}{{\tt hep-ex/0003022}}].

\bibitem{Friedland:2000rn}
A.~Friedland, {\it On the evolution of the neutrino state inside the sun},
  {\em Phys. Rev.} {\bf D64} (2001) 013008,
  [\href{http://xxx.lanl.gov/abs/hep-ph/0010231}{{\tt hep-ph/0010231}}].

\bibitem{Kachelriess:2001bs}
M.~Kachelriess and R.~Tomas, {\it Non-adiabatic level crossing in (non-)
  resonant neutrino oscillations},  {\em Phys. Rev.} {\bf D64} (2001) 073002,
  [\href{http://xxx.lanl.gov/abs/hep-ph/0104021}{{\tt hep-ph/0104021}}].

\bibitem{Kuo:1989qe}
T.~K. Kuo and J.~Pantaleone, {\it Neutrino oscillations in matter},  {\em Rev.
  Mod. Phys.} {\bf 61} (1989) 937.

\bibitem{usprep}
C.~Lunardini and A.Yu.~Smirnov, in preparation.

\bibitem{clayton}
D.~D.~Clayton, ``Principles of Stellar Evolution and Nucleosynthesis,'' {\it
  Chicago, USA: Univ. Pr. (1983) 612 p}.

\bibitem{shig}
K.~Shigeyama, ÊT.;~Nomoto, {\it Theoretical light curve of SN1987A and mixing
  of hydrogen and nickel in the ejecta},  {\em Astrophys. J.} {\bf 360} (1990)
  242--256.

\bibitem{Barger:2001yk}
V.~D. Barger {\em et.~al.}, {\it Neutrino oscillation parameters from Minos,
  Icarus and Opera combined},  {\em Phys. Rev.} {\bf D65} (2002) 053016,
  [\href{http://xxx.lanl.gov/abs/hep-ph/0110393}{{\tt hep-ph/0110393}}].

\bibitem{HyperK}
See e.g. M.~Shiozawa, talk given at the ``International Workshop on a Next
  Generation Long-Baseline Neutrino Oscillation Experiment"; transparecies
  available at http://neutrino.kek.jp/jhfnu/workshop2/ohp.html.

\bibitem{Jung:1999jq}
C.~K. Jung, {\it Feasibility of a next generation underground water cherenkov
  detector: UNO},  \href{http://xxx.lanl.gov/abs/hep-ex/0005046}{{\tt
  hep-ex/0005046}}.

\bibitem{UNO}
See also M.Vagins, talk given at the {\it "Conference on Underground Science"},
  Lead, South Dakota (USA), October 4-7, 2001, transparencies available at
  http://mocha.phys.washington.edu/int\_talk/NUSL/2001/People?Vagins\_M.

\bibitem{Suzuki:2001rb}
{\bf TITAND Working Group} Collaboration, Y.~Suzuki {\em et.~al.}, {\it
  Multi-megaton water cherenkov detector for a proton decay search: Titand
  (former name: Titanic)},  \href{http://xxx.lanl.gov/abs/hep-ex/0110005}{{\tt
  hep-ex/0110005}}.

\bibitem{Beacom:1998ya}
J.~F. Beacom and P.~Vogel, {\it Mass signature of supernova nu/mu and nu/tau
  neutrinos in Superkamiokande},  {\em Phys. Rev.} {\bf D58} (1998) 053010,
  [\href{http://xxx.lanl.gov/abs/hep-ph/9802424}{{\tt hep-ph/9802424}}].

\bibitem{Beacom:1998yb}
J.~F. Beacom and P.~Vogel, {\it Mass signature of supernova nu/mu and nu/tau
  neutrinos in the Sudbury neutrino observatory},  {\em Phys. Rev.} {\bf D58}
  (1998) 093012, [\href{http://xxx.lanl.gov/abs/hep-ph/9806311}{{\tt
  hep-ph/9806311}}].

\bibitem{waltham}
See e.g. C.~Waltham, proceedings from the International Conference on Cosmic
  Rays (ICRC) 2001, available at
  www.copernicus.org/icrc/papers/ici7144\_p.pdf~.

\bibitem{Aglietta:2001jf}
M.~Aglietta {\it et al.},
{\it Effects of neutrino oscillations on the supernova signal in LVD},
{\em Nucl. Phys. Proc. Suppl.} {\bf 110}, 410 (2002)
[\href{http://xxx.lanl.gov/abs/astro-ph/0112312}{{\tt astro-ph/0112312}}].

\bibitem{Arneodo:2001tx}
F.~Arneodo {\it et al.}  [ICARUS collaboration],
{\it The ICARUS experiment, a second-generation proton decay experiment and  neutrino observatory at the Gran Sasso Laboratory},
\href{http://xxx.lanl.gov/abs/hep-ex/0103008}{{\tt hep-ex/0103008}}. 

\bibitem{Strumia:bx}
A.~Strumia and F.~Vissani,
{\it Massive Neutrinos And Theoretical Developments},
{\em Int. J. Mod. Phys. A} {\bf 17}, 1755 (2002).

\bibitem{Cline:2001pt}
D.~B.~Cline, F.~Sergiampietri, J.~G.~Learned and K.~McDonald,
{\it LANNDD: A massive liquid argon detector for proton decay, supernova and  solar neutrino studies, and a neutrino factory detector},
\href{http://xxx.lanl.gov/abs/astro-ph/0105442}{{\tt astro-ph/0105442}}.


\bibitem{Nakamura:2000vp}
S.~Nakamura, T.~Sato, V.~Gudkov, and K.~Kubodera, {\it Neutrino reactions on
  deuteron},  {\em Phys. Rev.} {\bf C63} (2001) 034617,
  [\href{http://xxx.lanl.gov/abs/nucl-th/0009012}{{\tt nucl-th/0009012}}].


\bibitem{Vogel:1999zy}
P.~Vogel and J.~F. Beacom, {\it The angular distribution of the reaction $\bar
  \nu_e + p \rightarrow e^+ + n$},  {\em Phys. Rev.} {\bf D60} (1999) 053003,
  [\href{http://xxx.lanl.gov/abs/hep-ph/9903554}{{\tt hep-ph/9903554}}].

\bibitem{Beacom:2001hr}
J.~F. Beacom and S.~J. Parke, {\it On the normalization of the neutrino
  deuteron cross section},  {\em Phys. Rev.} {\bf D64} (2001) 091302,
  [\href{http://xxx.lanl.gov/abs/hep-ph/0106128}{{\tt hep-ph/0106128}}].


\bibitem{Strumia:2003zx}
A.~Strumia and F.~Vissani,
{\it Precise quasielastic neutrino nucleon cross section }, {\tt astro-ph/0302055}.
%%CITATION = ASTRO-PH 0302055;%%

\bibitem{Akhmedov:2002zj}
E.~K. Akhmedov, C.~Lunardini, and A.~Y. Smirnov, {\it Supernova neutrinos:
  Difference of nu/mu - nu/tau fluxes and conversion effects},  {\em Nucl.
  Phys.} {\bf B643} (2002) 339--366,
  [\href{http://xxx.lanl.gov/abs/hep-ph/0204091}{{\tt hep-ph/0204091}}].


\bibitem{Lunardini:2001pb}
C.~Lunardini and A.~Y. Smirnov, {\it Supernova neutrinos: Earth matter effects
  and neutrino mass spectrum},  {\em Nucl. Phys.} {\bf B616} (2001) 307--348,
  [\href{http://xxx.lanl.gov/abs/hep-ph/0106149}{{\tt hep-ph/0106149}}].

\bibitem{Takahashi:2000it}
K.~Takahashi, M.~Watanabe, and K.~Sato, {\it The earth effects on the supernova
  neutrino spectra},  {\em Phys. Lett.} {\bf B510} (2001) 189--196,
  [\href{http://xxx.lanl.gov/abs/hep-ph/0012354}{{\tt hep-ph/0012354}}].

\bibitem{Takahashi:2001dc}
K.~Takahashi and K.~Sato, {\it Earth effects on supernova neutrinos and their
  implications for neutrino parameters},  {\em Phys. Rev.} {\bf D66} (2002)
  033006, [\href{http://xxx.lanl.gov/abs/hep-ph/0110105}{{\tt
  hep-ph/0110105}}].

\bibitem{PREM}
A.~M.~Dzewonski and D.~L.~Anderson, Phys.\ Earth.\ Planet. \ Inter. {\bf 25}
  (1981) 297.

\bibitem{Takahashi:2002yj}
K.~Takahashi, K.~Sato, H.~E. Dalhed, and J.~R. Wilson, {\it Shock propagation
  and neutrino oscillation in supernova},
  \href{http://xxx.lanl.gov/abs/astro-ph/0212195}{{\tt astro-ph/0212195}}.

\bibitem{CAprep}
A.~Friedland and C.~Lunardini, in preparation.

\end{thebibliography}\endgroup

%%%%%%%%%%%%%%%%%%%%%%%%%%%%%%%%%%%%%%%%%%%%%%%%%%%%%%%%%%%%%%%%%%%%%%%%%%

%%%%%%%%%%%%%%% FIGURES %%%%%%%%%%%%%%%%%%%%%%%%%%%%%%%%%%%%

%%%%%%%%%%%%%%%%%%%%%%%%%%%%%%%%%%%%%

%%%%%%%%%%%%%%%%%%%%%%%%%%%%%%%%%%%%%%%%%%%%%%%%%%%%%%%%%%%%%%%%%%
%%%%%%%%%%%%%%%%%%%%%%%%%%%%%%%%%%%%%%%%%%%%%%%%%%%%
%%%%%%%%%%%%%%%%%%%%%%%%%%%%%%%%% 

%%%%%%%%%%%%%%%%%%%%%%%%%%%%%%%%%%%%%%%%%%%%%%%%%%%%%%%%%%%%%%%%%%
\end{document}